\documentclass[12pt]{article}
\pdfoutput=1

\usepackage{putex}
\usepackage{amsmath,amssymb,amsfonts}
\usepackage{psfrag}
\usepackage{footmisc}
\usepackage{url}
\usepackage{mathtools}
\usepackage{color}

\definecolor{darkblue}{rgb}{0.1,0.1,.7}
\definecolor{purple}{rgb}{0.6,0,0.6}
\definecolor{orange}{rgb}{0.9,0.6,0}
\usepackage[colorlinks, linkcolor=darkblue, citecolor=darkblue, urlcolor=darkblue, linktocpage]{hyperref} 
\usepackage[square, comma, compress,numbers]{natbib}
\usepackage[]{amsmath}
\usepackage[]{graphicx}
\usepackage[]{latexsym}
\usepackage[utf8]{inputenc}
\usepackage{geometry}
\usepackage{amscd}
\usepackage[all,cmtip]{xy}
\usepackage{mathrsfs}
\usepackage[margin=10pt,font=small,labelfont=bf]{caption}
\usepackage{dsdshorthand}
\usepackage{changepage}
\usepackage{setspace}
\usepackage{tikz}
\usepackage{subcaption}

\usepackage[titles]{tocloft}
\usetikzlibrary{arrows,positioning}

\def\ma{{\mathbf \alpha}}

\def\SL2{\widetilde{SL}(2,\mathbb R)}

\def\mC{\mathcal C}

\newcommand{\es}[2] {\begin{equation} \label{#1} \begin{split} #2 \end{split} \end{equation}}

\newcommand\mR{\mathbb{R}}

\numberwithin{equation}{section}

\newcommand{\grp}[1]{\mathrm{#1}}

\newcommand{\tie}{\tilde e}

\newcommand {\bes} {\begin {equation*}}
\newcommand {\ees} {\end {equation*}}

\renewcommand{\es}[2] {\begin{equation} \label{#1} \begin{split} #2 \end{split} \end{equation}}

\newcommand{\Sch}{\text{Sch}}

\numberwithin{equation}{section}

\def\<{\langle}
\def\>{\rangle}
\def\sT{{\sf T}}

\tikzset{
    >=stealth',
    punkt/.style={
           rectangle,
           rounded corners,
           draw=black, very thick,
           text width=15em,
           minimum height=2em,
           text centered},
    pil/.style={
           ->,
           thick,
           shorten <=2pt,
           shorten >=2pt,}
}

 \def\ie{\begin{equation}\begin{aligned}}
\def\fe{\end{aligned}\end{equation}}

\begin{document}

\begin{flushright}
\hfill{\tt PUPT-2598}
\end{flushright}

\institution{PU}{Joseph Henry Laboratories, Princeton University, Princeton, NJ 08544, USA}

\title{
On 2D gauge theories in Jackiw-Teitelboim gravity
}

\authors{Luca V. Iliesiu}

\abstract{
The low-energy behavior of near-extremal black holes can be understood from the near-horizon $AdS_2$ region. In turn, this region is effectively described by using Jackiw-Teitelboim gravity coupled to Yang-Mills theory through the two-dimensional metric and the dilaton field. We show that such a two-dimensional model of gravity coupled to gauge fields is soluble for an arbitrary choice of gauge group and gauge couplings. Specifically, we determine the partition function of the theory on two-dimensional surfaces of arbitrary genus and with an arbitrary number of boundaries. When solely focusing on the contribution from surfaces with disk topology, we show that the gravitational gauge theory is described by the Schwarzian theory coupled to a particle moving on the gauge group manifold. When considering the contribution from all genera, we show that the theory is described by a particular double-scaled matrix integral, where the elements of the matrix are functions that map the gauge group manifold to complex or real numbers.  Finally, we compute the expectation value of various diffeomorphism invariant observables in the gravitational gauge theory and find their exact boundary description.  
}
\date{}

\maketitle
\tableofcontents

\newpage

\section{Introduction }
\label{sec:intro}

 The geometry of the near-horizon region in near-extremal black holes is universal: as we approach the horizon there is an $AdS_2$ throat with a slowly varying internal space. The low-energy behavior of such black holes is expected to arise from the near-horizon region which, in turn, can be captured by a two-dimensional effective gravitational action coupled to Yang-Mills theory  \cite{Almheiri:2016fws, Anninos:2017cnw, Sarosi:2017ykf, Nayak:2018qej, Moitra:2018jqs, Hadar:2018izi, Moitra:2019bub, Sachdev:2019bjn}\footnote{It is instructive to consider how the action \eqref{eq:JT-action+yang-mills} arises from the dimensional reduction to $AdS_2$ in a specific example of near-extremal black holes. The dimensional reduction of the near-horizon region in  Reissner-Nordstr\"{o}m black holes in flat space is discussed extensively in the review \cite{Sarosi:2017ykf}. The inclusion  (in asymptotically flat or $AdS_4$ space) of the Maxwell field under which the black hole is charged is discussed in \cite{Almheiri:2016fws, Moitra:2018jqs,Sachdev:2019bjn}, while the addition of the massless gauge degrees of freedom appearing due to the isometry of the $S^2$ internal space is discussed in \cite{Anninos:2017cnw, Moitra:2018jqs}.
}
\be
\label{eq:JT-action+yang-mills}
S^E_{JTYM}= &-\frac{1}2\Phi_0 \int_\cM d^2 x \sqrt{g} \,\cR   -\frac{1}2 \int_{\cM} d^2x \sqrt{g}\,\Phi(\cR + 2) \nn\\ & - \int_\cM d^2 x  \sqrt{g}\, g^{\mu \eta} g^{\nu \rho} \left(\frac{1}{4e^2} + \frac{\Phi_0 + \Phi}{4e_{\Phi}^2\Phi_0 }\right)\tr F_{\mu \nu} F_{\eta \rho} +  S_\text{boundary}(g, \Phi, A)\,.
\ee
The action \eqref{eq:JT-action+yang-mills} captures all the massless degrees of freedom that can generically arise in such an effective description.\footnote{Through-out this paper we solely work with the action \eqref{eq:JT-action+yang-mills} written in Euclidean signature. Above, $g_{\mu \nu}$ is the metric, $\cR$ is the scalar curvature and $F_{\mu \nu}$ is the field strength associated to the gauge field $A_\mu$. Further details about the conventions in \eqref{eq:JT-action+yang-mills} will be discussed in the beginning of section \ref{sec:preliminaries}. Details about the integration contours for the fields are also discussed in that section and the meaning of the $\Phi$ integral contour in the context of the low energy effective action of near-extremal black holes is discussed in footnote~\footref{footnote:integration-contour-for-phi}. } The first line in \eqref{eq:JT-action+yang-mills} describe the bulk terms in pure Jackiw-Teitelboim (JT) gravity \cite{teitelboim1983gravitation, Jackiw:1984je}, with a cosmological constant normalized to $\Lambda = -2$. The dilaton $\Phi_0+\Phi$ parametrizes the size of the internal space and is split into two-parts: $\Phi_0$ parametrizes the size of the internal space at extremality, while $\Phi$ gives the deviation from this values. While generically, the dimensional reduction on the internal space gives rise to a more complicated dependence in the action of the dilaton field $\Phi_0 + \Phi$, because we are solely interested in describing the near-horizon region close to extremality, we may assume that $\Phi \ll \Phi_0$. Consequently, we can linearize the potential for the dilaton field to obtain the effective gravitational action \eqref{eq:JT-action+yang-mills} which is linear in the deviation $\Phi$. 

The gauge fields that appear in \eqref{eq:JT-action+yang-mills} through the field strength $F= dA - A \wedge A$ have two possible origins: (i) they are present in the higher dimensional gravitational theory, and the near-extremal black hole could, for instance, be charged under them; for example, for Reissner-Nordstr\"{o}m black holes in $AdS_4$ or in flat space, the $\grp{U}(1)$ Maxwell field under which the black hole is charged is also present in the dimensionally reduced theory; (ii) the fields can arise from the dimensional reduction on the internal space, in which case, the gauge group is given by the isometry of this space; including such degrees of freedom in the effective action describes the behavior of the black hole beyond the S-wave sector \cite{Moitra:2018jqs}.\footnote{ Depending on the origin of the gauge fields, the couplings $e$ and $e_\Phi$ can be related to the value of the dilaton at extremality $\Phi_0$.}

Beyond appearing in the effective action that describes the dimensional reduction of the near-horizon region of such black holes, pure JT gravity serves as a testbed for ideas in 2D/1D holography and quantum gravity \cite{kitaevTalks, Maldacena:2016hyu, Polchinski:2016xgd, Gross:2017vhb, Gross:2017aos, Stanford:2017thb,Mezei:2017kmw, Witten:2016iux, Klebanov:2016xxf, Maldacena:2016upp, Jensen:2016pah, Engelsoy:2016xyb, Bagrets:2016cdf,  Maldacena:2017axo,Mertens:2017mtv, Maldacena:2018lmt, Harlow:2018tqv, Lam:2018pvp, Kitaev:2018wpr, yang2018quantum, Saad:2019lba, Mertens:2019tcm, Maldacena:2019cbz, Iliesiu:2019xuh, Stanford:2019vob}. For instance, when solely isolating contributions from surfaces with disk topology, the quantization of pure JT gravity can be shown to be equivalent to that of the Schwarzian theory \cite{Maldacena:2016upp, Saad:2019lba}; in  turn, this 1d model arises as the low-energy limit of the SYK model \cite{Sachdev:1992fk, georges2001quantum,kitaevTalks,Maldacena:2016hyu}. When considering the quantization of the gravitational theory on surfaces with arbitrary topology, the partition function of the theory can be shown to agree with the genus expansion of a certain double-scaled matrix integral \cite{Saad:2019lba, Stanford:2019vob}. The solubility of pure JT gravity is due, in part, to the fact that the bulk action  can be re-expressed as a topological field theory \cite{Isler:1989hq, Chamseddine:1989yz,Jackiw:1992bw,Saad:2019lba,Iliesiu:2019xuh}. Consequently, all bulk observables in the purely gravitational theory are invariant under diffeomorphisms and can oftentimes be shown to be equivalent to boundary observables directly at the level of the path integral. The addition of the Yang-Mills term in \eqref{eq:JT-action+yang-mills} provides an additional layer of complexity for a theory of 2d quantum gravity since the bulk action is no longer topological. Consequently, there is a  richer set of diffeomorphism invariant observables that could be explored in the bulk.

In this paper, we present an exact quantization of the gravitational theory \eqref{eq:JT-action+yang-mills}, for an arbitrary choice of gauge group $G$ and gauge couplings, $e$, and $e_\Phi$. By combining techniques used to quantize pure JT gravity and the Schwarzian theory \cite{teitelboim1983gravitation, Jackiw:1984je, Saad:2019lba, Iliesiu:2019xuh}, together with known results from the quantization of 2D Yang-Mills \cite{migdal1975phase, Migdal:1984gj,Blau:1991mp, Witten:1991we, Fine:1991ux,  Witten:1992xu, Cordes:1994fc, ganor1995string}, we derive the partition function of the new gravitational gauge theory \eqref{eq:JT-action+yang-mills} for surfaces with arbitrary genus.  While in this paper we mainly focus on performing the gravitational path integrals over orientable manifolds, our derivation can be easily generalized to the unorientable cases discussed in \cite{Stanford:2019vob}, and we outline the ingredients necessary for this generalization. 

The derivation of the partition function depends on the choice of boundary conditions for the metric, dilaton and gauge field. In turn, this choice fixes the boundary term $S_\text{boundary}(g, \Phi, A)$ needed in order for \eqref{eq:JT-action+yang-mills} to have a well-defined variational principle. For the metric and dilaton field, we solely set asymptotically $AdS_2$ Dirichlet boundary conditions~\cite{Stanford:2017thb}, 
\be 
\label{eq:boundary-conditions-for-dilaton}
g_{uu}|_{\text{bdy.}} = \frac{1}{\epsilon^2}, \qquad \Phi|_{\text{bdy.}}= \frac{\Phi_b}\e\,,
\ee 
where $u \in [0, \,\b]$ is a variable that parametrizes the boundary, whose total proper length is fixed, $\int_0^\beta du \sqrt{g_{uu}} = \b/\e$. In this paper, we analyze the limit $\e \to 0$ which implies that we are indeed considering surfaces which are asymptotically $AdS_2$.\footnote{An analysis for any value of $\epsilon$ is forthcoming \cite{Iliesiu:toAppear}.} However, for the gauge field, we study a variety of boundary conditions for which the gravitational gauge theory \eqref{eq:JT-action+yang-mills} will prove to be dual to different soluble 1d systems.

Specifically, when solely focusing on the contribution to the path integral of surfaces with disk topology, we find that with the appropriate choice of boundary conditions for the gauge field, the theory \eqref{eq:JT-action+yang-mills} is equivalent to the Schwarzian theory coupled to a particle moving on the gauge group manifold. Based on symmetry principles, one expects such a theory to arise in the low energy limit of SYK or tensor models with global symmetries \cite{Sachdev:2015efa, Davison:2016ngz, Gross:2016kjj, Fu:2016vas, Narayan:2017qtw, Yoon:2017nig, Narayan:2017hvh, Klebanov:2018nfp, Liu:2019niv}. For instance, the low-energy limit of  the complex SYK model with a $\grp{U}(1)$ global symmetry can be described by the Schwarzian coupled to a $\grp{U}(1)$ phase-mode \cite{Sachdev:2015efa, Davison:2016ngz, Sachdev:2019bjn}; on the gravitational side, such a theory arises from \eqref{eq:JT-action+yang-mills} when fixing the gauge group to be $\grp{U}(1)$ \cite{ Anninos:2017cnw,  Moitra:2018jqs, Sachdev:2019bjn}. 

When considering the path integral over surfaces with arbitrary genus, we find that the  partition function of the gravitational gauge theory can equivalently be described in terms of a collection of double-scaled matrix integrals. Each matrix is associated with a unitary irreducible representation of the gauge group, and the size of that matrix is related to the dimension of its associated representation. Yet another equivalent description of this matrix integral, and consequently of the gravitational theory, can be obtained by considering Hermitian matrices whose elements are not regular complex numbers,\footnote{In the case in which the path integral is solely over orientable manifolds.} but instead are  functions which map group elements of $G$ to complex numbers.  Such matrix elements are given by the complex group algebra $\mathbb C[G]$.\footnote{We thank H. Verlinde for providing the unpublished pre-thesis work of A. Solovyov \cite{Solovyov} and for suggesting the useful mathematical references \cite{mulase2002generating, mulase2005non, mulase2007geometry}. While these works focus on an analysis of matrix integrals in the case of discrete groups, they proved to be a valuable source of inspiration for our analysis of gauge theories whose gauge groups are compact Lie groups. } This construction can easily be extended to include the contribution of unorientable manifolds by studying the same matrix integral, this time considering symmetric matrices whose elements are functions mapping group elements of $G$ to real numbers (i.e., the real group algebra $\mathbb R[G]$).

Beyond, our computation of partition functions, we construct several diffeomorphism invariant bulk observables, compute their expectation value in the weakly coupled limit and discuss their boundary dual. One such observable is obtained by coupling the gauge field to the world-line action of a charged particle (for instance, a quark) moving on the surface $\cM$ in \eqref{eq:JT-action+yang-mills}. The resulting operator is a generalization of the Wilson lines from pure Yang-Mills theory to a non-local diffeomorphism invariant operator in the gravitational gauge theory \eqref{eq:JT-action+yang-mills}.  Studying such observables is crucial for understanding the coupling of  \eqref{eq:JT-action+yang-mills} to charged matter. From the perspective of the effective theory describing the aforementioned black holes, such charged matter fields can arise from the Kaluza-Klein reduction on the internal space and can play an essential role in the low-energy behavior of near-extremal black holes \cite{Moitra:2018jqs}.

The remainder of this paper is organized as follows. In section \ref{sec:preliminaries} we discuss the preliminaries needed for the quantization of the theory with action \eqref{eq:JT-action+yang-mills}. As a warm-up problem which emphasizes the role of boundary conditions in the gauge theory, we start by discussing the simple case in which the gauge theory is weakly coupled. In section \ref{sec:genus-zero-part-function} we move on to discuss the case of general coupling, compute the partition function of the gravitational gauge theory on surfaces with disk topology, and describe the dual boundary theory. In section \ref{sec:higher-genus-partition-function}, we compute the partition function of the gravitational theory on surfaces with arbitrary genus, $g$, and an arbitrary number of boundaries, $n$. Next, we show how this result can be obtained from the genus expansion of the previously introduced matrix integrals. We discuss the construction of several diffeomorphism invariant observables in section \ref{sec:observables} and compute their expectation values in a variety of scenarios. Finally, in section \ref{sec:discussion} we summarize our results and discuss future research directions.

\textit{Note}: During the development of this paper we became aware about the ongoing work \cite{Kapec:toAppear} which has some overlap with our results.

\section{Preliminaries and a first example}

\label{sec:preliminaries}

Before, proceeding with the quantization of theory \eqref{eq:JT-action+yang-mills}, we first recast the bulk action into a more convenient form, by introducing the field $\phi$ as a $G$-adjoint valued zero-form \cite{Witten:1991we}. The path integral associated to the action  \eqref{eq:JT-action+yang-mills} can be rewritten as:
\be
\label{eq:JTandYMactions}
Z_{JTYM}& = \int Dg_{\mu\nu} D\Phi DA\, e^{-S_E[\Phi, g^{\mu\nu}, A]} \nn \\
&= \int Dg_{\mu \nu} D\Phi D\phi DA\,\,\, \exp\bigg[\frac{1}2\Phi_0 \int_\cM d^2x \sqrt{g}\, \cR   +\frac{1}2 \int_{\cM} d^2x \sqrt{g}\, \Phi(\cR + 2) \nn \\ &\hspace{0.6cm}+ \int_\cM i \tr \phi F + \frac{1}2\int_\cM d^2 x  \sqrt{g} \left( \frac{e^2 e_{\Phi}^2}{e^2 (1 + \frac{\Phi}{\Phi_0})+ e_{\Phi}^2}\right) \tr \phi^2 + S_\text{boundary}(g, \Phi, A) \bigg]\,.
\ee
Throughout the paper, we use $\tr(\dots)$  to denote the trace in the fundamental representation of the group $G$. The trace in the fundamental representation can be explicitly expressed in terms of the $G$ generators $T^i$, normalized such that $\tr(T^i T^j) = \cN\eta^{ij}$, where $\cN$ is the Dynkin index and $\eta^{ij}$ is chosen such that $\eta^{ij} = \diag(-1,\, \dots,\,-1)$. The scalar curvature $\cR$ should not be confused with the notation $R$ for unitary irreducible representations of $G$ which will be used shortly.  The trace in all representations $R$ of the gauge group $G$ is denoted by $\chi_R(\dots)$. 

After (once again) considering the limit in which $\Phi \ll \Phi_0$, the action appearing in \eqref{eq:JTandYMactions} can be rewritten as, 
\be 
\label{eq:JTandYMactionrewritten}
S^E_{JTYM} &= -\frac{1}2\Phi_0 \int_\cM \cR   -\frac{1}2 \int_{\cM} \Phi(\cR + 2) - \int_\cM i \tr \phi F -\frac{1}2 \int_\cM d^2 x  \sqrt{g} \left(\tilde e - \tilde e_\Phi \Phi \right) \tr \phi^2\nn \\  &+ S_\text{boundary}(g, \Phi, A) \,,
\ee
where, 
\be
\label{eq:coupling-definitions}
\tilde e \equiv \frac{e^2 e_\Phi^2}{e^2 + e_{\Phi}^2}\,, \qquad \tilde e_\Phi \equiv \frac{e_\Phi^2 e^4}{\Phi_0(e^2 + e_\Phi^2)^2}\,.
\ee 
In the remainder of this paper we solely use  $\tilde e $ and $\tilde e_\Phi$ and  we will quantize the theory \eqref{eq:JTandYMactionrewritten} without making any assumptions about these two gauge couplings.  

As previously mentioned, in order to compute the partition function \eqref{eq:JTandYMactions} we need to specify the boundary term $S_\text{boundary}(g, \Phi, A)$ which is needed in order for the theory to have a well-defined variational principle. When considering the boundary condition \eqref{eq:boundary-conditions-for-dilaton} for the metric and the dilaton field, one needs to include a Gibbons-Hawking term in $S_\text{boundary}(g, \Phi, A) \supseteq   - \left[ \Phi_0 \int_{\partial \cM} du \sqrt{g_{uu}}\,\cK + \int_{\partial \cM} du \sqrt{g_{uu}} \,\Phi(\cK-1)\right]$. Here, $\cK$ is the boundary extrinsic curvature.

For the gauge field, we can, for instance, consider Dirichlet boundary conditions, in which we fix the value of the gauge field along the boundary, $\delta A_{u} = 0$. Equivalently, due to the invariance of the partition function under large gauge transformations, instead of fixing  $A_{u}|_{\partial \cM}= \cA_u(u)$ all along the boundary,\footnote{Here, we take $\cA_u(u)$ to be an arbitrary periodic function on the thermal circle. } we solely need to fix the holonomy around the boundary\footnote{We, however, need to fix gauge transformations on the boundary in section \ref{sec:wilson-loops-in-BF-theory}, when discussing correlators of boundary anchored Wilson lines. }
\be
\label{eq:holonomy-definition}
h \equiv \cP \exp\left(\oint_{\partial \cM}\cA^a \,T_a\right)\qquad \qquad \text{(Dirichlet)} \,.
\ee
As we will explain shortly, the states obtained by performing the path integral on surfaces with disk topology and fixed boundary holonomy $h$, span the entire Hilbert space associated to Yang-Mills theory; as we exemplify shortly, we can always compute correlators in the presence of a different set of boundary conditions for the gauge field, by inserting a boundary condition changing defect \cite{Iliesiu:2019xuh} in the theory with Dirichlet boundary. 

With Dirichlet boundary conditions for the gauge field and the boundary conditions \eqref{eq:boundary-conditions-for-dilaton} for the metric and dilaton, no other boundary term besides the Gibbons-Hawking term is needed in order for the theory to have a well-defined variational principle. Thus, the action \eqref{eq:JTandYMactionrewritten} can finally be recasted as, 
\be 
\label{eq:JTandYMaction-with-bdy} 
S^E_{\substack{\text{JTYM}\\\text{Dirichlet}}} = &- 2\pi\,\Phi_0\chi(\cM) -\left[\frac{1}2 \int_{\cM} d^2 x \sqrt{g}\, \Phi(\cR + 2)+\int_{\partial \cM} du \sqrt{g_{uu}} \Phi(\cK-1)\right] \nn \\ &-\left[ \int_\cM i \tr \phi F + \frac{1}2\int_\cM d^2 x  \sqrt{g} \left(\tilde e - \tilde e_\Phi \Phi \right) \tr \phi^2\right]\,,
\ee 
where $\chi(\cM)$ is the Euler characteristic of the manifold $\cM$, which appears due to the Gauss-Bonnet relation $\frac{1}2 \int_\cM \sqrt{g} \cR + \int_{\partial \cM} \cK = 2\pi \chi(\cM)$.  From here on, we denote $S_0 = 2\pi \Phi_0$ and $e^{S_0}$  serves as the genus expansion parameter when discussing path integral over surfaces with arbitrary genus.

Our goal is thus to quantize the theory with action \eqref{eq:JTandYMaction-with-bdy} and theories related to  \eqref{eq:JTandYMaction-with-bdy}  by a change of boundary conditions for the gauge field. Towards that scope, it is first useful to discuss the symmetries of the problem in the weak gauge coupling limit $\tilde e$ and $\tilde e_\Phi \to 0$. In this case the theory becomes topological: the third-term in the action \eqref{eq:JTandYMactionrewritten} describes a BF topological theory and in fact, as previously mentioned, the bulk JT gravity action itself, can also be recast as a BF theory whose gauge algebra is $\mathfrak{sl}(2, \mR)$ \cite{Isler:1989hq, Chamseddine:1989yz,Jackiw:1992bw,Saad:2019lba,Iliesiu:2019xuh}. This limit proves useful for understanding the boundary dual of the gravitational theory in a simpler setting and for the computation of various diffeomorphism invariant observables in section \ref{sec:observables}. Therefore, as a warm-up, we discuss it first in the next subsection.

\subsection{A warm up: the weakly coupled limit on the disk topology}
\label{sec:genus-zero-BF-plus-defect}

Because in the weakly coupled limit, the gauge theory is topological, we can proceed by separately computing the path integral for the pure JT sector and the gauge theory sector. Thus, we first review the computation of the path integral in JT gravity following \cite{Maldacena:2016upp, Saad:2019lba}. By integrating out the dilaton field $\Phi$ along the contour $\Phi =\Phi_b/\e  + i \mR$,\footnote{To understand the meaning of this contour in the context of the near-extremal black hole effective action it is useful to review how the integral over $\Phi$ behaves in Lorentzian signature. In that case, the contour for $\Phi$ is restricted from $-\Phi_0$ to $\infty$, due to the fact that the internal space should have a positive volume $(\Phi+\Phi_0 > 0)$.  In the limit considered in this paper, $\Phi_0 \to \oo$, the integral over $\Phi$ indeed converges to $\delta(\cR+2)$ in a distributional sense. To make this statement precise we could keep track of the higher powers of the dilaton in the action, whose coefficients are suppressed in $\Phi_0$, and vanish in the limit $\Phi \to \pm \infty$. Then, the path integral over $\Phi$ would be peaked around the configurations where $\cR = -2+O(1/\Phi_0)$. When in Euclidean signature, we have to analytically continue $\Phi$ along the complex axis in order to get a convergent answer, still peaked around $\cR = -2+ O(1/\Phi_0)$.  While such a contour for $\Phi$ does not have a nice geometric meaning when relating $\Phi+\Phi_0$ to the volume of the internal space, it isolates the same type of constant curvature configurations in Euclidean signature as those that dominate in the Lorentzian path integral. We thank R.~Mahajan and D.~Kapec for useful discussions about this point. \label{footnote:integration-contour-for-phi}} we find that the curvature of the surfaces considered in the path integral is constrained:
\be
\label{eq:JT-path-integal-R-constrain}
Z_{JT} = \int Dg_{\mu \nu} e^{\int_{\partial{\cM}} du \sqrt{g_{uu}} \frac{\Phi_b}{\epsilon} \cK[g_{\mu\nu}]}\, \delta(\cR+2)\,.
\ee
The remaining path integral is thus solely over the boundary degrees of freedom of $AdS_2$ patches. In order to simplify the path integral over the boundary degrees of freedom, we consider parametrizing the $AdS_2$ patches by using Poincar\'e coordinates, under which the boundary condition for the metric becomes
\be
\label{eq:metric-and-boundary-condition-for-metric}
ds^2 = \frac{d F^2 + dz^2}{z^2}\,, \qquad g_{uu}|_\text{bdy.} = \frac{ (F')^2 + (z')^2}{z^2} = \frac{1}{\e^2}\,,  
\ee
where  the boundary is parametrized using the variable $u$, with $F' = \partial F/\partial u$.  Solving the latter equation to first order in $\epsilon$, we find $z = \e F' + O(\epsilon^2)$. Since $z(u)$ is small in the $\e\to 0$ limit, the path integral is thus indeed dominated by asymptotically $AdS_2$ patches. In this set of coordinates, the extrinsic curvature can be expressed as 
\be 
\label{eq:extrinsic-curvature-Schw-field}
\cK[F(u), z(u)] = \frac{F'(F'^2 +z'^2 + z z'')-z z' F''}{(F'^2+z'^2)^{3/2}} = 1+ \e^2 \,\text{Sch}(F, u) + O(\epsilon^3)\,.
\ee
Thus, \eqref{eq:JT-path-integal-R-constrain} can be rewritten as a path integral over the boundary coordinate $F(u)$
\be 
\label{eq:Schwarzian-path}
Z_\text{JT}^\text{disk}(\Phi_b, \b)  &=  Z_\text{Schw.}(\Phi_b, \b)= e^{S_0}\int DF\, e^{{\Phi_b}\int_0^\beta\{F(u), u\}}\,, \qquad   D F = \prod_{u \in \partial \cM} \frac{dF(u)}{F'(u)}\,.
\ee 
where the measure $DF $ is obtained by using the symplectic form over flat gauge connections in the $\mathfrak{sl}(2, \mR)$ BF theory rewriting of JT gravity \cite{Saad:2019lba}. The path integral  \eqref{eq:Schwarzian-path} can be computed by using localization and has been found to be one-loop exact \cite{Stanford:2017thb}. The solution obtained from localization is given by
\be 
\label{eq:JT-path}
Z_\text{JT}^\text{disk}(\Phi_b, \b)  &=  Z_\text{Schw.}(\Phi_b, \b)= e^{S_0}\int ds\, \frac{ s}{2\pi^2} \sinh(2\pi s) e^{-\frac{\beta s^2}{2\Phi_b}}=  e^{S_0}\frac{\Phi_b^{3/2} e^{\frac{2\pi^2 \Phi_b}\b}}{(2\pi)^{1/2} \b^{3/2}}\,,
\ee 
where one can consequently read-off the density of states for the Schwarzian theory:
\be 
\rho_0(E) = \frac{\Phi_b}{2\pi^2} \sinh(2\pi \sqrt{2 \Phi_b E})\,.
\ee

We now move on to describing the gauge theory side. With Dirichlet boundary conditions, the disk partition function is trivial, $Z_{BF}(h) = \delta(h)$ and, consequently, $Z_{JTBF}(h) = Z_\text{Schw.} \delta(h)$. In order to obtain a non-trivial result, the boundary conditions imposed on the gauge field need to explicitly break invariance under arbitrary diffeomorphisms in the topological theory.  One such boundary condition is obtained by relating the value of the gauge field on the boundary to the zero-form field $\phi$
\be 
\label{eq:diff-boundary-cond-A}
A_u|_{\partial \cM}  - \sqrt{g_{uu}}   i \e \tilde e_b \phi |_{\partial \cM} = \cA_u \qquad \qquad \text{(mixed)} \,,
\ee
for some constant $\cA_u$. We label this class of boundary conditions as ``mixed''. 

In order for the action to have a well-defined variational principle, one needs to add
\be
\label{eq:gauge-field-boundary-term}
 S_\text{boundary}^\text{gauge} [\phi,  A] =\frac{i}2 \int_{\partial\cM}du \, \tr  \phi A_u \,,
\ee
to the aforementioned Hawking-Gibbons term specified in \eqref{eq:JTandYMaction-with-bdy}. As in pure JT gravity, we can reduce the BF path integral to an integral over boundary degrees of freedom, whose action is given by \eqref{eq:gauge-field-boundary-term}. The integral over the zero-form field $\phi$ in the bulk, restricts the path integral to flat gauge connections, with $A =  q^{-1} d  q$, where $q$ is a function mapping $\cM$ to group elements of $G$. Plugging in this solution for $A$ into the boundary term \eqref{eq:gauge-field-boundary-term} and using the boundary condition \eqref{eq:diff-boundary-cond-A}, we find that
\be 
\label{eq:BF-theory-path-integral-to-particle-on-G}
& Z_{\substack{\text{BF}\\ \text{mixed}}}^\text{disk}(\beta, h)= Z_G(\beta, h) =  \int Dq\, e^{\frac{1}{2\e \tilde e_b}\int_0^\b du \sqrt{g_{uu}} g^{uu}   \tr\left[ (q^{-1} \partial_u q)^2 + \cA_u  (q^{-1} \partial_u q) \right]}\,,\nn \\ 
& Z_{\substack{\text{JTBF}\\ \text{mixed}}}^\text{disk}(\Phi_b, \b, h) = Z_\text{Schw.}(\Phi_b, \b) Z_G(\b, h) \,.
\ee
Just like in the case of the pure JT gravity path integral, the measure for the boundary degree of freedom $Dh$ is obtained from the symplectic form in the BF theory with gauge group $G$.

The path integral in  \eqref{eq:BF-theory-path-integral-to-particle-on-G} describes a particle moving on the $G$ group manifold, whose partition function we denote as  $Z_G(\b,  \cA_u)$; as we will explain shortly, $\cA_u$ serves as a background gauge field for one of the $G$ symmetries present in this theory.

\subsection{Reviewing the quantization of a particle moving on a group manifold}
\label{sec:quantization-particle-on-G}

To proceed, we briefly review the quantization of a particle moving on a group manifold $G$ \cite{marinov1979dynamics, picken1989propagator, chu1994quantisation}, in the presence of an arbitrary 1d background metric and of a $G$ background gauge field.  In order to do so it is again useful to introduce a Lagrange multiplier  $\pmb \ma$, valued in the adjoint representation of $G$. The path integral \eqref{eq:gauge-field-boundary-term} can be rewritten  as 
\be
 \label{eq:introduce-Lagr-multiplier-a}
 Z_G(\b, h) = \int Dq D\pmb \ma\, e^{\int_0^\b du  \left(i \tr (\pmb \ma\, q^{-1}D_\cA q) + \sqrt{g_{uu}} \frac{\e\,\tilde e_b}2\tr \,\pmb \ma^2 \right)}\,, \qquad D_\cA q = \partial_u q+ q \cA_u\,.
\ee
At this point it proves useful to turn-off the background $ \cA_u$ and analyze the symmetries of the action appearing in \eqref{eq:introduce-Lagr-multiplier-a}. Firstly, we note that \eqref{eq:introduce-Lagr-multiplier-a} is invariant under reparametrizations, $u \to F(u)$ and thus, instead of using the variable $u$ we can also use the $AdS_2$ boundary coordinate $F(u)$ to describe the action in \eqref{eq:introduce-Lagr-multiplier-a}.\footnote{This is oftentimes done when discussing the low energy behavior of SYK models with global symmetries. For instance, this appears when coupling the Schwarzian to a phase mode \cite{Anninos:2017cnw,Yoon:2017nig, Moitra:2018jqs, Sachdev:2019bjn, Liu:2019niv}.}  Furthermore, for an arbitrary choice of parametrization of the boundary, such that $g_{uu}(u)$ is an arbitrary function of $u$, we can always perform a diffeomorphism and assume a constant boundary metric $g_{uu}$, as in the boundary condition \eqref{eq:boundary-conditions-for-dilaton}. Invariance under such diffeomorphisms also implies that the temperature dependence of the partition function appears as $Z_G(\tilde e_b,\, \b, \, \cA_u) = Z_G(\tilde e_b \b, \, \cA_u)$.

 Expanding $q(u)$ around a base-point, with $q(u) = e^{x^a(u)T_a} q(u_0) $ we find that the canonical momenta associated to $x^a(u)$ in the action in \eqref{eq:introduce-Lagr-multiplier-a} are given by
\ie
\pi_{x^i} = \tr (T_i q {\pmb \alpha} q^{-1})\,,
\fe
which are in fact the generators of the $G$ symmetry which acts by left multiplication on $q$, as $q \rightarrow U q$ and $\pmb \alpha \to \pmb \alpha$. Similarly, one finds that the generators of the $G$ symmetry that acts by right multiplication on $q$, as $q \to q \, U$  and $\pmb \a \to  U^{-1}\pmb \a U$ are simply given by $\pmb \alpha_i$. The background $\cA_u$, which appeared in the choice of mixed boundary conditions \eqref{eq:diff-boundary-cond-A}, gauges the right acting copy of the symmetry group $G$ (alternatively, we could choose to background gauge the left acting copy).  

The Hamiltonian is time dependent and is given by $H(u) =\e \tilde e_b \sqrt{g_{uu}}\tr \pmb\a^2/4$. In turn, this is proportional to the quadratic Casimir associated to $G$, given by
\be 
\label{eq:Hamiltonian-particle-on-a-group}
\frac{\hat C_2}{\cN}= - \frac{\eta^{ij} \pi_{x^i} \pi_{x^j}}\cN=  \tr( \pmb \alpha^2)  = \frac{4\,H(u)}{\e\tilde e_b\sqrt{g_{uu}}}\,.
\ee

  The Hilbert space  of the theory, $\cH^G$,  is given by normalizable functions on the group manifold that are spanned by the matrix element of all unitary irreducible representations $R$, $U_{R, m}^n(h)$. By definition, such states of course transform correctly under the action of the left- and right- acting $G$ symmetry groups. Namely, we take the generators of the $G$ symmetry that acts by left multiplication to act on the left index, $n$, and those of the right-acting symmetry to act on $m$. Such states are also eigenstates of the Hamiltonian with  $\hat C_2 \,U_{R, m}^n(h) = C_2(R) U_{R, m}^n(h)$. Thus, the thermal partition function at inverse-temperature $\b$ associated to the action \eqref{eq:introduce-Lagr-multiplier-a} is given by,\footnote{Note that the path ordering which is needed in \eqref{eq:partition-function-particle-on-a-group-manifold} does not affect the exponentiated integral since the Hamiltonian is always proportional to the Casimir of $G$ and, therefore, commutes with itself at any time.  }
  \be 
  \label{eq:partition-function-particle-on-a-group-manifold}
  Z_{G}(\b) = \tr_{\cH^G} e^{-\int_0^\b H(u) du} = \sum_R (\dim R)^2 e^{-\frac{\e \tilde e_b C_2(R)}{4\,\cN} \int_0^\beta du \sqrt{g_{uu}}}  = \sum_R (\dim R)^2 e^{-\frac{\b \tilde e_b C_2(R)}{4\,\cN}} \,.
  \ee
  Here, the sum is over all unitary irreducible representations $R$ of the gauge group $G$.
 Because we will encounter this situation when discussing the boundary dual of gravitational Yang-Mills theory, we note that if we replace $\tr \pmb \a^2$ by a general function $\hat V(\pmb \a)$ (that preserves the $G$ symmetries by being a trace-class function) in the action in \eqref{eq:introduce-Lagr-multiplier-a}, the resulting theory has a Hamiltonian that can always be expressed in terms of the Casimirs of the group $G$. Thus, in the partition function, the eigenvalue  $C_2(R) $ of the quadratic Casimir is replaced by a function $V(R)$ that can be easily be related to $\hat V(\pmb \a)$.\footnote{For instance, when $G=\text{SU}(2)$ or $\text{SO}(3)$, all higher-order Casimirs can be expressed in terms of powers of the quadratic Casimir and, consequently, the potential can always be expressed as $\hat V(\pmb \a) \equiv \tilde V(\tr \pmb \a^2)$. In this case $V(R) = \tilde V(C_2(R))$. }
  
 We now re-introduce the background gauge field $\cA$ which appeared through the boundary condition \eqref{eq:diff-boundary-cond-A}, to obtain the partition function of \eqref{eq:introduce-Lagr-multiplier-a} in the more general case. Just like in the case of Yang-Mills theory with Dirichlet boundary conditions, the action in \eqref{eq:introduce-Lagr-multiplier-a} is invariant under background gauge transformations and, consequently, the partition function depends solely on the holonomy of the background $\cA$, $h = \cP \exp(\oint \cA)$ through trace-class functions. The insertion of such a background is equivalent to adding a chemical potential for the left-acting $G$-symmetry, that exponentiates the associated charges of the left $G$-symmetry to a $G$ group element in the same conjugacy class as $h$.  Thus, the partition function \eqref{eq:introduce-Lagr-multiplier-a} becomes 
  \be
    \label{eq:partition-function-particle-on-a-group-manifold-w-bkgrd}
  Z_G(\b, \,h) = \tr_{\cH}\left( h \, e^{-\int_0^\b H(u) du} \right) =  \sum_R (\dim R) \, \chi_R(h)  e^{-{ \frac{\tilde e_b \b C_2(R)}{4\cN}}}\,,
\ee  
where $\chi_R(h)$ are the characters of the group  element $h$ associated to the representation $R$. Similarly, in the theory whose potential is $\hat V(\pmb \ma) $, the partition function is given by
  \be
    \label{eq:partition-function-particle-on-a-group-arb-function-manifold-w-bkgrd}
  Z_G^{\hat V}( \b,\, h)  =  \sum_R (\dim R) \, \chi_R(h)   e^{-{ \tilde e_b V(R)} \int_0^\b du \sqrt{g_{uu}}} = \sum_R (\dim R) \, \chi_R(h)   e^{-{\tilde e_b  \beta V(R)}}\,.
\ee  
Thus, to summarize, in the weak gauge coupling limit, we have found that the gravitational gauge theory \eqref{eq:JT-action+yang-mills} is equivalent to the Schwarzian theory decoupled from a particle moving on the gauge group manifold. Its partition function, with boundary conditions \eqref{eq:boundary-conditions-for-dilaton} for the metric and dilaton and \eqref{eq:diff-boundary-cond-A} for the gauge field, is given by
\be 
\label{eq:JT+BF-path-integral-result}
Z_{\substack{\text{JTBF}\\\text{mixed}}}^\text{disk}(\Phi_b, \b, h)  =e^{S_0}\left(\int ds\,  \frac{s}{2\pi^2} \sinh(2\pi s) e^{-\frac{\beta s^2}{2\Phi_b}}\right) \left[\sum_R \dim R \, \chi_R\left(\cP e^{\int \cA_u}\right)  e^{-{ \frac{\tilde e_b  \b C_2(R)}{4\,\cN}}}\right]\,.
\ee 
\subsection{Reviewing the quantization of 2d Yang-Mills}

While in the weakly coupled limit we were able to directly reduce the bulk path integral to a boundary path integral, since the theory is not topological at non-zero gauge coupling, this cannot be easily done more generally. Thus, it  proves instructive to reproduce the partition function \eqref{eq:JT+BF-path-integral-result} by performing the path integral directly in the bulk. 

Before performing the bulk path integral, it is useful to review the well known quantization of the gauge theory \cite{migdal1975phase, Migdal:1984gj,Blau:1991mp, Witten:1991we, Fine:1991ux,  Witten:1992xu, Cordes:1994fc, ganor1995string}, when fixing the metric $g_{\mu \nu}$ and the dilaton as backgrounds. Thus, we seek to quantize Yang-Mills theory, $S_{YM}^E =  -\int_\cM i \tr \phi F - \frac{1}2 \int_\cM d^2 x  \sqrt{g} j(x) \tr \phi^2 $, where $j(x) \equiv  \tilde e - \tilde e_\Phi \Phi(x)$ is an arbitrary source for the operators $\tr \phi^2$.\footnote{In this paper we omit the possibility of adding a $\theta$-angle for the gauge field. This will be discussed in the study of the weak gauge coupling limit \cite{Kapec:toAppear}.} The source $j(x)$ can be absorbed by changing the surface form $ d^2 x  \sqrt{g}$. Due to the fact that the theory is invariant under local area preserving diffeomorphisms, the partition function can thus solely depend on the dimensionless quantity $a = \int_\cM d^2 x  \sqrt{g}\, j(x)$. It is therefore sufficient to review the quantization of the theory on a flat manifold with area $\tilde a$ and coupling $e_\text{YM}^2$, such that $a  =  e_\text{YM}^2\tilde a$.

The quantization of this theory is similar to that of the particle moving on the gauge group manifold discussed in the previous subsection and, for pedagogical purposes, it is useful to emphasize these similarities. When using the Dirichlet boundary conditions \eqref{eq:holonomy-definition} the partition function of the gauge theory is a trace-class function of $h$ and thus it is spanned by characters of the group $\chi_R(h)$. Consequently, the characters  $\chi_R(h)$ can be viewed as a set of wavefunctions which span the Hilbert space $\cH^\text{YM}$ of the gauge theory.

The partition function on a manifold with arbitrary genus $g$ and an arbitrary number of boundaries $n$ can be computed using the cutting and gluing axioms of quantum field theory and by solely using the partition function of the gauge theory on the disk with the Dirichlet boundary condition \eqref{eq:holonomy-definition}.  As previously mentioned, in the limit  $a \to 0$ the gauge theory becomes topological.  In this limit, the integral over $\phi$ imposes the condition that $A$ is a flat connection, which yields $h= e$ (where $e$ is the identity element of $G$), so \cite{Witten:1991we}
\es{ZSmallLimit}{
	\lim_{a \to 0} Z_\text{YM}^\text{disk}(a, h) = \delta(h)
	= \sum_R \dim R \, \chi_R(h) \,,
}
where $\delta(h)$ is the delta-function on the group $G$ defined with respect to the Haar measure on $G$, which enforces that $\int dh\, \delta(h) x(h) = x(e)$.  This is the same as the partition function of the particle moving on the $G$ group manifold \eqref{eq:partition-function-particle-on-a-group-manifold} in the limit $\tilde e_b \to 0$. 

For non-zero $a$, note that the canonical momentum conjugate to the space component of the gauge field $A_1^i(x)$ is $\phi_i(x)$, and thus the Hamiltonian density  is just $H = \frac{e_\text{YM}^2}{4} \tr (\phi_i T^i)^2$.  It then follows, from $\pi_i = -i \cN \phi_i$, that $H =- \frac{e_\text{YM}^2}{4\cN}\eta^{ij} \pi_i \pi_j$. Using $\pi_j = \frac{\delta}{\delta A_1^j}$, each momentum acts on the wavefunctions $\chi_R(g)$ as $\pi_i \chi_R(h) = \chi_R (T_i h)$.  It follows that the Hamiltonian density acts on each basis element of the Hilbert space $\chi_R(g)$ diagonally with eigenvalue $e_\text{YM}^2 C_2(R)/(4\cN)$ \cite{Cordes:1994fc}, where $C_2(R)$ is the quadratic Casimir, with $C_2(R) \geq 0$ for compact groups. Note that the Hamiltonian of the gauge theory is therefore closely related to that of the particle moving a group manifold \eqref{eq:Hamiltonian-particle-on-a-group}. One then immediately finds
\es{ZdiskYM}{
	Z_\text{YM}^\text{disk}(a, h) 
	&= \sum_R \dim R \, \chi_R(h) e^{-\frac{e_\text{YM}^2 \tilde a C_2(R)}{4 \cN}}  =  \sum_R \dim R \, \chi_R(h) e^{-\frac{C_2(R)}{4\cN} \int d^2 x \sqrt{g} j(x)} \,.
}
Following from the relation between the Hamiltonian of the gauge theory and that of a particle moving on the $G$ group manifold, we of course find that  \eqref{ZdiskYM} agrees with \eqref{eq:partition-function-particle-on-a-group-manifold-w-bkgrd} for the appropriate choice of $\tilde e_b$ or $j(x)$.

The partition function of Yang-Mills theory on an orientable manifold $\cM_{g, n}$ of genus $g$, with $n$ boundaries, can be obtained by gluing different segments on the boundary of the disk  \cite{Blau:1991mp, Witten:1991we, Fine:1991ux,  Witten:1992xu, Cordes:1994fc}. This is given by
\be 
\label{eq:general-YM-partition-function}
Z_\text{YM}^{(g, n)}(a, h_1,\, \dots\,, h_n) = \sum_R (\dim R)^{\chi(\cM_{g, n})} \chi_R(h_1) \chi_R(g_2) \dots \chi_R(h_n) e^{-\frac{C_2(R)}{4\cN} \int d^2 x \sqrt{g} j(x)}\,.
\ee
With these results in mind, we can therefore proceed with the analysis of the simplified case of obtaining the contribution to the path integral of the disk topology in the weakly coupled limit by directly performing the path integral in the bulk.
 
\subsection{Quantization with a boundary condition chancing defect}
\label{sec:quantization-with-defect}

To determine the partition function with the boundary condition \eqref{eq:diff-boundary-cond-A} we consider a boundary changing defect 
\be
\label{eq:defect-action-simple}
S_\text{Defect}^E[g, \phi] = -\frac{\e \tilde e_b}2 \int_{I} du \sqrt{g_{uu}} \tr \phi^2\,,
\ee
which we can insert along a contour $I$ which is arbitrarily close to the boundary $\partial \cM$. We now show that the boundary condition changing defect indeed implements the change of boundary conditions from Dirichlet to those listed in \eqref{eq:diff-boundary-cond-A}. By integrating the equation of motion obtained from the variation of $\phi$ at the location of defect on an infinitesimal interval in the direction perpendicular to the defect we find, 
\be 
\label{eq:simple-eq}
A_u|_{\partial \cM} - A_u|_{I} = - i \sqrt{g_{uu}}\e \tilde e_b \phi |_{I} \,,
\ee
 where $A_u|_{\partial \cM}$ is the gauge field on the boundary on $\cM$ that is fixed when using Dirichlet boundary conditions for the action, $ A_u|_{I}$ is the gauge field in the immediate neighborhood inside of the defect and $ \phi |_{I}$ is the value of the zero-form field on the defect. Moving $ A_u|_{I}$ to the RHS and setting $A_u|_{\partial \cM} = \cA_u$,  we reproduce the boundary condition \eqref{eq:diff-boundary-cond-A}. Thus, the theory with the defect and Dirichlet boundary conditions should reproduce the results in the theory without the defect and with the boundary condition  \eqref{eq:diff-boundary-cond-A} for the gauge field.  

As we further exemplify in section \ref{sec:observables}, the advantage of using the description of the BF theory in the presence of the defect \eqref{eq:defect-action-simple} is that the expectation value of any observable can easily be computed  by using standard techniques in 2d Yang-Mills theory. For example, when computing the partition function of the theory with the defect \eqref{eq:defect-action-simple} on a disk, we can use \eqref{ZdiskYM} setting $j(x)\sim \delta(x-x_I)$ and $h = \cP \exp(\int_{\partial\cM} \cA)$, to find that
\be
\label{eq:BF-partition-function-on-a-disk}
Z_{\substack{\text{BF}\\ \text{mixed}}}^\text{disk}(\b, h)  = \sum_{R} \dim(R) \chi_{R}(h) e^{- \frac{\tilde e_b \beta C_2(R)}{4 \cN} } \,.
\ee
Using this result together with the reduction of the JT gravity path integral on a disk to that of the Schwarzian, we find the result \eqref{eq:JT+BF-path-integral-result}.  Moving forward, we fix the normalization of the Casimir by fixing the Dynkin index, $\cN \equiv 1/2$.

More generally, we can consider adding a defect which depends on a general gauge invariant potential $\hat V(\phi)$, $S_\text{Defect}[g, \phi] = -  \int_{I} du \sqrt{g_{uu}}\, \e\hat V(\phi)$. In this case, the boundary condition which the gauge field needs to satisfy is again given by the $\phi$ equation of motion, which implies that $(A_u-i \e \partial \hat V(\phi)/\partial \phi)|_{\partial \cM} = \cA_u$.
The quantization of Yang-Mills theory with such a general potential was discussed in \cite{Witten:1992xu, ganor1995string} and closely follows the quantization of a particle moving on a group manifold with the general potential $\hat V(\pmb \a)$ discussed in the previous subsection. In fact the result for the bulk partition function
\be 
\label{eq:general-mixed-bdy-cond-partition-function}
Z_{\substack{\text{BF}\\ \text{mixed}\,\hat V(\phi)}}^\text{disk}(\b, h)  = \sum_{R} \dim(R) \chi_{R}(h) e^{- \tilde e_b \beta V(R) } 
\ee 
agrees with the partition function \eqref{eq:partition-function-particle-on-a-group-arb-function-manifold-w-bkgrd} obtained by considering a particle moving on the $G$ group manifold with a potential $\hat V(\pmb \a)$ and in the presence of the background gauge field $\cA_u$.  Therefore, we obtain the first general equivalence which we schematically present in figure \ref{fig:schematic-dualtiy-BF}.
  \begin{center}
  \vspace{-0.3cm}
  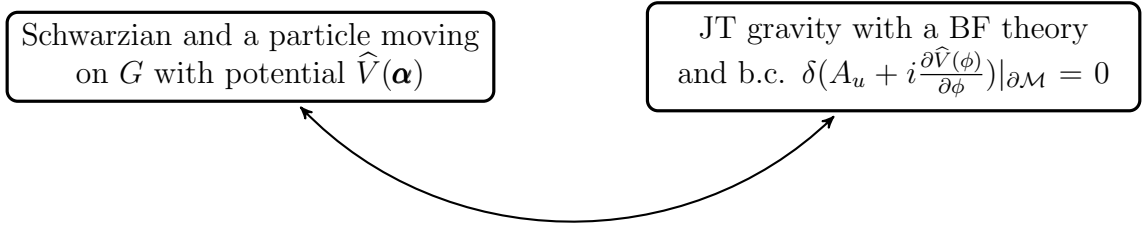
\begin{figure}[h!]
\begin{tikzpicture}[node distance=1cm, auto]
 \node[punkt] (first) {Schwarzian and a particle moving \\ on $G$ with potential $\hat V(\pmb \a)$};
 \node[punkt, inner sep=5pt,right=2cm of first](second) {JT gravity with a BF theory  \\ and b.c.  $\delta(A_u+i\frac{\partial \hat V(\phi)}{\partial \phi})|_{\partial \cM} = 0$ }
 edge[pil,<->, bend left=45]  (first);
\end{tikzpicture}\\
\vspace{-0.5cm}
	\caption{\label{fig:schematic-dualtiy-BF} Schematic representation of the equivalence between the  gravitational gauge theory  at weak gauge coupling and the Schwarzian decoupled from a particle moving on the group manifold $G$.  }
\end{figure}
\vspace{-1.5cm}
\end{center}

\section{Disk partition function}
\label{sec:genus-zero-part-function}

\subsection{2D Yang-Mills theory with Dirichlet boundary conditions}
\label{sec:genus-zero-YM-theory}

We finally arrive at the quantization of the theory \eqref{eq:JT-action+yang-mills} for arbitrary gauge group and gauge couplings, when fixing the boundary conditions to \eqref{eq:boundary-conditions-for-dilaton} for the metric and dilaton and when using Dirichlet boundary conditions for the gauge field $A_{u}|_\text{bdy.} =  \cA_u$.  Using \eqref{eq:general-YM-partition-function} for $\chi(\cM) =1$, $j(x) = \tilde e - \tilde e_\Phi \Phi(x)$ and setting $h = \cP \exp(\int_{\partial \cM} \cA)$, we find that after integrating out the gauge field $A_\mu$ and the zero-form field $\phi$, the partition function is given by\footnote{Here we assume the path integral over the gauge degrees of freedom can always be made convergent with the proper choice of integration contour for the field $\phi$.}
\be
\label{eq:JT+YM-path-integral}
Z_{\substack{\text{JTYM}\\\text{Dirichlet}}}^\text{ disk}(\Phi_b, \b, h)  & =  \int Dg_{\mu\nu}D\Phi e^{-S_{JT}[g_{\mu \nu},\, \Phi]} \left(\sum_{R} \dim(R) \chi_{R}(h) e^{- \frac{C_2(R) \int_\cM d^2 x \, \sqrt{g} \left[\tilde e - \tilde e_\Phi \Phi\right]}{2} } \right)\hspace{-0.2cm} \nn \\ &= e^{S_0}\sum_{R} \dim(R) \chi_{R}(h) \int Dg_{\mu\nu}D\Phi  \,e^{\frac{1}2 \int_{\cM} d^2 x \sqrt{g}\, \Phi(\cR + 2 +  \tilde e_\Phi C_2(R))}\nn\\ & \qquad \qquad \times e^{ - \frac{ \tilde e C_2(R)}2 \int_\cM d^2 x \, \sqrt{g}+\int_{\partial \cM} du \sqrt{g_{uu}} \Phi(\cK-1) },
\ee
where the couplings $\tilde e$ and $\tilde e_\Phi$ are related to the initial couplings by  \eqref{eq:coupling-definitions}. We can now view the terms in the exponent in \eqref{eq:JT+YM-path-integral} as coming from an effective action for each representation $R$ of the gauge group. 

Integrating out the dilaton field $\Phi$, we once again find that the path integral localizes to $AdS_2$ patches, whose cosmological constant is now given by $\tilde \Lambda = -2 - \tilde e_\Phi C_2(R) $ and whose boundary degrees of freedom is the sole remaining dynamical degrees of freedom in the path integral. Thus, we are summing over $AdS_2$ patches whose curvatures depend on the representation sector from the sum in \eqref{eq:JT+YM-path-integral}. 

After  integrating out the dilaton field $\Phi$ one can rewrite the remaining area term $\tilde e \int_\cM d^2 x \sqrt{g}$ using the Gauss-Bonnet theorem
\be 
\label{eq:Euler-relation-1}
\tilde e  \int_{\cM}d^2 x \sqrt{g} = -\frac{\tilde e}{2 + {\tilde e_\Phi C_2(R)}}\int d^2 x \sqrt{g}\,\cR = \frac{\tilde e}{1  + \frac{\tilde e_\Phi C_2(R)}{2}} \left[\int_{\partial \cM} \sqrt{h} \,\cK - \chi(\cM)\right]\,,\ee
where for the disk, the Euler characteristic is $\chi(\cM) = 1$. Thus, the path integral becomes, 
\be 
\label{eq:disk-partition-function-path-integral}
Z_{\substack{\text{JTYM}\\\text{Dirichlet}}}^\text{ disk}(\Phi_b,\b, h)   &= e^{S_0} \sum_{R} \dim(R) \chi_{R}(h)  \int D\mu[F] \exp\bigg[\frac{\tilde e\, C_2(R)}{2 +{\tilde e_\Phi C_2(R)}} \\ &\nn +\left(\frac{\Phi_b}{\epsilon}- \frac{\tilde e\, C_2(R)}{2 + {\tilde e_\Phi C_2(R)}}\right) \int_{\partial \cM} du \sqrt{g_{uu}} \cK[F(u)] - \frac{\Phi_b}{\epsilon}\int_{\partial \cM} du \sqrt{g_{uu}} \bigg] \,.
\ee 
where we have used the fact that the path integral over the gauge degrees of freedom does not affect the measure for the Schwarzian field, $D\mu[F]$, and we have added a counter-term $- \frac{\Phi_b}{\epsilon}\int_{\partial \cM} du \sqrt{g_{uu}}$ to cancel the leading divergence appearing in the exponent. It is convenient to define a ``renormalized'' Casimir  
\be 
\label{eq:defintion-C2-tilde}
\tilde C_2(R) \equiv \frac{C_2(R)}{2\left(1 + \frac{\tilde e_\Phi C_2(R)}{2}\right)}\,,
\ee 
to capture the dependence on the $G$-group second-order Casimir appearing in \eqref{eq:disk-partition-function-path-integral}. The origin of this modified Casimir comes from the $R$ dependence of the  cosmological constant that can be seen through \eqref{eq:Euler-relation-1}. Note that for compact Lie groups, when choosing the coupling $e$ and $e_\Phi$ to be real, $\tilde C_2(R)$ is a real positive function of $R$, which for representations with growing dimensions, asymptotes to a constant value.

The path integral can then be rewritten using the relation \eqref{eq:extrinsic-curvature-Schw-field} between the extrinsic curvature and the Schwarzian derivative 
\be 
\label{eq:JT+YM-disk-path-integral-interm-step}
Z_{\substack{\text{JTYM}\\\text{Dirichlet}}}^\text{ disk} (\Phi_b,\b, h)  = \sum_{R} \dim(R) \chi_{R}(h)  \int D\mu[F] e^{\left[\tilde e \tilde C_2(R)-\frac{\Phi_b \beta}{\epsilon^2}+\left({\Phi_b}- \epsilon \tilde e\tilde C_2(R)\right) \int_0^\beta du \left(\frac{1}{\epsilon^2} + \Sch(F, u) \right)\right]}\,.
\ee
For now, let's ignore the fact that the coupling in front of the Schwarzian might be negative for sufficiently large $\epsilon$ and assume that $\Phi_b > \e \tilde e \tilde C_2(R)$. Once again using the computation for the Schwarzian path integral, which is one-loop exact, we find
\be 
\label{eq:JT+YM-disk-part-function}
Z_{\substack{\text{JTYM}\\\text{Dirichlet}}}^\text{ disk}(\Phi_b,\b, h)  &=   \sum_{R} \dim(R) \chi_{R}(g)  \int ds \frac{s}{2\pi^2} \sinh(2\pi s) e^{ - \frac{\beta}{(\Phi_b - \epsilon\,\tilde e \tilde C_2(R))} s^2 + \tilde e \tilde C_2(R) \left(1-\frac{\beta}{\epsilon}\right)}\nn\\ & =  \sum_{R} \dim(R) \chi_{R}(h) \frac{1}{(2\pi)^{1/2}}\left(\frac{\tilde \Phi_b(R) }{  \beta}\right)^{3/2}e^{\frac{\pi^2\tilde \Phi_b(R)}{\beta} + \tilde e \tilde C_2(R)\left(1 -  \frac{\beta}{\epsilon} \right)}\,,
\ee
where we have defined
\be 
\label{eq:definition-phi-b-R}
\tilde \Phi_b(R) \equiv \Phi_b - \epsilon\, \tilde e \,\tilde C_2(R)\,,
\ee
which can be seen as the ``renormalization'' of the boundary value of the dilaton $\Phi_b$. Thus, the addition of the Yang-Mills term to the JT gravity action has the effect of ``re-normalizing'' all the dimensionful quantities appearing in JT gravity by a representation dependent factor.

As previously mentioned, our result is reliable only in the regime in which $ \Phi_b > \epsilon\tilde e \tilde C_2(R) $ for which the coupling in the Schwarzian action in \eqref{eq:JT+YM-disk-path-integral-interm-step} is positive. If this was not the case than the path integral over the field $F(u)$ would no longer be convergent, at least when considering a contour along which $F(u)$ is real. From the perspective of near-extremal black holes, this inequality is indeed obeyed: namely, for representations with very large dimensions one expects $C_2(R) \rightarrow \infty$ and thus $\tilde C_2(R) \to 2/\tilde e_\Phi$.   Since $\tilde e_\Phi > 0$ when the couplings $e$ and $e_\Phi$ are real in \eqref{eq:JT-action+yang-mills} , $\tilde C_2(R)$ asymptotes to a negative constant and therefore satisfies $ \Phi_b > \epsilon\tilde e \tilde C_2(R)$ for sufficiently small $\e$.

In the $(\epsilon/\tilde e \to 0, \,\tilde e_\Phi \to 0)$ limit the singlet representation dominates in the sum in \eqref{eq:JT+YM-disk-part-function}. This $1/\epsilon$ divergence in the exponent appears due to a divergence in the area of the nearly $AdS_2$ patches that dominate in the gravitational gauge theory path integral. In the upcoming subsection, we show that such a divergence can be eliminated using a change in boundary conditions for the gauge field, which amounts to adding the appropriate boundary counter-term that cancels the divergence in the action. In the limit $(\epsilon \to 0, \, \tilde e_\Phi \to 0)$, with $\epsilon/\tilde e $ kept finite, the partition function of the theory matches the one we have found in section \ref{sec:preliminaries} when coupling JT gravity to a BF theory.  

Going away from the strict $\epsilon \to 0$ limit and instead viewing \eqref{eq:JT+YM-disk-part-function} in an $\e$ expansion we note that if we keep the next order terms in $\epsilon$ in the extrinsic curvature in \eqref{eq:extrinsic-curvature-Schw-field} they would only contribute $O(\e^2)$ in the exponent.\footnote{This can be easily seen by computing the next order in the $\e$ expansion in the solution of \eqref{eq:metric-and-boundary-condition-for-metric}, $\tau = \epsilon F' + \epsilon^3 \frac{(F'')^2}{2F'}  + O(\epsilon^5)$. Plugging this result in the extrinsic curvature formula \eqref{eq:extrinsic-curvature-Schw-field}, we find that 
	\be 
	\cK[F(u)] = 1+ \e^2 \text{Sch}(F, u)  + \e^4 \left(\frac{27}8 \frac{(F'')^4}{(F')^4} + \frac{(F^{(3)})^2}{(F')^2}+ \frac{F^{(4)}F''}{(F')^2}-\frac{11(F'')^2 F^{(3)}}{2 (F')^3}\right) + O(\e^6)
	\ee
	Consequently the first correction on the gravitational side coming from $\Phi_b \cK[F(u)]/\e^2$ is $O(\e^2)$. Work on computing the partition function in pure JT gravity to all perturbative orders in $\e$ is currently underway \cite{Iliesiu:toAppear}. A similar perspective can be gained by studying an analog of the $T\bar T$ deformation in 1d \cite{Gross:2019ach}.  } Thus, the Casimir dependent terms shown in \eqref{eq:JT+YM-disk-part-function}, which are $O(1/\e)$ to $O(\epsilon)$, are the most important contributions in the $\epsilon$ expansion of the partition function of the gravitational gauge theory \eqref{eq:JT-action+yang-mills}.

\subsection{Counter-terms from a change in boundary conditions}
\label{sec:counter-terms-from-a-change-in-boundary-cond}

As is typical when analyzing theories in $AdS$ in the holographic context, the action of the theory under consideration is generically not finite on-shell and needs to be supplemented by boundary terms, a procedure referred to as holographic renormalization. Given the appropriate boundary terms, one could then use the variational principle to check what boundary conditions can be consistently imposed in order for the variational problem to be well defined and in order for the overall on-shell action to be finite. Although various boundary terms supplementing the Maxwell or Yang-Mills actions have been considered in the past in the context of 2d/1d holography (for example, see \cite{Castro:2008ms, Grumiller:2014oha, Grumiller:2015vaa, Cvetic:2016eiv, Jensen:2016pah, Mezei:2017kmw}), here we take a different approach and show that, in order to cancel the divergence in the exponent in \eqref{eq:JT+YM-disk-part-function}, it is sufficient to add a boundary condition changing defect similar to the one considered in section \ref{sec:quantization-with-defect}. After stating the proper form of the boundary condition changing defect, we can immediately derive the necessary boundary conditions that the gauge theory needs to satisfy.

Namely, we consider adding 
\be
\label{eq:add-tr-phi2-defect}
S_\text{defect} = \frac{1}2 \int_I du \sqrt{g_{uu}}  \left[\frac{\tilde e \,\Tr \phi^2}{1 + \frac{\tilde e_\Phi\, \Tr \phi^2}2} - \e \, \tilde e_b \Tr \phi^2\right]\,,
\ee
to the action \eqref{eq:JTandYMaction-with-bdy} where, once again, $I$ is a contour which is arbitrarily close to the boundary $\partial \cM$ and $\tilde e_b$ is an arbitrary constant. Similar to our analysis in subsection \ref{sec:quantization-with-defect}, multiplying $\tilde e_b$ by $\tr \phi^2$ instead of a more general trace-class function $V(\phi)$ is an arbitrary choice that is only meant to regularize the sum over all irreducible representations appearing in the partition function.   Integrating the equation of motion on the defect yields
\be 
\label{eq:eq-with-gc2-non-zero}
A_u|_{\partial \cM} - A_u|_{I} = -i\, \sqrt{g_{uu}} \, \left[\frac{\tilde e \phi}{1 + \frac{\tilde e_\Phi}2\, \Tr \phi^2} - \frac{\tie\, \tilde e_\Phi \,\phi\, \Tr \phi^2}{2\left(1+\frac{\tilde e_\Phi}2 \,\Tr \phi^2\right)^2} - \e \tilde e_b  \phi  \right] \bigg|_{I}\,.
\ee
Once again moving $A_u|_I$ to the right hand side and denoting $A_u|_{\partial\cM} = \cA_u$, we find that by inserting the defect the new ``mixed'' boundary condition in the resulting theory is given by
\be
\label{eq:boundary-condition-that-makes-things-finite}
\delta \left(A_u - i \sqrt{g_{uu}} \,\left[\frac{\tilde e \phi}{1 + \frac{\tilde e_\Phi}2\, \Tr \phi^2} - \frac{ \tilde e\, \tilde e_\Phi \,\phi\, \Tr \phi^2}{2\left(1+\frac{\tilde e_\Phi}2 \,\Tr \phi^2\right)^2}  - \e \tilde e_b \phi \right]\right)\bigg|_I = 0\,. 
\ee
Adding this defect modifies the path integral computation at the step  \eqref{eq:JT+YM-disk-path-integral-interm-step}. Following the procedure  presented in subsection \ref{sec:quantization-with-defect}, we find that after integrating out the gauge field degrees of freedom we get
\be 
\label{eq:JT+YM-disk-path-integral-interm-step-with-defect}
Z_{\substack{\text{JTYM},\\ \text{mixed}}}^{\text{disk}}(\Phi_b, \b, h)  &= \sum_{R} \dim(R) \chi_{R}(g)\nn \\ &\times  \int DF e^{\left[\tilde e \tilde C_2(R) \left(1+\frac{\b}\e \right)-  \b \tilde e_b C_2(R)-\frac{\Phi_b \beta}{\epsilon^2}+\left({\Phi_b}- \epsilon \tilde C_2(R)\right) \int_0^\beta du \left(\frac{1}{\epsilon^2} + \{F, u\} \right)\right]}\,.
\ee 
After performing the integral over $F(u)$ by following the steps in \eqref{eq:JT+YM-disk-part-function}, we find 
\be
\label{eq:partition-function-with-mixed-bc-YM}
Z_{\substack{\text{JTYM},\\ \text{mixed}}}^{\text{disk}}(\Phi_b, \b, h)  = \sum_{R} \dim(R) \chi_{R}(g) \frac{1}{(2\pi)^{1/2}} \left(\frac{\tilde \Phi_b(R) }{  \beta}\right)^{3/2} e^{\frac{\pi^2\tilde \Phi_b(R)}{\beta} + \tilde e \tilde C_2(R) - \tilde e_b \beta C_2(R)}\,.
\ee
Note that, the $1/\e$ divergence present in the exponent in \eqref{eq:JT+YM-disk-part-function} has vanished, the singlet representation is no longer the dominating representation and the sum over all irreducible representations $R$ is generically convergent for $\tilde e_b \geq 0$. With these results in mind, we now discuss the boundary dual of the 2d gravitational Yang-Mills theory \eqref{eq:JT-action+yang-mills}, both with  Dirichlet boundary conditions and the mixed conditions discussed in this subsection.

\subsection{Equivalent boundary theory}
\label{sec:equiv-bdy-theory}

As extensively discussed in subsections \ref{sec:genus-zero-BF-plus-defect}--\ref{sec:quantization-with-defect}, when adding a BF theory to the JT gravity action, and using mixed boundary conditions between the gauge field and the zero-form scalar $\phi$, the gravitational theory can be equivalently expressed as the Schwarzian theory decoupled from a particle moving on the group manifold $G$. Here, we show how, by going to finite gauge coupling, the two boundary theories become coupled. 

To find the dual of JT gravity coupled to Yang-Mills theory it is useful to interpret the partition functions \eqref{eq:JT+YM-disk-part-function} (Dirichlet) or \eqref{eq:partition-function-with-mixed-bc-YM} (mixed) in terms of the path integral of a particle moving on a group manifold with a time dependent metric $g_{uu}$. Towards that aim, we use this particle's path integral to reproduce  the  intermediate steps \eqref{eq:JT+YM-disk-path-integral-interm-step} and \eqref{eq:JT+YM-disk-path-integral-interm-step-with-defect} in which we have integrated out the gauge degrees of freedom, but have not yet integrate out the Schwarzian field $F(u)$. To do this we set  $\sqrt{g_{uu}(u)}\equiv j(u)$  for the particle moving on the group manifold $G$:\footnote{One should not be concerned about the invertibility of the 1d metric in \eqref{eq:cases-source-metric}. Rather one can view this metric as an arbitrary source for the potential $\hat V(\pmb \a)$ in the path integral of the particle moving on the $G$ group manifold.  } 
\be 
\label{eq:cases-source-metric}
\begin{cases}
	j_\text{Dirichlet}(u) = \frac{1}\e -\frac{1}{\b}+\e\,\Sch(F, u)\,, \qquad &\text{for dual of Dirichlet b.c. from \eqref{eq:JT+YM-disk-part-function}}\,,\\ j_\text{mixed}(u)=-\frac{1}{\b}+ \e \,\Sch(F, u) \qquad  &\text{for dual of mixed b.c. from \eqref{eq:partition-function-with-mixed-bc-YM}}\,.
\end{cases}
\ee
Fixing the action of the particle moving on a group manifold coupled to the Schwarzian theory to be given by
\be
\label{eq:boundary-theory-for-YM}
\begin{cases}
	S_{\substack{\text{Schw}\rtimes G\\ \text{Dirichlet}}}\equiv \int_0^\b du \left[  \left(\frac{\Phi_b}2 - \frac{  \e \,\tilde e \,\tr \pmb\a^2}{2\left(1+{\tilde e_\Phi}\tr\pmb \a^2\right)} \right)\Sch(F, u) -  i \tr ( \pmb \ma\, h^{-1}D_A h)+\frac{\tilde e \,\tr \pmb\a^2}{2\b\left(1+{\tilde e_\Phi}\tr\pmb \a^2\right)} - \frac{\tilde e_b}2 \tr\pmb \a^2\right] \,,\\
	S_{\substack{\text{Schw}\rtimes G\\ \text{mixed}}}\equiv \int_0^\b du \left[ \left(\frac{\Phi_b}2 - \frac{  \e \,\tilde e \,\tr \pmb\a^2}{2\left(1+{\tilde e_\Phi}\tr\pmb \a^2\right)} \right)\Sch(F, u) -  i \tr ( \pmb \ma\, h^{-1}D_A h)+\frac{\tilde e  \left(\frac{1}\e - \frac{1}\b\right)\,\tr \pmb\a^2}{2\left(1+{\tilde e_\Phi}\tr\pmb \a^2\right)} \right] \,.
\end{cases}
\ee 
After integrating out $h$  and $\pmb \a$ that the partition function of this theory is given by, 
\be
Z_{\substack{\text{Schw}\rtimes G \\ j(u)}}(\b, h) = \sum_R (\dim R) \chi_R(h)\int D\mu[F] \,e^{- \frac{\b \tilde e_b  C_2(R)}{2}-\left(\tilde e \tilde C_2(R)\int_0^\beta du \,j(u)\right)+\left({\Phi_b} \int_0^\beta du \, \Sch(F, u) \right)}\,.
\ee  
where $j(u)$ is the source in \eqref{eq:cases-source-metric}. Comparing this partition function to \eqref{eq:JT+YM-disk-path-integral-interm-step} for Dirichlet boundary conditions in the bulk or with \eqref{eq:JT+YM-disk-path-integral-interm-step-with-defect} for mixed boundary conditions, we conclude that the partition function of the particle moving on the group manifold coupled to the Schwarzian theory matches the partition function of gravitational Yang-Mills theory, for an arbitrary $G$ holonomy $h$: $Z_{\substack{\text{JTYM},\\ \text{Dirichlet}}}^{\text{disk}} (h)= Z_{\substack{\text{Schw}\rtimes G,\\ \text{Dirichlet}}}(h)$ and 
$Z_{\substack{\text{JTYM},\\ \text{mixed}}}^{\text{disk}} (h)= Z_{\substack{\text{Schw}\rtimes G,\\ \text{mixed}}}(h)$. Based on this result we conjecture the result presented in figure \ref{fig:schematic-dualtiy}.
\begin{center}
	\vspace{-0.2cm}
	\begin{figure}[h!]
		\begin{tikzpicture}[node distance=1cm, auto]
		\node[punkt] (first) {Schwarzian coupled to a particle moving \\ on $G$ with potential \\ \vspace{0.2cm}$\hat V(\pmb \a) = \frac{\tilde e \,\tr \pmb\a^2}{2\left(1+ \frac{\tilde e_\Phi}2\,\tr\pmb \a^2\right)}$};
		\node[punkt, inner sep=5pt,right=2cm of first](second) {JT-gravity coupled to YM theory  \\ with Dirichlet or mixed b.c. }
		edge[pil,<->, bend left=45]  (first);
		\end{tikzpicture}\\
		\vspace{-0.5cm}
		\caption{\label{fig:schematic-dualtiy} Schematic representation of the equivalence between the gravitational gauge theory and the Schwarzian coupled to a particle moving on the group manifold $G$.  }
	\end{figure}
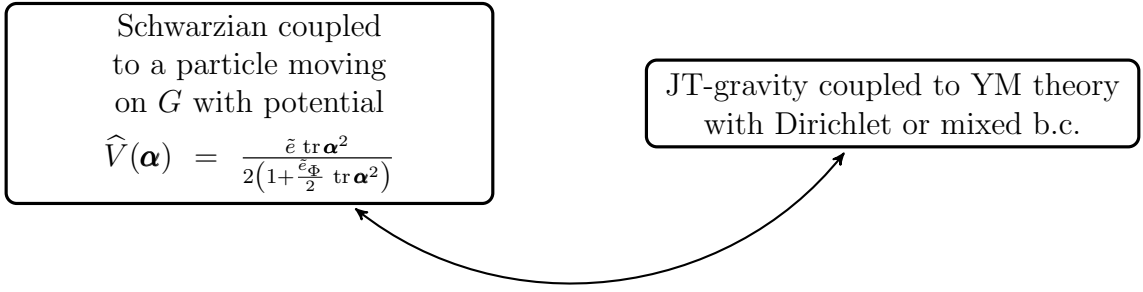
	\vspace{-1.0cm}
\end{center} 

More generally, one can replace $\tilde e \tr \phi^2$ and $\tilde e_\Phi \tr \phi^2$ in the action \eqref{eq:JTandYMaction-with-bdy} by generic gauge-invariant functions of $\phi$.\footnote{Such functions could appear when keeping tracks of higher field-strength powers in the effective action for higher-dimensional near-extremal black holes. } In such a case we expect that the dual quantum mechanical theory be given by
\be
\label{eq:action-1D-Schw-ltimes-G}
S_{\substack{\text{Schw} \rtimes G\\ \text{General}}} =  \int_0^\b du \left[ - i  \tr ( \pmb \ma\, h^{-1}D_A h)-\hat \cW(\pmb \a) + \hat \cV(\pmb \a)\, \Sch(F(u), u)  \right]\,.
\ee
The functions $\hat\cV(\pmb \a)$ and $\hat\cW(\pmb \a)$ are invariant under adjoint transformations of $\pmb \a$ and can be straightforwardly related to the functions of $\phi$ that appear in the generalization of the action \eqref{eq:JTandYMaction-with-bdy}.\footnote{Explicitly if considering replacing the terms in the action of the gravitational gauge theory \eqref{eq:JTandYMactionrewritten}
	\be 
	S_\text{JTYM} \supseteq \frac{1}2 \int_\cM d^2 x  \sqrt{g} \left(\tilde e - \tilde e_\Phi \Phi \right) \tr \phi^2 \qquad \to \qquad \int_\cM d^2 x \sqrt{g}\,\left( \cV_1(\phi) - \Phi \cV_2(\phi)\right)
	\ee
	and considering the boundary condition $\delta(A_u +i \sqrt{g_{uu}}\, \hat \cV_b(\phi)) = 0$, we find that the the functions  $\hat\cV(\pmb \a)$ and $\hat\cW(\pmb \a)$ in \eqref{eq:action-1D-Schw-ltimes-G} are given by
	\be 
	\hat\cV(\pmb \a)  =  \Phi_b - \frac{\e \hat \cV_1(\pmb \a)}{1+2{\hat \cV_2(\pmb \a)}}\,,\qquad \hat\cW(\pmb \a) =\left( \frac{1}{\e} -\frac{1}{\b}\right) \frac{ \hat \cV_1(\pmb \a)}{1+2\hat \cV_2(\pmb \a)} -\frac{\hat \cV_b(\pmb \a)}{\e} \,.
	\ee
} 

The action \eqref{eq:action-1D-Schw-ltimes-G} is a generic effective action with a $G \times SL(2, \mR)$ symmetry.\footnote{In fact, the global symmetry group in this action is enhanced to $G\times G\times SL(2, \mR)$.} Based on symmetry principles, we expect that such an effective action, preserving $G \times SL(2, \mR)$, appears in the low energy limit of a modification of SYK models which have a global symmetry $G$ \cite{Sachdev:2015efa, Davison:2016ngz, Gross:2016kjj, Fu:2016vas, Narayan:2017qtw, Yoon:2017nig, Narayan:2017hvh, Klebanov:2018nfp, Liu:2019niv}. For instance, when $G = \grp{U}(1)$, \eqref{eq:action-1D-Schw-ltimes-G} should appear in  the low-energy limit of the complex SYK model studied in \cite{Sachdev:2015efa, Davison:2016ngz}; it would be interesting to derive the functions $\cV(\pmb \a)$ and $\cW(\pmb \a)$ directly in this model. 

\section{Higher genus partition function}
\label{sec:higher-genus-partition-function}

Following the same strategy of firstly integrating out the gauge field degrees of freedom and rewriting the resulting area dependence from the Yang-Mills path integral in terms of the extrinsic curvature, we determine the partition function of the gravitational gauge theory for surfaces of arbitrary genus. 

\subsection{The building blocks}
\label{sec:building-blocks}

In computing the contribution of the gravitational degrees of freedom to the higher genus partition function, we follow the strategy presented in \cite{Saad:2019lba}. The basic building blocks needed in order to obtain the genus expansion of the gravitational gauge theory are given by \cite{Saad:2019lba}: 
\begin{itemize}
	\item The disk partition functions computed in sections \ref{sec:preliminaries} or \ref{sec:genus-zero-part-function}.
	\item The path integral  over a  ``trumpet'', $\cM_T$, which on one side has  asymptotically $AdS_2$ boundary conditions specified by \eqref{eq:boundary-conditions-for-dilaton} and, on the other side, ends on a geodesic of length $b$. For the gauge field, we first consider Dirichlet boundary conditions by fixing the holonomy on both sides of ``trumpet'': we denote $h_{nAdS_2}$ to be the holonomy of the side with asymptotically   $AdS_2$ boundary conditions and $h_b$ to be the holonomy on the other side. Following our analysis in section \ref{sec:counter-terms-from-a-change-in-boundary-cond} we then consider mixed boundary conditions on the asymptotically   $AdS_2$ boundary. 
	
	\item The path integral over a bordered Riemann surfaces of constant negative curvature that has $n$ boundaries and genus $g$. For such surfaces, we  fix the holonomies $h_1$, $h_2$, $\dots$, $h_n$ and the  lengths of the geodesic boundaries $b_1$, \dots, $b_n$, across all $n$ boundaries. 
\end{itemize}
By gluing the above geometries along the side where the boundary is a geodesic, we are able to obtain any constant negative curvature geometry that is orientable (with arbitrary genus $g$ and an arbitrary number of boundaries $n$) and has asymptotically $AdS_2$ boundaries. 

We start by computing the path integral over the trumpet geometry, by integrating out the gauge field. Using \eqref{eq:general-YM-partition-function} we find 
\be
\label{eq:JT+YM-path-integral-trumpet}
Z_{\substack{\text{JTYM}\\\text{Dirichlet}}}^\text{trumpet}  =  \int Dg_{\mu\nu}D\Phi e^{-S_{JT}[g_{\mu \nu},\, \Phi]} \left(\sum_R \chi_{R}(g_{nAdS_2}) \chi_{R}(g_b) e^{- \frac{C_2(R) \int_{\cM_\text{T}} d^2x \sqrt{g}\left[\tilde e -\tilde e_\Phi \Phi(x) \right]}{2} } \right)
\ee
where the area term depends on the bulk metric configuration. Integrating out the dilaton field $\Phi$ in each representation sector $R$, we localize over trumpets with constant negative curvature (once again, with $\tilde \Lambda = -2 - \tilde e_\Phi C_2(R)$), whose boundary degrees of freedom are given by Schwarzian field describing the wiggles on the nearly-$AdS_2$ boundary. The trumpet area term is given by Gauss-Bonnet:
\be
\int_{\cM_T}d^2 x \tilde e \sqrt{g} = -\frac{ \tilde e }{2 + {\tilde e_\Phi C_2(R)}}\int_{\cM_T} d^2 x \sqrt{g}\cR = \frac{\tilde e}{1 + \frac{\tilde e_\Phi C_2(R)}{2}}\int_{\partial \cM_T} du \sqrt{g_{uu}}\, \cK\,,\ee
where, for the trumpet, we have used the Euler characteristic $\chi(\cM_T) = 0$a and the fact that the extrinsic curvature has $\cK =0$ along the geodesic boundary. Above we have denoted $\partial \cM_T$ to be the boundary of the trumpet with asymptotically $AdS_2$ boundary conditions. Thus, the path integral becomes
\be 
\label{eq:trumpet-iterm-step1}
Z_{\substack{\text{JTYM}\\\text{Dirichlet}}}^\text{trumpet}= \sum_{R}\chi_{R}(h_{nAdS_2}) \chi_{R}(h_b)  \int \frac{d\mu(\tau)}{\grp{U}(1)} e^{\left(\frac{\Phi_b}{\epsilon}- \tilde e \tilde C_2(R)\right) \int_{\partial \cM_T} du\, {\sqrt{g_{uu}}} \,\cK - \frac{\Phi_b}{\epsilon}\int_{\partial \cM_T} du\, \sqrt{g_{uu}}}\,, 
\ee 
The metric can be parametrized as  $ds^2 = d\sigma^2 + \cosh^2(\sigma) d\tau^2$, with the periodic identification $\tau(u) \sim \tau(u)+b$. Writing the extrinsic curvature \eqref{eq:extrinsic-curvature-Schw-field} in these coordinates, the path integral becomes \cite{Saad:2019lba} 
\be 
Z_{\substack{\text{JTYM}\\\text{Dirichlet}}}^\text{trumpet} = \sum_{R}\chi_{R}(h_{nAdS_2}) \chi_{R}(h_b)  \int \frac{d\mu(\tau)}{\grp{U}(1)} e^{-\frac{\Phi_b \beta}{\epsilon^2}+\left({\Phi_b} - \epsilon \,\tilde e \,\tilde C_2(R) \right) \int_0^\beta du \left(\frac{1}{\epsilon^2} + \{\exp[-\tau(u)], u\} \right)}\,,
\ee
where we note that the periodic identification of $\tau$ breaks the $SL(2, \mR)$ isometry of the disk down to $\grp{U}(1)$ translations of $\tau$. Once again performing the one-loop exact path integral over the Schwarzian field $\tau(u)$ \cite{Stanford:2017thb, Saad:2019lba}, we find 
\be 
\label{eq:JT+YM-trumpet-part-function}
Z_{\substack{\text{JTYM}\\\text{Dirichlet}}}^\text{trumpet} &= \pi \sum_{R}\chi_{R}(h_{nAdS_2}) \chi_{R}(h_b) e^{-\frac{ \tilde C_2(R) \beta}{\epsilon}}  \int \frac{ds}{\pi^{1/2}}  \cos( b s) e^{- \frac{\beta}{2\left(\Phi_b - \epsilon \tilde C_2(R)\right)} s^2}\nn\\&=  
\sum_{R} \chi_{R}(h_{nAdS_2}) \chi_{R}(h_b) \left(\frac{\Phi_b - \epsilon \tilde C_2(R)}{ 2\pi\beta}\right)^{1/2}e^{-\frac{\Phi_b b^2}{2\beta} - {\tilde C_2(R)}\left(  \frac{\beta}{\epsilon} - \frac{\epsilon b^2}{2\beta}\right)}\nn\\&=\sum_{R} \chi_{R}(h_{nAdS_2}) \chi_{R}(h_b) \left(\frac{\tilde \Phi_b(R)}{2\pi \beta}\right)^{1/2}e^{-\frac{\tilde \Phi_b(R) b^2}{2\beta} -\frac{\beta \,\tilde e\, {\tilde C_2(R)}   }{\epsilon} } \,,
\ee
where $\tilde C_2(R)$ is given by \eqref{eq:defintion-C2-tilde} and $\tilde \Phi_b(R)$ is given by \eqref{eq:definition-phi-b-R}. We again encounter a $1/\e$ divergence appearing in the exponent in \eqref{eq:JT+YM-trumpet-part-function} which is due to the divergence of the area of the trumpet at finite values of $b$. 

In order to eliminate such a divergence we consider the change of boundary conditions for the gauge field given by \eqref{eq:eq-with-gc2-non-zero} at the nearly-$AdS_2$ boundary.  As explained in section \ref{sec:counter-terms-from-a-change-in-boundary-cond} this change can be  implemented by inserting the boundary condition changing defect.  The insertion of such a defect indeed leads to a convergent term in the exponent  in \eqref{eq:JT+YM-trumpet-part-function}, as can be seen from the resulting partition function
\be 
\label{eq:JT+YM-trumpet-part-function-mixed}
Z_{\substack{\text{JTYM}\\\text{mixed}}}^\text{trumpet}(\Phi_b, \beta, b, h_{nAdS_2}, h_b)  = \sum_R  \chi_{R}(h_{nAdS_2}) \chi_{R}(h_b) \left(\frac{\tilde \Phi_b(R)}{2\pi \beta}\right)^{1/2} e^{ \, - \frac{\tilde \Phi_b(R) b^2}{2\beta} } e^{-\tilde e_b \b C_2(R)}\,.
\ee

We now compute the partition function associated to the $n$-bordered Riemann surface of genus $g$, which we denote by $Z^{(g, n)}_{\substack{\text{JTYM}\\\text{Dirichlet} }}(b_j, \, h_j) $. Integrating out the gauge field by using \eqref{eq:general-YM-partition-function} and then integrating out the dilaton, we find
\be 
Z^{(g, n)}_{\substack{\text{JTYM}\\\text{Dirichlet}}}(b_j, \, h_j) &=   \sum_R (\dim R)^{2-n-2g} \chi_R(h_1)\dots \chi_R(h_n)  \,e^{\chi(\cM_{g, n})S_0}\nn\\&\times  \int Dg^{\mu\nu}\delta\left(R+2+{\tilde e_\Phi C_2(R)}\right) \, e^{- \frac{ \tilde e\, C_2(R) \int_{\cM_{g,\,n}} d^2x \sqrt{g}}{2} }\,,
\ee
where $\int_{\cM_{g,\,n}} d^2x \sqrt{g}$ is the area of the constant curvature manifold. From Gauss-Bonnet, we find 
\be 
\int_{\cM_{g,\,n}}d^2x \sqrt{g} =-\frac{1}{2 + {\tilde e\, C_2(R)}}\int_{\cM_{g,\,n}}d^2x \sqrt{g} \cR = \frac{2g+n-2}{1  + \frac{ \tilde e  C_2(R)}{2}} \,,
\ee
where we have used $\chi(\cM_{g,n})=  2-2g-n$ and have used the fact that the extrinsic curvature vanishes on the geodesic borders of this Riemann surface. Thus, the partition function of the $n$-bordered Riemann surface is given by
\be 
\label{eq:JT+YM-riemann-n-bordered}
Z^{(g, n)}_{\substack{\text{JTYM}\\\text{Dirichlet}}}(b_j, \, h_j) &= \sum_R\chi_R(h_1)\dots \chi_R(h_n)  \Vol_{g, n}(b_1,\, \dots,\,b_n)  \left(\dim R\, e^{S_0}\,e^{{\tilde e \tilde C_2(R)}}\right)^{\chi(\cM_{g, n})} \,,
\ee
where $\Vol_{g, n}(b_1,\, \dots,\,b_n)$ is the volume of the moduli space of $n$-bordered Riemann surfaces with constant curvature. A recursion relation for these volumes was found in \cite{mirzakhani2007simple} (see \cite{Dijkgraaf:2018vnm} for a review). It was later showed that this recursion relation can be related to the ``topological recursion'' seen in the genus expansion of a double-scaled matrix integral \cite{Eynard:2007fi}. As we discuss later, this relation proves important when discussing the matrix integral interpretation of the genus expansion in pure and gauged JT gravity.

Using \eqref{eq:JT+YM-trumpet-part-function} or \eqref{eq:JT+YM-trumpet-part-function-mixed}, together with \eqref{eq:JT+YM-riemann-n-bordered} we now determine the partition function on surfaces with arbitrary genus. 

\subsection{The genus expansion}
\label{sec:genus-expansion}

Using the gluing rules outlined above, the partition function when summing over all orientable manifold is given by the genus expansion, 
\be
Z_{\substack{\text{JTBF}\\\text{mixed}}}^{n=1}(\Phi_{b}, \beta, h) =  Z_{\substack{\text{JTBF}\\\text{mixed}}}^{\text{disk}}(\Phi_{b}, \beta, h)  + \sum_{g=1}^\infty \int d\tilde h  \int db \, b\, Z_{\substack{\text{JTBF}\\\text{mixed}}}^{\text{trumpet}}(\Phi_{b}, \beta, b, h, \tilde h)  Z_{\substack{\text{JTBF}\\\text{Dirichlet}}}^{\text{(g, 1)}}(b, \tilde h)\,.
\ee
Putting \eqref{eq:JT+YM-disk-part-function}, \eqref{eq:JT+YM-trumpet-part-function} and \eqref{eq:JT+YM-riemann-n-bordered} together, we find the genus expansion for the gravitational partition function for surfaces with a single boundary on which  we fix Dirichlet boundary conditions for the gauge field: 
\be 
\label{eq:YM-genus-exp-partition-function}
Z_{\substack{\text{JTYM}\\\text{Dirichlet}}}^{n=1}(\Phi_b\,, \beta, \,h) &= \sum_R \chi_R(h) e^{-\frac{\tilde C_2(R) \beta}{\e} }\bigg[\left(\dim(R)e^{ \tilde e \tilde C_2(R)} e^{S_0}\right) \, \frac{1}{(2\pi)^{1/2}}\left(\frac{\tilde \Phi_b(R) }{ \beta}\right)^{3/2} e^{\frac{2\pi^2 \tilde \Phi_b(R)}\b }\,  \\ &+ \sum_{g=1}^\infty  \left(\dim(R)e^{\tilde e \tilde C_2(R)} e^{S_0}\right)^{\chi(\cM_{g,\,1})}  \left(\frac{\tilde \Phi_b(R)}{2\pi  \beta}\right)^{\frac{1}2}  \int_0^\infty db\, b\, e^{- \frac{\tilde\Phi_b(R) b^2} {2\b}} \Vol_{g, 1}^\a(b) \bigg]\nn\,.
\ee
It is instructive to express this result in terms of $ Z_{g,1}(\Phi_{b_1}, \dots, \Phi_{b_n}, \b_1, \dots, \b_n)$, the contribution of surfaces of genus $g$ with $n$ asymptotically $AdS_2$ boundaries to the pure JT gravity partition function. Thus \eqref{eq:YM-genus-exp-partition-function} can be compared to the result in pure JT gravity:
\be 
\label{eq:YM-genus-exp-partition-function-epsilon-0}
&Z_{\text{JT}}^{n=1}(\Phi_{b}, \b)= \sum_{g=0}^\infty e^{S_0\chi(\cM_{g, 1})} Z_{g, 1}( \b/\Phi_b) \nn \\&\xRightarrow[\substack{\text{adding Yang-Mills} \\ \text{term}}]{} \,\,\,\,
Z_{\substack{\text{JTYM}\\\text{Dirichlet}}}^{n=1}(\Phi_b\,, \beta, \,h) = \sum_R \chi_R(h) e^{-\frac{\tilde C_2(R) \beta}{\e} }\\\ & \hspace{5.9cm} \times \left[\sum_{g=0}^\infty \left(\dim(R)e^{ \tilde e \tilde C_2(R)} e^{S_0}\right)^{\chi(\cM_{g,1})} Z_{g,1}\left(\b/\tilde \Phi_{b}(R)\right)\right]\nn \,,
\ee
where  we have absorbed the entropy dependence $e^{\chi(\cM_{g, n}) S_0}$, in $Z_{g,n}(\Phi_{b_1}, \dots, \Phi_{b_n}, \b_1, \dots, \b_n)$: $ Z^{(g,n)}_\text{JT}(\Phi_{b_1}, \dots, \Phi_{b_n}, \b_1, \dots, \b_n) \equiv e^{\chi(\cM_{g, n})  S_0}  Z_{g,n}( \b_1/\Phi_{b_1}, \dots, \b_n/\Phi_{b_n})$ (from the partition function on trumpet geometries, one immediately deduces that $Z_{g, n}$ solely depends on the ratios $\b_j/\Phi_{b_j}$). The coefficients $  Z_{g,n}(\b_j/\Phi_{b_j}) \equiv Z_{g,n}(\b_1/\Phi_{b_1}, \dots, \b_n/\Phi_{b_n})$ are in fact those encountered in the genus expansion of correlators of the partition function operator in the double-scaling of the certain  matrix integral that we have previously mentioned. 

We can also determine the partition function of the space which has $n$ boundaries, 
\be 
\label{eq:YM-genus-exp-partition-function-n-boundaries}
&Z_{\substack{\text{JTYM}\\\text{Dirichlet}}}^{n}(\Phi_{b,j}, \beta_j, h_j) =\sum_R\chi_R(h_1)\dots \chi_R(h_n) e^{- \frac{ \tilde e \tilde C_2(R) \sum_{j=1}^n \beta_j}{\e}} \bigg[ \sum_{g = 0}^\infty (\dim R e^{-\tilde e \tilde C_2(R)}e^{S_0})^{\chi(\cM_{g,\,n})}\nn \\  &\times \left(\frac{\tilde \Phi_{b,1}(R) \dots \tilde \Phi_{b,1}(R)}{\pi^{n} \beta_1\dots \b_n }\right)^{\frac{1}2}   \int_0^\infty db_1  b_1 \dots \int_0^\infty db_n b_n  \, \Vol_{g, n}^\a(b_{1, \dots, n}) e^{- \sum_{i=1}^n \frac{\tilde\Phi_{b, i}(R) {b^2_i}}{\b_i}} \bigg] \,.
\ee
In terms of the coefficients $Z_{g,n}(\b_j/\Phi_{b_j})$, this becomes
\be 
\label{eq:YM-genus-exp-partition-function-n-boundaries-epsilon-0}
Z_{\substack{\text{JTYM}\\\text{Dirichlet}}}^{n}(\Phi_{b_j}\,, \beta_j, \,h_j) &= \sum_R \chi_R(h_1) \dots \chi_R(h_n)  e^{- \frac{ \tilde e \tilde C_2(R) \sum_{j=1}^n \beta_j}{\e}} \nn \\&\qquad \times\left[\sum_{g=0}^\infty \left(\dim(R)e^{ \tilde e \tilde C_2(R)} e^{S_0}\right)^{\chi(\cM_{g,n})} Z_{g,n}\left(\b_j/\tilde \Phi_{b_j}(R)\right)\right] \,.
\ee
In the $\e\to0$ limit, $\tilde \Phi_{b_j}(R) = \Phi_{b_j}$ for all $j$ and, in the square parenthesis in \eqref{eq:YM-genus-exp-partition-function-epsilon-0} and \eqref{eq:YM-genus-exp-partition-function-n-boundaries-epsilon-0}, the dependence  on the irreducible representation $R$ can be absorbed in the overall entropy on the disk $S_0 \to S_0 - \tilde e \tilde C_2(R) -  \log \dim R$; thus, the density of states associated to each representation sector is the same as in pure JT gravity. As we explain shortly, this serves as a useful guide in determining the matrix integral derivation of \eqref{eq:YM-genus-exp-partition-function}.

With Dirichlet boundary conditions and in the limit $\e  \to 0$, the singlet representation dominates in the sum over representations due to the $1/\e$ divergence in the first exponent of \eqref{eq:YM-genus-exp-partition-function} or \eqref{eq:YM-genus-exp-partition-function-n-boundaries}. This behavior can be altered by the change of boundary conditions \eqref{eq:boundary-condition-that-makes-things-finite} presented in section \ref{sec:counter-terms-from-a-change-in-boundary-cond} or, equivalently, by the addition of a defect close to each one of the $n$ boundaries of the manifold.  When using the boundary condition changing defect, the result in each representation sector gets regularized such that
\be 
\label{eq:w-defect}
Z_{\substack{\text{JTYM}\\\text{mixed}}}^{n}(\Phi_{b_j}, \beta_j, h_j) &= \sum_R  Z_{\substack{\text{JTYM}\\\text{Dirichlet}}}^{n}(\Phi_{b_j}, \beta_j, h_j)_R\,\,\, e^{\left(\frac{\tilde e \tilde C_2(R)}{\e}  \right)\left(\sum_{j=1}^n \beta_j \right) - \frac{1}2 C_2(R)\left(\sum_{j=1}^n \tilde e_{b_j}\beta_j \right)}  \,,
\ee 
where $Z_{\substack{\text{JTYM}\\\text{Dirichlet}}}^{n}(\Phi_{b_j}, \beta_j, h_j)_R\,$ is the contribution of the representation $R$ to the sum in \eqref{eq:YM-genus-exp-partition-function-n-boundaries-epsilon-0}. Above, the mixed boundary condition obtained from \eqref{eq:boundary-condition-that-makes-things-finite} with a coupling $\tilde e_{b_j}$ is considered for each of the $n$ boundaries.

The result \eqref{eq:w-defect}  simplifies further in the (topological) weak gauge coupling limit
\be 
\label{eq:BF-genus-exp-partition-function-n-boundaries}
Z_{\substack{\text{JTBF}\\\text{mixed}}}^{n} 
&(\Phi_{b_j}, \beta_j, h_j)  = \sum_R\chi_R(h_1)\dots \chi_R(h_n) e^{- \frac{ C_2(R) \sum_{i=j}^n \tilde e_{b_j} \beta_j }{2}} \nn\\ &\qquad \qquad \qquad \times\left[\sum_{g=0}^\infty \left(\dim(R) e^{S_0}\right)^{\chi(\cM_{g,n})}Z_{g,n}(\b_j/\Phi_{b_j})\right]  \,,
\ee
where  we have used the boundary condition \eqref{eq:simple-eq}
\be
\label{eq:BF-boundary-cond-genus-exp}
\delta(A_u + i \sqrt{g_{uu}} \, e_{b_j} \,\phi)|_{(\partial \cM)_j} = 0  \,,
\ee
for each of the $n$-boundaries. 

It is worth pondering the interpretation of \eqref{eq:BF-genus-exp-partition-function-n-boundaries}. While for the disk contribution to the partition function \eqref{eq:BF-partition-function-on-a-disk}, the gravitational and topological theories were fully decoupled, the topological theory of course couples to JT gravity through the genus expansion.

One case in which the sum over $R$ can be explicitly computed is when $\tilde e_b = 0$, for which the sum over irreducible representations evaluates to the volume of flat $G$ connection on each surface of genus $g$. For instance, in the case when $G= \text{SU}(2)$ all such volumes have been computed explicitly in \cite{Witten:1991we}. More generally for any $G$, when focusing on surfaces with a single boundary ($n=1$) and setting $h \neq e$, the contribution from surfaces with disk topology to \eqref{eq:BF-genus-exp-partition-function-n-boundaries} vanishes, and the leading contribution is given by surfaces with the topology of a  punctured torus. In this limit, the contribution of non-trivial topology is, in fact, visible even at large values of $e^{S_0}$. In the limit in which $h \to e$, the contribution from surfaces with the topology of a disk or a punctured torus are divergent; in the case when $G=\text{SU}(2)$ such divergences behave as $O(1/\tilde e_b^{3/2})$ and $O(1/\tilde e_b^{1/2})$ respectively. The leading contribution for all other surfaces behaves as $O(1)$. In other words, this limit further isolates the contribution of surfaces with disk and punctured torus topology in the partition function.

\subsection{Matrix integral description} 
\label{sec:matrix-model-description}

\subsubsection*{Reviewing the correspondence between pure JT gravity and matrix integrals}

In order to understand how to construct the matrix integral that  reproduces the genus expansion in the gravitational gauge theory \eqref{eq:JT-action+yang-mills} we first briefly review this correspondence in the case of pure JT gravity, following \cite{Saad:2019lba}. Consider a Hermitian matrix integral over $N \times N$ Hermitian matrices with some potential $S[H]$:
\be 
\label{eq:standard-matrix-partition-function}
\cZ = \int dH e^{-S(H)}\,, \qquad S[H]\equiv N\left( \frac{1}2 \Tr_N H^2 + \sum_{j\geq 3} \frac{t_j}{j} \Tr_N  H^j\right)\,,
\ee
where $\Tr_N$ is the standard trace over $N \times N$ matrices. An observable that proves important in the genus expansion of the gravitational theory is the correlator of the thermal partition function operator, $Z(\b) = \Tr_N \,e^{-\b H}$. Correlators of such operators have an expansion in $1/N$, where each order in $N$ can be computed by looking at orientable double-line graphs of fixed genus \cite{Brezin:1977sv, tHooft:1973alw} (for a review see \cite{brezin2006applications}). Consequently, this is known as the genus expansion of the matrix model \eqref{eq:standard-matrix-partition-function}.

For a general set of potentials $S[H]$, each order in the expansion can be determined in terms of a single function $\rho_0(E)$. This function is simply the leading density of eigenvalues in matrices with $N \to \oo$. Consider the double-scaling limit of \eqref{eq:standard-matrix-partition-function}, in which the size of the matrix $N \to \oo$ and in which 
we focus on the edge of the eigenvalue distribution of the matrix $H$, where the eigenvalue density remains finite and is denoted by $e^{S_0}$. The expansion of the correlators mentioned above can now be expressed in terms of $e^{S_0}$ instead of the size of the matrix $N$. In this double-scaled limit the density of eigenvalues $\rho_0(E)$ is not necessarily normalizable and with an appropriate choice of potential $S[H]$, $\rho_0(E)$ can be set to be equal to the energy density in the Schwarzian theory \eqref{eq:Schwarzian-path}
\be
\label{eq:density-eigenvalues-matrix-model}
\rho_0(E) =  \frac{\Phi_b}{2\pi^2} \sinh(2\pi \sqrt{2 \Phi_b E})\,.
\ee
 In the remainder of this subsection, we follow  \cite{Saad:2019lba} and normalize 
\be
\label{eq:simplication-normalization}
\Phi_{b_j} \equiv  1/2\,, \qquad Z_{g,n}(\b_j) \equiv Z_{g,n}(\b_j/\Phi_{b_j})\,,
\ee 
for all the $n$ boundaries of the theory, and use the short-hand notation in \eqref{eq:simplication-normalization}. As previously emphasized, choosing  \eqref{eq:density-eigenvalues-matrix-model} determines all orders (in the double scaled limit) in the $e^{-S_0}$ perturbative expansion for correlators of operators such as  $Z(\b) = \Tr_N\,e^{-\b H}$ \cite{Eynard:2004mh}. The result found by \cite{Saad:2019lba}, building on the ideas of \cite{Eynard:2007fi}, is that the genus expansion in pure JT gravity agrees with the $e^{S_0}$ genus expansion of the double-scaled matrix integral whose eigenvalue density of states is given by \eqref{eq:density-eigenvalues-matrix-model}:
\be
Z_\text{JT}^{n}(\b_1, \dots, \b_n) =  \<Z(\b_1)\dots Z(\b_n)\>= \sum_{g} Z_{g, n}(\b_j) e^{-S_0 \chi(\cM_{g, n})}\,.
\ee

The density of states \eqref{eq:density-eigenvalues-matrix-model} was shown to arise when considering the matrix integral associated to  the $(2, p)$ minimal string.  Specifically, this latter theory was shown to be related to a matrix integral whose density of eigenvalues is given by \cite{Moore:1991ir, Edwards:1991jx, Seiberg:2003nm, Seiberg:2004at, Seiberg:toAppear}
\be 
\label{eq:(2,p)-density-of-states}
\rho_0(E) \sim \sinh \left(\frac{p}2 \text{arccosh}\left(1+\frac{E}{\kappa}\right)\right)\,,
\ee
where $\kappa$ is set by the value of $p$ and by the value of $\mu$ from the Liouville theory which is coupled to the $(2, p)$ minimal model \cite{Polyakov:1981rd}. Taking the $p \to \infty$ limit in \eqref{eq:(2,p)-density-of-states} and rescaling $E$ appropriately, one recovers the density of states \eqref{eq:density-eigenvalues-matrix-model}. Consequently, one can conclude that the double-scaled matrix integral which gives rise to the genus expansion in pure JT gravity is the same as the matrix integral which corresponds to the $(2, \oo)$ minimal string.

Our goal is to extend this analysis and find a modification of the matrix integral presented in \eqref{eq:standard-matrix-partition-function} such that the partition function includes the contributions from the gauge field that appeared in the genus expansion of JT gravity coupled to Yang-Mills theory. As we will show below, there are two possible equivalent modifications of the matrix integral \eqref{eq:standard-matrix-partition-function}:
\begin{itemize}
	\item As shown in subsection \ref{sec:genus-expansion}, in the $\e \to 0$ limit, the contribution of the gauge degrees of freedom to the partition function can be absorbed in each representation sector $R$ by an $R$-dependent shift of the entropy $S_0$. This indicates that instead of obtaining the gravitational gauge theory partition function from a single double-scaled matrix integral, one can obtain the contribution of the gauge degrees of freedom from a collection of double-scaled matrix integrals, where each matrix $H^R$ is associated to a different irreducible representation $R$ of $G$. The size of $H^R$ is proportional to the dimension of the representation $R$. 
	
	\item In order to obtain such a collection of random matrix ensembles in a natural way, we consider a different modification of the matrix integral  \eqref{eq:standard-matrix-partition-function}. Specifically, instead of considering a Hermitian matrix whose elements are complex, we rather consider matrices whose elements are complex functions on the group $G$ (equivalently, they are elements of the group algebra $\mathbb C[G]$). Equivalently, as we will discuss shortly, one can consider matrices that in addition to the two discrete labels characterizing the elements, have two additional labels in the group $G$ and are invariant under $G$ transformations. By defining the appropriate traces over such matrices, we show that such matrix integrals are equivalent to the previously mentioned collection of matrix integrals, which in turn reproduce the genus expansion in the gravitational gauge theory. This latter model serves as our starting point.
\end{itemize}
In our analysis, we first consider the necessary modifications of the matrix integral \eqref{eq:standard-matrix-partition-function} which reproduce the results from the weak gauge coupling limit and, afterward, we discuss the case of general coupling.

\subsubsection*{Modifying the matrix integral: the weakly coupled limit}

We start by modifying the structure of the Hermitian matrix $H$, by supplementing the discrete indices $i, j \in 1, \dots, N$ that label the elements $H_{ij}$, by two additional elements $g, h\in G$.\footnote{Here we consider the case when $G$ is a compact Lie group, while the past discussion of matrix integrals of this type focused solely on the case when  $G$ is a finite group \cite{Solovyov,mulase2002generating, mulase2005non, mulase2007geometry}. } Thus, elements of the matrix are given by $H_{(i, g), (j, h)}$. For such matrices, their multiplication is defined by 
\be 
\label{eq:matrix-multiplication-G-matrices}
(H M)_{(i, g), (j, \tilde g)} = \sum_{k=1}^N\int d h H_{(i, g), (k, h)} M_{(k, h), (j, \tilde g)} \,
\ee
where $dh$ is the Haar measure defined on the group, normalized by the volume of group such that $\int dh = 1$.

The (left) action of the group element $f\in G$ on the matrix $H_{(i, g), (j, h)}$ is defined as $H_{(i, g), (j, h)} \to H_{(i, fg), (j, fh)}$, where we emphasize that the integer indices remain unaltered. In order to reproduce the  collection of matrix integrals that we have previously mentioned, in this work  we are interested in $G$-invariant matrices \cite{Solovyov}, defined by the property
\be 
\label{eq:G-invariant-definition}
H_{(i, g), \,(j, h)} = H_{(i, f g), \,(j, f h)}\,, 
\ee
for any $f \in G$. For such matrices one can therefore, define $H_{i, j}(g)$ by using \cite{Solovyov} 
\be
\label{eq:G-invariant-algebra-definition}
H_{(i, g),\,(j, h)}= H_{(i, e),\,(j, g^{-1} h)} \equiv H_{i, j}(g^{-1}h) \in \mathbb C[G]
\ee
where $ \mathbb C[G]$ is the complex group algebra associated to $G$. In other words, each element $H_{i, j}$, instead of being viewed as a complex element, can be viewed as a function on the group $G \to \mathbb C$.  For $G$-invariant matrices, the product \eqref{eq:matrix-multiplication-G-matrices} simplifies to  
\be
(H M)_{i j}(g)= \sum_{k=1}^N \int dh H_{ik}(h) M_{kj}(h^{-1}g)\,,
\ee
where the integral over $h$ simply gives the convolution of functions defined on the group $G$. 

We wish to understand the free energy of a matrix model whose action is given by \cite{Solovyov}
\be
\label{eq:weakly-coupled-matrix-model-action}
S[H] =  N \left[ \frac{1}2 \,  \chi_\text{el}(H^2) + \sum_{j\geq 3} \frac{t_j}{j} \chi_\text{el}(H^j)\right]\,,
\ee
where $H$ is a $G$-invariant matrix defined through \eqref{eq:G-invariant-algebra-definition} and $\chi_\text{el}$ is the trace which, at first, we take to be in the elementary representation of the group $G$. The trace in the (reducible) elementary representation of the group is given by evaluating the $H$ in \eqref{eq:G-invariant-algebra-definition} on the identity element $e$ of the group $G$,\footnote{One might contemplate whether \eqref{eq:definition-elemetary-trace} is indeed a well-defined trace. We, in fact, show that the trace is still valid when replacing $\delta(\tilde h)$ in \eqref{eq:trace-cyl} by an arbitrary trace-class function, $\sigma(\tilde h^{-1})$. This can, of course, be viewed as a trace in an arbitrary (most often) reducible representation of $G$. To show this, we have
	\be
	\chi_{f}(HM) &= \int d\tilde h\, dh \, \sigma(\tilde h^{-1}) \sum_{i, k=1}^n H_{ik}(h)M_{ki}(h^{-1}\tilde h) =\sum_R \sigma_R  \int dh \sum_{i, k=1}^n H_{ik}(h) U_R(h^{-1})(M_{ki})_R \nn  \\  
	&= \sum_R \sigma_R \sum_{i, k=1}^n \sum_{m, p=1}^{\dim R} (H_{ik})_{R, m}^p (M_{ki})_{R, p}^m = \chi_f( M H) \,\, \Longrightarrow \,\,  \chi_f([H, M]) =0\,, 
	\ee
	which indeed implies that $\chi_\text{el}(\dots)$ is a well-defined trace. Above, we have used the fact that for trace-class function $\sigma(\tilde h^{-1})$, there is a decomposition $\sigma(\tilde h^{-1}) = \sum_R \sigma_R \,\chi_R(\tilde h^{-1})$ .
} 
\be
\label{eq:definition-elemetary-trace}
\chi_{\text{el}}(H) &\equiv \sum_{i = 1}^N  H_{i, i}(e)= \int d\tilde h \,\delta(\tilde h) \sum_{i = 1}^N H_{i,i}(\tilde h) = \sum_{i=1}^N \sum_R \int d\tilde h\,(\dim R) \chi_R(\tilde h^{-1})  H_{i,i}(\tilde h) \nn \\  &=\sum_R (\dim R) \sum_{i=1}^{ \,N} \sum_{j=1}^{ \dim R}  (H_{i, i})_{R, j}^j = \sum_R (\dim R) \Tr_{(\dim R)\,N}(H_R)\,,
\ee
Here, we have used the decomposition $H_{i, j}(g) = \sum_R \sum_{k, l= 1}^{\dim R}(\dim R) U_{R, l}^k(g) (H_{i, j})_{R, l}^k$ where $ U_{R, l}^k(g)$ are the matrix elements of $G$.\footnote{Note that $H_{i, j}(g)$ is generically not trace class since $H_{i, j}(h^{-1}g h) \neq H_{i, j}(g)$, for generic group elements $g$ and $h$. Thus, $H_{i, j}(g)$ should be decomposed in the matrix elements of $G$, $ U_{R, l}^k(g)$, instead of its~characters~$\chi_R(g)$.} Thus, we can view $H_R$ as an $(\dim R\, N)\times (\dim R\, N)$ matrix and, above, $\Tr_{\dim R\, N}(\dots)$ is the standard trace over such matrices.  Furthermore, to evaluate the trace in the elementary representation for products of such matrices we can use
\be 
\label{eq:product-decomposition-matrix}
(H^k)_{i_1, i_{k+1}}(h) =\sum_R&\sum_{\substack{j_1,\,\dots ,\\ j_{k+1}= 1}}^{\dim R} (\dim R)  \sum_{\substack{i_2,\,\dots ,\\ i_{k}= 1}}^{N}   (H_{i_1, i_2})_{R, j_2}^{j_1}   \dots  (H_{i_{k}, i_{k+1}})_{R, j_{k+1}}^{j_{k}} U_{R, j_{k+1}}^{j_1}(h)\,,
\ee
which yields
\be 
\label{eq:convolution-CG}
\chi_\text{el}(H^k) = \sum_R (\dim R) (H_{i_1, i_2})_{R, j_1}^{j_2}    \dots  (H_{i_{k}, i_{1}})_{R, j_k}^{j_1}= \sum_R (\dim R)\Tr_{(\dim R) N} (H_R^k)\,.
\ee

Thus, the action \eqref{eq:weakly-coupled-matrix-model-action} becomes \cite{Solovyov}
\be
\label{eq:weakly-coupled-action-ensemble-of-matrices}
S[H] =  \sum_R N(\dim R) \left[ \frac{1}2 \,\Tr_{(\dim R) N}(H^2_R) + \sum_{j\geq 3} \frac{t_j}{j} \,\Tr_{(\dim R) N }(H^j_R)\right]\,,
\ee
which is the same as a collection of decoupled GUE-like matrix integrals, where each matrix $H_R$ is Hermitian, is associated to the representation $R$, and has dimension $(\dim R\, N)\times (\dim R\, N)$.  Such matrix integrals are truly decoupled if the measure for the path integral in \eqref{eq:weakly-coupled-matrix-model-action} associated to $H(g)$ is chosen such that it reduces to the standard measure for GUE-like matrix integrals associated to $dH_R$. To summarize, this result simply comes from the harmonic decomposition onto different representation sectors of our initial Hermitian matrices whose elements were in $\mathbb C[G]$.

We now compare correlation functions in the standard Hermitian matrix model with $N \times N$ matrices, to those in the model whose matrix elements are part of the group algebra $\mathbb C[G]$, when having the same couplings in both models. Equivalently, we can compare such correlators to those in the collection of matrix models in \eqref{eq:weakly-coupled-action-ensemble-of-matrices}. In order to do this we compare correlation functions of the trace of $ e^{-\b H}$ to the gravitational answer. When   $H$ is an $N \times N$ Hermitian matrix the trace is the standard $\Tr_N  e^{-\b H}$. However, when $H$ has elements in $\mathbb C[G]$ the trace needs to be modified :
\be
\label{eq:new-trace-def-cyl}
Z(\b)= \Tr_N\left( e^{-\b H}\right)\,\qquad\Rightarrow \qquad Z_\text{cyl.}(h, E) = \chi_{\text{cyl.},\,\, h}(e^{-\b H})\,,
\ee 
where,
\be
\label{eq:trace-cyl}
\chi_{\text{cyl.},\,\, h}(H) &= \int d\tilde h  \, Z_{\substack{\text{BF}\\\text{mixed}}}^\text{(0, 2)}(\tilde h ^{-1}, h) \sum_{i=1}^N H_{i,i}(\tilde h) = \int d\tilde h \sum_R  \chi_R(\tilde h^{-1}) \chi_R(h) e^{-\frac{\tilde e_b \beta C_2(R)}{2}} \sum_{i=1}^N H_{i,i}(\tilde h)\nn \\ &= \sum_R  \chi_R(h) e^{-\frac{\tilde e_b \beta C_2(R)}{2}}  \Tr_{(\dim R)N }(H_R)\,,
\ee
where $Z_{\substack{\text{BF}\\\text{mixed}}}^\text{(0, 2)}(g^{-1}, h)$ is the partition function of BF theory on the cylinder given by \eqref{eq:general-YM-partition-function}, where on one of the edges we use Dirichlet boundary conditions and on the other we impose the mixed boundary conditions discussed for BF theory. 

Consequently, using the multiplication properties for the $\mathbb C[G]$ matrices \eqref{eq:convolution-CG}, we find
\be 
\label{eq:trace-cyl-exp}
\chi_{\text{cyl.},\,\, h}(e^{-\b H})= \sum_R  \chi_R(h) e^{-\frac{\tilde e_b \beta C_2(R)}{2}}  \Tr_{(\dim R)N }(e^{-\b H_R})\,.
\ee
When $h=e$ and $\tilde e_b = 0$, one finds that $\chi_{\text{cyl.},\,\, h}(H)=  \chi_\text{el}(H)$ and this will correspond to imposing Dirichlet boundary conditions on the boundary on the gravitational gauge theory. The role of the trace \eqref{eq:new-trace-def-cyl} is to reproduce results when setting mixed boundary conditions  for each boundary of $\cM_{g, n}$ in the genus expansion of the partition function in the gravitational gauge theory.

We start by checking that by using the matrix ensemble given by \eqref{eq:weakly-coupled-matrix-model-action}, or equivalently \eqref{eq:weakly-coupled-action-ensemble-of-matrices}, together with the new definition of the trace we are able to reproduce this expansion for surfaces with a single boundary ($n=1$).  Using \eqref{eq:weakly-coupled-action-ensemble-of-matrices}, we find that in comparison to the initial regular matrix integral the one-point function of $Z_\text{cyl.}(h, \b)$ becomes
\be
\label{eq:Z-series-expansion-weakly-coupled}
\<Z(\b)\>&_\text{conn.} \simeq \sum_{g = 0}^\infty  \frac{\tilde Z_{g, 1}(\b)}{ N^{\chi(\cM_{g,n})}} \nn  \\ &  \xRightarrow[\substack{\,H_{ij} \to H_{(i, g), (j, h)}\, \\ \Tr(\dots) \to \chi_\text{el}(\dots)}]{} \,\,\,\,   \<Z_\text{cyl.}(h, \b)\>_\text{conn.} = \sum_{R}  \chi_R(h) e^{-\frac{\tilde e_b \beta C_2(R)}{2}} \<\Tr_{(\dim R) N} e^{-\b H_R} \>\nn \\&\hspace{3.1cm}  \simeq \sum_{g = 0}^\infty\, \sum_{R}  (\dim R\,N)^{-\chi(\cM_{g,1})} { \chi_R(h) e^{-\frac{\tilde e_b \beta C_2(R)}{2}} \tilde Z_{g, 1}(\b)}\,,
\ee
where $\tilde Z_{g,n}(\b_j)$ are the factors appearing in the genus expansion of the regular matrix integral \eqref{eq:standard-matrix-partition-function}. Replacing $N \to e^{S_0}$ as the expansion parameter in the double-scaling limit, and using the matrix integral discussed in \cite{Saad:2019lba}, the coefficients $\tilde Z_{g,1}(\b_j)$  in \eqref{eq:Z-series-expansion-weakly-coupled} become $Z_{g,1}(\b_j)$ which gives the contribution of surfaces of genus $g$ with $n$-boundaries to the JT gravity path integral. Thus,  we find that in the double-scaling limit the perturbative expansion \eqref{eq:Z-series-expansion-weakly-coupled} matches the genus expansion in the weakly coupled gravitational gauge theory \eqref{eq:BF-genus-exp-partition-function-n-boundaries} when $n=1$.  

Next, we check that the genus expansion of the gravitational gauge theory and the matrix integral matches for surfaces with an arbitrary number of boundaries. In order to obtain a match, we need to specify what to do with the holonomies appearing in the traces \eqref{eq:new-trace-def-cyl}. The procedure is to associate each holonomy to the boundary of a separate disk; in order to obtain a single surface with $n$-boundaries it is necessary to glue the boundaries of the $n$-disks, such that the holonomy of the resulting $n$-boundaries are $h_1$, \dots, $h_n$. This is precisely the same procedure used to glue $n$ disks into an $n$-holed sphere in Yang-Mills or BF-theory.  Such a gluing implies that instead of having a separate sum over irreducible representations for each insertion of $Z_\text{cyl.}(h_j, \b_j)$, we obtain a unique sum over $R$. We denote correlation functions after performing such a gluing as $\<\dots\>^\text{glued}(h_1, \dots, h_n)$. 

Thus, we find that the matrix integral results from pure JT gravity are modified such that\footnote{In \eqref{eq:Z-series-expansion-weakly-coupled-n-boundaries} when referring to the correlator $\<Z_\text{cyl.}(\b_1)\dots Z_\text{cyl.}(\b_n)\>_\text{conn.}^\text{glued}$ we have omitted to specify the holonomies associated to the traces $\chi_\text{cyl.}(\dots)$ appearing in $Z_\text{cyl.}$. That is because there are multiple gluing procedures that can be chosen to obtain a surface with the topology of the  $n$-holed sphere starting from $n$-disks. We thus only specify the final holonomies $h_1$, \dots, $h_n$ along the $n$-boundaries of $\cM_{g, n}$. \label{footnote:why-we-ommit-holo}}
\be
\label{eq:Z-series-expansion-weakly-coupled-n-boundaries}
&\xRightarrow[\substack{\,H_{ij} \to H_{(i, g), (j, h)}\, \\ \Tr(\dots) \to \chi_\text{el}(\dots)}]{} \,\,\,\,   \<Z_\text{cyl.}(\b_1)\dots Z_\text{cyl.}(\b_n)\>_\text{conn.}^\text{glued}(h_1, \, \dots,\, h_n) \simeq \nn \\ &\hspace{1.5cm}\simeq\sum_{R}  \chi_R(h_1) \dots  \chi_R(h_n) \,e^{-\frac{ C_2(R)\sum_{i=j}^n \tilde e_{b_j} \beta_j}{2}} \<\Tr_{(\dim R) N} e^{-\b_1 H_R}  \dots  \Tr_{(\dim R) N} e^{-\b_n  H_R} \>\nn \\&\hspace{1.5cm}= \sum_{g = 0}^\infty \sum_{R} (\dim R\,e^{S_0})^{\chi(\cM_{g,n})}  {\chi_R(h_1) \dots  \chi_R(h_n) \,e^{-\frac{ C_2(R)\sum_{i=j}^n \tilde e_{b_j} \beta_j}{2}} Z_{g, n}(\b_1, \dots, \b_n)} \,,
\ee
where  the dependence on $\Phi_{b_j}$ is realized through the overall re-scaling of the proper length $\b_j$ associated to each boundary. Of course one can use the second line in \eqref{eq:Z-series-expansion-weakly-coupled-n-boundaries} as the definition of the observable in the collection of matrix integrals \eqref{eq:weakly-coupled-action-ensemble-of-matrices}.

Thus, if we consider the matrix integral associated to the $(2, p)$ minimal string \cite{Moore:1991ir} in the $p \to \oo$ limit \cite{ Saad:2019lba} and if we promote the matrix $H$ to be of the form \eqref{eq:G-invariant-definition}, we find we can reproduce  the genus expansion in the gravitational gauge theory with the mixed boundary conditions \eqref{eq:BF-boundary-cond-genus-exp} for the gauge field (or with Dirichlet boundary conditions when $\tilde e_{b_j} = 0$ for all $j$). 

\subsubsection*{Modifying the matrix integral: arbitrary gauge couplings}

Similarly, we can reproduce the genus expansion with arbitrary gauge couplings $\tilde e$ and $\tilde e_\Phi$ for asymptotically $AdS_2$ ($\e \to 0$) boundaries by modifying the matrix integral \eqref{eq:weakly-coupled-matrix-model-action}. We start by considering mixed boundary conditions for the gauge field. Instead of taking the trace in the elementary representation we can consider the more general trace for the matrix $H$:
\be
\label{eq:definition-YM-trace}
\chi_{\text{YM}}(H) &\equiv  \int dg \, Z_\text{YM}^\text{disk}(g^{-1})\sum_{i = 1}^N (H)_{i,i}(g) = \sum_{i=1}^N \sum_R \int dg\,(\dim R) \chi_R(g^{-1})  (H)_{i,i}(g)\, e^{\tilde e \tilde C_2(R)}\nn \\  &=\sum_R (\dim R)\, e^{\tilde e \tilde C_2(R)} \,\Tr_{(\dim R)N }(H_R)\,,
\ee
where $\tilde C_2(R)$ is given by \eqref{eq:defintion-C2-tilde}. In such a case the action of the associated matrix model can be rewritten as, 
\be
\label{eq:matrix-model-YM-dual}
S[H] &= N \left[ \frac{1}2 \,  \chi_\text{YM}(H^2) + \sum_{j\geq 3} \frac{t_j}{j} \chi_\text{YM}(H^j)\right]\nn\\ &= \sum_R N(\dim R) e^{\tilde e \tilde C_2(R)}\left[ \frac{1}2 \,\Tr_{(\dim R)N }(H^2_R) + \sum_{j\geq 3} \frac{t_j}{j} \,\Tr_{(\dim R)N }(H^j_R)\right]\,,
\ee
Once again, this is a collection of decoupled matrix models, whose expansion parameter is given by  $ N(\dim R) e^{\tilde e \tilde C_2(R)}$.  In order to produce correlators with mixed boundary conditions, we again use the operator insertion $\chi_{\text{cyl,}\, h}(e^{-\b_j H})$. Thus, compared to the standard $(2, p)$ double-scaled matrix integral in the $p\to \oo$ limit, correlation functions of $Z_{\text{cyl.}}(\b_j)$ become
\be  &  \xRightarrow[\substack{\,H_{ij} \to H_{(i, g), (j, h)}\, \\ \Tr(\dots) \to \chi_\text{YM}(\dots)}]{} \,\,\,\,   \<Z_\text{cyl.}(\b_1)\dots Z_\text{cyl.}( \b_n)\>_\text{conn.}^\text{glued}(h_1, \, \dots,\,h_n) \simeq \nn \\ &\hspace{0.2cm}\simeq \sum_{g = 0}^\infty \sum_{R} (\dim R \,e^{\tilde e \tilde C_2(R)}e^{S_0})^{\chi(\cM_{g,n})} {\chi_R(h_1) \dots  \chi_R(h_n) \,e^{-\frac{ C_2(R)\sum_{i=j}^n \tilde e_{b_j} \beta_j}{2}} Z_{g, n}(\b_1,\, \dots, \,\b_n)}\,.
\ee

Thus, the matrix integral \eqref{eq:matrix-model-YM-dual} together with the cylindrical trace \eqref{eq:trace-cyl}, describe the partition function of JT gravity coupled to Yang-Mills on surfaces whose boundaries are asymptotically $AdS_2$ ($\e \to 0$). However, in section \ref{sec:genus-expansion} we have computed the first order correction in $\e$ which has led to the renormalization of the dilaton boundary value \eqref{eq:definition-phi-b-R}, $\Phi_b \Rightarrow  \tilde \Phi_b(R) = \Phi_b -\e\,\tilde e \,\tilde C_2(R)$. This renormalization changes the density of states that appears in the contribution of disk topologies in each representation sector $R$, $\rho_0(E) = \frac{\Phi_b}{2\pi^2} \sinh(2\pi \sqrt{2 \Phi_b E}) \Rightarrow \rho_0^R(E) = \frac{\tilde\Phi_b(R)}{2\pi^2} \sinh \left(2\pi \sqrt{2\tilde\Phi_b(R) E}\right) $. This implies that when setting $\Phi_b \equiv 1/2$, if rescale the temperature $\b_j$ in each representation sector, such that in the cylindrical trace \eqref{eq:trace-cyl-exp} we replace $ \Tr_{R(\dim N)} e^{-\b H_R} \Rightarrow   \Tr_{R(\dim N)} e^{-\b\frac{H_R}{1 - \e \tilde e \tilde C_2(R)} }$, we can reproduce the  genus expansion of the partition functions \eqref{eq:YM-genus-exp-partition-function} and \eqref{eq:YM-genus-exp-partition-function-n-boundaries}; as previously mentioned, this accounts for the first order correction in $\e$ to correlators of $Z_{\text{cyl.},\,h}(\b)$. Therefore, including this correction in $\e$ simply amounts to correcting the trace \eqref{eq:trace-cyl} for the matrix integral operator insertion.

Thus, the equivalence between the genus expansion of correlators in the gravitational gauge theory and the genus expansion of the matrix integral is schematically summarized in figure \ref{fig:schematic-dualtiy-matrix-1}.
  \begin{center}
  \vspace{0.1cm}
  \begin{figure}[h!]
\begin{tikzpicture}[node distance=1cm, auto]
\node[punkt] (schw) {JT gravity coupled to Yang-Mills 
in the genus expansion with Dirichlet or mixed b.c.};
\node[above=1.75cm of schw](dummy) {};
 \node[punkt, inner sep=5pt,left=1cm of dummy] (first) {A collection of GUE-like matrix integrals, with matrices $\prod_R^\otimes H_{N_R\times N_R}$ with $N_R = (\dim R)\, N$}
 edge[pil,<->, bend right=35] node[left] {} (schw);
 \node[punkt, inner sep=5pt,right=1cm of dummy]
 (second) {Matrix integral for matrices with elements in $\mathbb C[G]$ }
 edge[pil,<->, bend left=35] node[right]{} (schw)
edge[pil,<->, bend right=35] node[above]{} (first) ;

\end{tikzpicture}\\
\vspace{0em}
	\caption{\label{fig:schematic-dualtiy-matrix-1} Schematic representation of the equivalence between the  gravitational gauge theory in the genus expansion, a collection of Hermitian random matrix ensembles  $\prod_R^\otimes H_{N_R\times N_R}$  and a single Hermitian random matrix ensemble with elements in $\mathbb C[G]$. }
\end{figure}
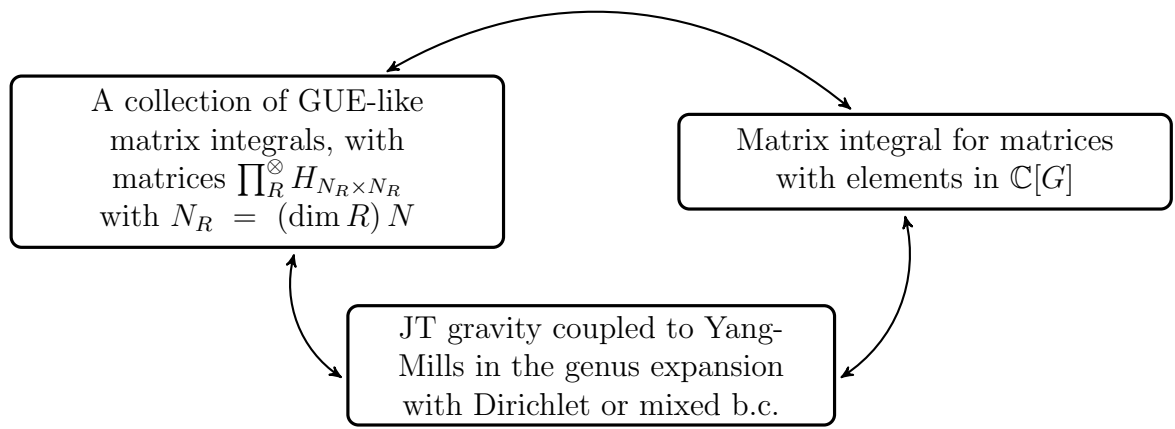
\vspace{-1.2cm}
\end{center}

\subsection{An interlude: the theory on orientable and unorientable manifolds }

In subsection \ref{sec:matrix-model-description} we have reviewed the relation between the gravitational genus expansion on orientable manifolds and matrix integrals over complex Hermitian matrices \cite{Saad:2019lba}, for which the symmetry group that acts on the ensemble of such matrices is $\grp{U}(N)$ (this is known as the $\b=2$ Dyson-ensemble \cite{Dyson:1962es}, also referred to as GUE). Furthermore, we have shown how these matrix integrals account for the gauge degrees of freedom when considering Hermitian matrices with elements in $\mathbb C[G]$ (i.e.,~G-invariant matrices \eqref{eq:G-invariant-definition} whose complex elements are labeled by two discrete labels and two group elements). 

To conclude our discussion about the equivalence between the genus expansions in the gravitational gauge theories and the random matrix ensemble, it is worth schematically mentioning how the results in the previous sections can be modified when also summing over unorientable manifolds. Considering such manifolds in the path integral is relevant whenever the boundary theory has time-reversal symmetry, $\sT$ \cite{Stanford:2019vob}. Thus, for pure JT gravity, the matrix integral which reproduces the correct genus expansion should be over matrices in which time-reversal is assumed. The contribution of such surfaces to the partition function and the relation to matrix integrals with time-reversal was studied in \cite{Stanford:2019vob}. Depending on the way in which one accounts for cross-cap geometries, one obtains two different bulk theories (whose partition function differs by a factor of $(-1)^{c}$ factor for the contribution of surfaces that include $c$ cross-caps)\footnote{As mentioned in \cite{Stanford:2019vob}, the gravitational computation in fact involves the factor $(-1)^{\chi(\cM)}$, however, it is convenient to replace the factor $(-1)^{\chi(\cM)}$ by $(-1)^c$. As noted in \cite{Stanford:2019vob}, the factors $(-1)^{\chi(\cM)}$ by $(-1)^c$ differs by a minus sign for each boundary component, since $2-2g$ is always an even number. This replacement serves to make a more clear map between JT gravity and random matrix resolvents.}  which are related to two different random matrix ensembles \cite{Dyson:1962es}: (i) if $\sT^2 = 1$ then the integral was shown to be over real symmetric matrices ($H_{ij} = H_{ji}$) for which the associated group is $\grp{O}(N)$ (labeled as the $\b =1$ Dyson-ensemble or as GOE-like); (ii) if  $\sT^2 = -1$ then the associated group is $\grp{Sp}(N)$ (labeled as the $\b =4$ Dyson-ensemble or as GSE-like). 

As was shown in \cite{norbury2008lengths, gendulphe2017s, Stanford:2019vob}, the volume of the moduli space of unorientable manifolds has a divergence appearing from the contribution of geometries that include small cross-caps. A similar divergence is found in the relevant double-scaled matrix integral, predicting the correct measure for the cross-caps, but impeding the study of arbitrary genus correlators \cite{Stanford:2019vob}. Nevertheless, when coupling the gravitational theory to Yang-Mills theory, we can still determine the contribution of the gauge degrees of freedom in the genus expansion of partition function even if the volume of the moduli space is divergent. On the matrix integral side, we can also understand how to modify the random matrix ensembles (i) or (ii) to account for this contribution (however, for matrix integrals we will focus on (i)).

We start by analyzing the path integral in the gravitational gauge theory over both orientable and unorientable surfaces. As before, the contribution of the gauge degrees of freedom to the partition function of the gravitational gauge theory is simply given by dressing the gravitational contribution $Z_{\cM}^{(\b=1,4)}$ by the appropriate representation dependent factors. Here, $Z_{\cM}^{(\b=1,4)}$ is the contribution of manifolds with the topology of $\cM$ to the pure JT gravity path integral. Since we are also summing over orientable manifolds, the partition function already includes all the terms in \eqref{eq:YM-genus-exp-partition-function-n-boundaries}, but also includes the contributions from unorientable manifolds which can always be obtained by gluing together surfaces with the topology of trumpets, three-holed spheres, punctured Klein bottles and  cross-cap geometries (punctured $\mathbb{RP}^{\,2}$) \cite{Witten:1991we}. Thus, we label such surfaces by $\cM_{g, n, s, c}$, where $s$ is the number of Klein bottles and $c$ is the number of cross-caps. 

When gluing together only trumpets, three-holed spheres, and Klein bottles, the contribution of the gauge fields exactly follows from \eqref{eq:general-YM-partition-function} \cite{Witten:1991we}, accounting for the contribution of the Klein bottles to the Euler characteristic and only including the sum over  representations that are isomorphic to their complex conjugates, $R = \bar R$ (real or quaternionic). The non-trivial contribution comes from the gluing of cross-cap geometries. Therefore, we first consider the example of a trumpet geometry, glued to a cross-cap and will then generalize our derivation to surfaces with arbitrary topology. To understand the contribution to the path integral in pure Yang-Mills theory of a surface with the topology of a cross-cap, it is useful to understand how to construct such a surface by gluing a 5-edged polygon \cite{Witten:1991we}. Specifically, introducing the holonomies $h_1$ and $h_2$, the cross-cap can be constructed by gluing the edges of the polygon  \cite{Witten:1991we}:
\be 
\begin{tikzpicture}[scale=.6, baseline={([yshift=0cm]current bounding box.center)}]
\label{eq:polygon}
\draw[arrows=->,thick] (-1.0,-1.0) -- (1.0,-1.0);
\draw[arrows=<-, thick] (-1.0,-1.0) -- (-1.4,0.6);
\draw[arrows=->, thick] (1.0,-1.0) -- (1.4,0.6);
\draw[arrows=<-, thick] (-1.4,0.6) -- (0,2);
\draw[arrows=->, thick] (1.4,0.6) -- (0,2);
\draw (0,-1.35) node {\scriptsize $h$};
\draw (-1.7,-0.2) node {\scriptsize $h_1^{-1}$};
\draw (1.6,-0.2) node {\scriptsize $h_1$};
\draw (-0.89,1.59) node {\scriptsize $h_2$};
\draw (0.89,1.59) node {\scriptsize $h_2$};
\end{tikzpicture}
\ee 
Above, $h$ is the holonomy on the resulting boundary of the cross-cap. 
Thus,  the contribution of a single cross-cap glued to a trumpet whose boundary is asymptotically  $AdS_2$ is schematically given by
\be
\label{eq:cross-cap-Yang-Mills-contribution}
Z^{(0, 1, 0, 1)}_{\substack{\text{JTYM}\\ \text{mixed}}}(\Phi_b,\b,  h)&=e^{S_0 \chi(\cM_{0,1,0, 1})} \int Dg^{\mu\nu} \delta\left(\cR+2+{\tilde e_\Phi C_2(R)}\right)e^{\int du \sqrt{g_{uu}}
	\Phi \cK} \nn \\ & \times \bigg(\sum_R (\dim R) \, e^{-\tilde e_b \b C_2(R)- \frac{ \tilde e\, C_2(R) \int_{\cM_{0, 1, 0,1}} d^2x \sqrt{g}}{2 } } \int dh_1 dh_2\chi_R(h h_1 h_2^2 h_1^{-1}) \bigg)    \,,
\ee
where $\cM_{0,1,0, 1}$ are surfaces with cross-cap topology (equivalent to $\mathbb{RP}^2$ with a puncture) that has genus $0$, $1$ boundary, $0$ Klein bottles and, of course, 1 cross-cap component. Consequently, $\chi(\cM_{0,1,0, 1})= 0$. Above, the measure over the gravitational degrees of freedom of course depends on whether the bulk theory is defined to weight cross-cap geometries by a factor of $(-1)^c$. 

After integrating out $h_1$ we are left with the group integral $\int dh_2 \chi_R(h_2^2)$. Thus, in order to compute \eqref{eq:cross-cap-Yang-Mills-contribution} we need to identify the Frobenius-Schur indicator for the representations $R$ of the compact Lie group $G$:
\be
f_R = \int dh \, \chi_R(h^2)\,, \qquad f_R = \begin{cases}
	1 \qquad & \exists\, \text{ symm.~invar.~bilinear form } R \otimes R \to \mathbb C \,,\\ 
	-1 \qquad & \exists\, \text{ anti-symm.~invar.~bilinear form } R \otimes R \to \mathbb C\,,\\
	0 \qquad & \not\exists\, \text{ invar.~bilinear form } R \otimes R \to \mathbb C\,.
\end{cases}
\ee
Such an invariant bilinear form exists if and only if $R = \bar R$. The representation is real, $R \in \hat G_1$, if $f_R=1$ and quaternionic (equivalent, to a pseudo-real irreducible representation), $R \in \hat G_4$, if $f_R = -1$. When the representation $R$ is complex, $R \in \hat G_2$ and $f_R = 0$. 

Integrating out the the gauge field we thus find that the contribution of a single cross-cap-trumpet, with holonomy $h$, is given by
\be
\label{cross-cap-path-integral}
Z^{(0, 1, 0, 1)}_{\substack{\text{JTYM}\\ \text{Dirichlet}}}(\Phi_b,\b,  h)&= \sum_R f_R \,\chi_R(h)\left(\dim R e^{S_0} e^{\tilde e \tilde C_2(R)}\right)^{\chi(\cM_{0,1,0, 1})}  e^{-\tilde e_b \b C_2(R)}Z_{0, 1, 0, 1}( \b/\Phi_b)   \,,
\ee
where $  Z_{0, 1, 0, 1}( \b/\Phi_b)  $ is the (divergent) contribution of the cross-cap topologies to the partition function \cite{Stanford:2019vob}. As previously mentioned, depending on the definition of the bulk theory $Z_{0, 1, 0, 1}( \b/\Phi_b)$ could differ by an overall sign for this cross-cap geometry.

Thus, when gluing this cross-cap geometry to other surfaces, we dress the gravitational results by the factors appearing in \eqref{cross-cap-path-integral}. Thus, the result in the gravitational gauge theory can be obtained from the result in pure JT gravity, by introducing a sum over representations, dressing the entropy factor $e^{S_0} \to \dim R e^{S_0} e^{\tilde e C_2(R)}$, introducing a factor $(f_R)^c$ for geometries with $c$ cross-caps, replacing the boundary value of the dilaton $\Phi_b \to \tilde \Phi_b(R)$ and adding the terms corresponding to the introduction of the boundary condition changing defect (or to the use of mixed bounday conditions) introduced in section \ref{sec:genus-zero-part-function}. Thus, the result from pure JT gravity over orientable and unorientable manifolds becomes
\be
\label{eq:unorientable-part-function}
&Z_{\text{JT}}^{n, \, (\b = 1, 4)}(\Phi_{b_j}, \b_j)= \sum_{\substack{\cM_{g, n, s, c}\\ n \text{ fixed}}} e^{S_0\chi(\cM_{g, n, s, c})} Z_{g, n, s, c}^{(\b=1, 4)}( \b_j/\Phi_{b_j}) \nn \\&\xRightarrow[\substack{\text{adding Yang-Mills} \\ \text{term}}]{} \,\,\,\,Z_{\text{JTYM}}^{n, \, (\b = 1, 4)}(\Phi_{b_j}, \b_j)=\sum_R\bigg[\bigg\{ \sum_{\substack{\cM_{g, n}\\ n \text{ fixed}}} ( \dim R e^{\tilde e \tilde C_2(R)}e^{S_0})^{\chi(\cM_{g, n})} e^{-\frac{C_2(R)}2 \left(\sum_{j=1}^n \tilde e_{b_j}\b_j\right)} \nn \\ & \qquad \qquad \qquad \qquad \qquad \times Z_{g, n, s, c}^{(\b=1, 4)}(\b_j/\tilde \Phi_{b_j}(R)) \bigg\}+ \bigg\{\sum_{\substack{\cM_{g, n, s, c}\\ \text{unorientable}\\n \text{ fixed}}} (f_R)^{c} ( \dim R e^{\tilde e \tilde C_2(R)}e^{S_0})^{\chi(\cM_{g, n, s, c})} \nn \\ & \qquad \qquad \qquad \qquad \qquad \times e^{-\frac{C_2(R)}2 \left(\sum_{j=1}^n \tilde e_{b_j}\b_j\right)} Z_{g, n, s, c}^{(\b=1, 4)}(\b_j/\tilde \Phi_{b_j}(R))\bigg\}\bigg]\,,
\ee
where the first sum in the first parenthesis is over all orientable manifold $\cM_{g, n}$ and the sum in the second parenthesis is over all distinct topologies among the  manifolds $\cM_{g, n, s, c}\,$ which are unorientable. Above, the number of boundaries $n$ is kept fixed.

Only real and quaternionic representations appear in the contribution of unorientable manifolds to the path integral since $f_R = 0$ for complex representations.  In fact, due to the factor $(f_R)^c$, switching between the $\b=1$ and $\b=4$ bulk definitions is equivalent to switching the role of real and quaternionic representations.

As mentioned previously, the contributions from all geometries which contain a cross-cap have a divergence appearing from small cross-caps, and thus, in practice, the contribution of higher genus or demigenus unorientable surfaces is impossible to compute. Nevertheless, we can still formally reproduce the genus expansion over orientable and unorientable surfaces from matrix integrals. For simplicity, we only discuss the limit $\e \to 0$, in which we consider $\tilde \Phi_b(R) = \Phi_b \equiv 1/2$. Once again, for this normalization, we use the shorthand notation $Z_{g, n, s, c}^{(\b=1, 4)}(\b_j) \equiv Z_{g, n, s, c}^{(\b=1, 4)}(\b_j/\Phi_{b_j}) $.  We also focus on the case in which we start from a GOE-like matrix integral ($\b=1$), for which matrices are real and symmetric.

Our starting point is once again the same general matrix potential from subsection \ref{sec:matrix-model-description}, however, we now consider matrices whose elements are real functions on the group manifold $G$ (describing the real group algebra, $\mathbb R[G]$), instead of complex functions;  i.e.~they are G-invariant matrices \eqref{eq:G-invariant-definition} that have real elements which are labeled by two discrete labels and two group elements.  Similar to our derivation for $\mathbb C[G]$, we wish to decompose $\mathbb R[G]$, accounting for the contribution of each representation $R$. Using the trace \eqref{eq:definition-YM-trace}, we conclude that the decomposition is given by\footnote{Once again, \cite{mulase2002generating, mulase2005non, mulase2007geometry} list a similar decomposition to \eqref{eq:definition-YM-trace-real} for finite groups.  } 
\be 
\label{eq:definition-YM-trace-real}
\chi_{\text{YM}}(H) &\equiv  \int dh \, Z_\text{YM}^\text{disk}(h^{-1})\sum_{i = 1}^N (H)_{i,i}(h) = \sum_{\substack{R_i \in \hat G_1\\i=1, 2, 4}} (\dim R_i)\, e^{\tilde e \tilde C_2(R_i)} \,\Tr_{(\dim R_i)N}(H_{R_i})\,,
\ee
where  $\hat G_1$ are all the real unitary irreducible representations of $G$,  $\hat G_2$ are all the complex ones and $\hat G_4$ are all the quaternionic (pseudo-real) representations of $G$. Consequently, the symmetry groups associated to the matrices $H_{R_i}$ follow from the properties of $U_{R_i, l_i}^{k_i}(h)$: $H_{R_1}$ is GOE-like, $H_{R_2}$ is GUE-like and $H_{R_4}$ is GSE-like (also known as a quaternionic matrix) \cite{mulase2002generating, mulase2005non, mulase2007geometry, mulase2003duality}.  Similarly, the same decomposition follows for any power of $H$, following the convolution properties \eqref{eq:convolution-CG}. The matrix model \eqref{eq:matrix-model-YM-dual} thus becomes
\be
\label{eq:matrix-model-YM-dual}
S[H] &= N \left[ \frac{1}2 \,  \chi_\text{YM}(H^2) + \sum_{j\geq 3} \frac{t_j}{j} \chi_\text{YM}(H^j)\right]\nn\\ &= \sum_{\substack{R_i \in \hat G_i\\ i=1, 2, 4}} N(\dim R_i) e^{\tilde e \tilde C_2(R_i)}\left[ \frac{1}2 \,\Tr_{(\dim R_i)N }(H^2_{R_i}) + \sum_{j\geq 3} \frac{t_j}{j} \,\Tr_{(\dim R_i)N }(H^j_{R_i})\right]\,.
\ee
The appropriate choice of measure for the initial path integral $dH(g)$ decomposes to give the standard GOE-like matrix integral measure for $H_{R_1}$, the GUE-like measure for $H_{R_2}$ and the GSE-like measure for $H_{R_4}$. Once again we find that the matrix integral over $H_{ij}(g)$ is equivalent to a collection of matrix integrals, where each integral is associated to a unitary irreducible representation $R$ and the associated symmetry group to each matrix is set by the reality of this representation. As was the case for $\mathbb C[G]$, all the results presented so far in this subsection are due to the harmonic decomposition of our matrices whose elements in $\mathbb R[G]$.

Compared to the (formal) topological expansion of correlators of $Z(\b_j)$ in the matrix integral associated to pure JT gravity, the expansion of correlators of the thermal partition sum $Z_\text{cyl.}(\b_j) = \chi_{\text{cyl.}}(e^{-\b_j H})$ becomes, 
\be 
\label{eq:matrix-integral-decomp-reps}
& \<Z(\b_1) \dots Z(\b_n)\>^{(\b=1)} = \sum_{\substack{\cM_{g, n, s, c}\\ n \text{ fixed}}} e^{S_0\chi(\cM_{g, n, s, c})} Z_{g, n, s, c}^{(\b=1)}(\b_j/\Phi_{b_j}) \nn \\&\xRightarrow[\substack{\,H_{ij} \to H_{(i, g), (j, h)}\, \\ \Tr(\dots) \to \chi_\text{YM}(\dots)}]{} \,\,\,\,   \<Z_\text{cyl.}(\b_1)\dots Z_\text{cyl.}( \b_n)\>_\text{conn.}^{\text{glued}, \,  (\b=1)}(h_1, \, \dots,\,h_n) \simeq \nn \\ &\hspace{0.2cm}\simeq  \bigg[ \sum_{\substack{\cM_{g, n, s, c}\\ \text{orientable \&}\\ \text{unorientable}}} \,\,\, \sum_{\substack{R_i \in \hat G_i \\ i = 1, 4}} (f_{R_i})^{c}(\dim R_i \,e^{\tilde e \tilde C_2(R_i)}e^{S_0})^{\chi(\cM_{g,n, s, c})} \chi_{R_i}(h_1) \dots  \chi_{R_i}(h_n) \,\nn\\ &  \hspace{0.2cm} \times  e^{-\frac{ C_2(R_i)\sum_{i=j}^n \tilde e_{b_j} \beta_j}{2}} Z_{g, n, s, c}^{\b=1}(\b_1,\, \dots, \,\b_n)\bigg]+ \bigg[\sum_{\cM_{g, n}} \sum_{\substack{R_2 \in \hat G_2}} (\dim R_i \,e^{\tilde e \tilde C_2(R_i)}e^{S_0})^{\chi(\cM_{g,n})} \,\nn\\ &  \hspace{0.2cm} \times   \chi_{R_2}(h_1) \dots  \chi_{R_2}(h_n)e^{-\frac{ C_2(R_2)\sum_{i=j}^n \tilde e_{b_j} \beta_j}{2}} Z_{g, n, s, c}^{\b=1}(\b_1,\, \dots, \,\b_n)\bigg]\,.
\ee
Since the matrix integrals over $H_{R_1}$ and $H_{R_4}$ are GOE-like and GSE-like  respectively, the sum in the first parenthesis is over all distinct topologies among both the orientable and unorientable manifolds $\cM_{g, n, s, c}\,$. The factor of $ (f_{R_i})^{c}$  precisely accounts for the $(-1)^c$ factor for the GOE and GSE ensembles associated to the integrals over  $H_{R_1}$ and, respectively, $H_{R_4}$. Because $H_{R_2}$ is hermitian, the sum in the second square parenthesis is solely over orientable manifolds. Noting that $f_{R_2} = 0$, for complex representation $R_2$ it is straightforward to realize that the sums in \eqref{eq:matrix-integral-decomp-reps} reduce to those in \eqref{eq:unorientable-part-function}, in the limit in which $\tilde \Phi_b(R)=\Phi_b$. Thus, we indeed find a (formal) agreement between the matrix integral and the gravitational gauge theory genus expansion. A similar proof is straightforward to derive when starting with a GSE-like matrix integral (and, consequently, using the other definition for the bulk theory). 

Thus, we suggest the equivalence between the Euler characteristic expansion of correlators in the gravitational gauge theory,  on both orientable and unorientable surfaces, and the expansion in the matrix integral discussed above. This relation is summarized through  diagram \ref{fig:schematic-dualtiy-matrix-2}. With this generalization in mind, we now return to the usual situation in which we sum solely over orientable manifolds, with the goal to analyze the diffeomorphism and gauge-invariant operators of the theory. 
\begin{center}
	\vspace{0.1cm}
	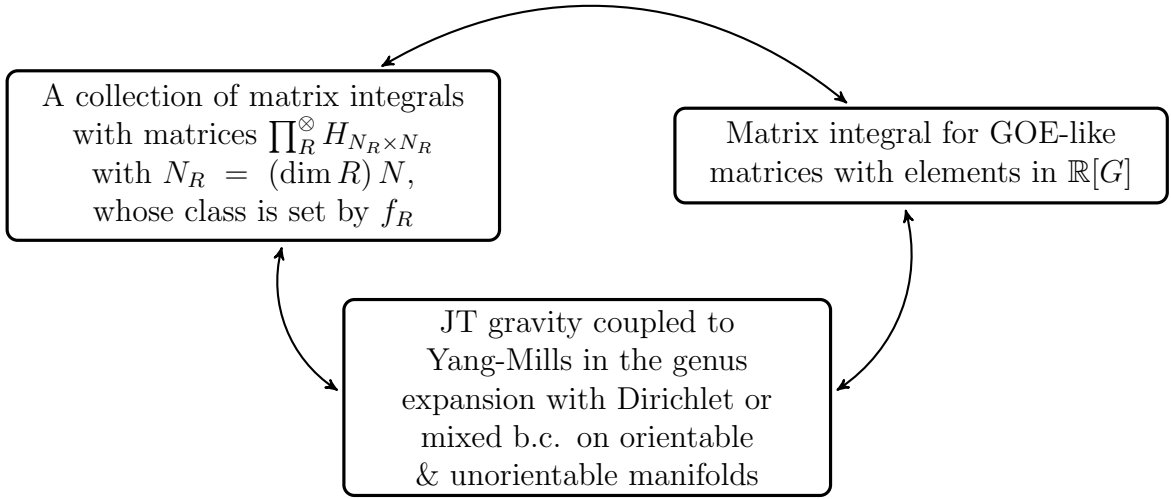
\begin{figure}[h!]
		\begin{tikzpicture}[node distance=1cm, auto]
		\node[punkt] (schw) {JT gravity coupled to Yang-Mills 
			in the genus expansion with Dirichlet or mixed b.c. on orientable \& unorientable manifolds};
		\node[above=1.75cm of schw](dummy) {};
		\node[punkt, inner sep=5pt,left=1cm of dummy] (first) {A collection of matrix integrals with matrices $\prod_R^\otimes H_{N_R\times N_R}$ with $N_R = (\dim R)\, N$, whose class is set by $f_R$ }
		edge[pil,<->, bend right=35] node[left] {} (schw);
		\node[punkt, inner sep=5pt,right=1cm of dummy]
		(second) {Matrix integral for GOE-like matrices  with elements in $\mathbb R[G]$ }
		edge[pil,<->, bend left=35] node[right]{} (schw)
		edge[pil,<->, bend right=35] node[above]{} (first) ;
		
		\end{tikzpicture}\\
		\vspace{0em}
			\caption{\label{fig:schematic-dualtiy-matrix-2} Schematic representation of the equivalence between the  gravitational gauge theory in the genus expansion on orientable and unorientable surfaces, a collection of random matrix ensembles  $\prod_R^\otimes H_{N_R\times N_R}$ whose class is specified by $f_R$ and a single GOE-like random matrix ensemble with elements in $\mathbb R[G]$. }
	\end{figure}
	\vspace{-1.2cm}
\end{center}

\section{Observables}
\label{sec:observables}

\subsection{Diffeomorphism and gauge invariance}

The goal in this section is to define a set of diffeomorphism and gauge invariant  observables in the gravitational BF or Yang-Mills theories. In order to do this it useful to first review how diffeomorphisms act on the zero-form and one-form fields in the theory. Under a diffeomorphism defined by an infinitesimal vector field $\xi$, the zero form field and the one form field transform as,
\be 
\label{eq:action-of-diffeo}
\phi &\to \phi+  i_\xi d \phi\,, \nn \\ 
A &\to A + i_\xi dA + d(i_\xi A) = A+ i_\xi F + D_A(i_\xi A) \,,
\ee
where $i_\xi$ represents the standard map from a $p$-form to a $(p-1)$-form. Since we are fixing the metric along the boundary, we fix diffeomorphisms on $\partial \cM$ to vanish, $\xi|_{\partial \cM}= 0$. 

To start, we first analyze the possible set of local operators. In Yang-Mills theory, the local operator $\Tr \,\phi^2(x)$ (which is also proportional to the quadratic Casimir of the gauge group $G$) is indeed a good diffeomorphism invariant operator since $d \,\Tr \phi^2(x) = 0$ (also valid as an operator equation). Similarly, all other local gauge-invariant operators are given by combinations of Casimirs of the group $G$. Since all other Casimirs are constructed by considering the trace of various powers of $\phi$, they are also conserved on the entire manifold. Consequently, they also serve as proper diffeomorphism and gauge-invariant observables in the gravitationally coupled  Yang-Mills theories. 

We also analyze the insertion of non-local operators of co-dimension 1: i.e.~Wilson lines and loops, 
\be
\label{eq:standard-Wilson-loop}
\cW_R(\cC) = \chi_R\left(\cP e^{\int_\mC A}\right) 
\ee
where the meaning of the contour $\mC$ will be specified shortly. 

Before moving forward with the analysis of correlators for \eqref{eq:standard-Wilson-loop}, we have to require that non-local observables are also diffeomorphism invariant. In the weak gauge coupling limit (BF theory) the path integral localizes to the space of flat connections, and thus the infinitesimal diffeomorphism \eqref{eq:action-of-diffeo} is, in fact, equivalent to an infinitesimal gauge transformation with the gauge transformation parameter given by $\L = i_\xi A$. Since Wilson loops or lines are invariant under bulk gauge transformations, in BF theory they are also invariant under diffeomorphisms (which, of course, also follows from the fact that in BF theory the expectation value of Wilson loops or lines only depends on their topological properties rather than on the exact choice of contour). When computing such correlators in the genus expansion, one has to also specify the homotopy class of the Wilson line or loop.  Since the manifolds that we are summing over in the genus expansion, have different fundamental groups and, therefore, different homotopy classes for the Wilson loops(or lines), there is no way to specify the fact that the contour of the loop or line belongs to a particular class within the genus expansion. Of course, the exceptions are the trivial classes in which the contour can always be smoothly contracted to a segment of the boundary (for boundary anchored lines) or to a single point (for closed loops).

An even more pronounced problem appears in Yang-Mills theory where the observable \eqref{eq:standard-Wilson-loop} is not diffeomorphism invariant,  even when placing the theory on a disk; because the path integral no longer localizes to the space of flat connections, the infinitesimal diffeomorphism in \eqref{eq:action-of-diffeo} is no longer equivalent to a gauge transformation. Rather, the expectation value of a Wilson line or loop is affected by performing the infinitesimal diffeomorphism \eqref{eq:action-of-diffeo}. Therefore, we are forced to consider generalizations of \eqref{eq:standard-Wilson-loop} which should be diffeomorphism invariant. Thus, we define the generalized Wilson loops, by summing over all contours (either closed or anchored at two boundary points) on the manifolds $\cM_{g, n}$, included in the genus expansion in  \eqref{eq:YM-genus-exp-partition-function} or \eqref{eq:YM-genus-exp-partition-function-n-boundaries}:
\be 
\label{eq:generalized-W-loop}
\cW_R \equiv \int [d\mC]\,\chi_R\left(\cP e^{\int_\mC A}\right) \, , \qquad \cW_{\l, R} &\equiv  \int [dC]\, e^{i m\int_\mC ds \sqrt{g_{\mu \nu} \dot x^\mu \dot x^\nu} } \chi_R\left(\cP e^{\int_\mC ds \dot x^\mu A_\mu }\right) \,.
\ee
where $m^2 = \l(1-\l)$ the measure  $[d\mC]$ is chosen such that  \eqref{eq:generalized-W-loop} is diffeomorphism invariant.\footnote{Instead of expressing our results in terms of the mass $m$ of the particle, it proves convenient to use the $\SL2$ representation $\l$ \cite{Kitaev:2017hnr, Iliesiu:toAppear},  which is the charge of the particle under $AdS_2$ isometries.} When considering lines that are anchored, we can fix gauge transformations on the boundary in order for \eqref{eq:generalized-W-loop} to be gauge invariant. When fixing gauge transformations on the boundary, we can consider the more general diffeomorphism and gauge invariant operators\footnote{In fact, one only needs to fix gauge transformations at the anchoring points in order for \eqref{eq:generalized-W-loop} and \eqref{eq:generalized-W-loop-II} to be gauge invariant. The expectation value of such operators in depends on the group elements $h_{j, j+1} = \cP e^{\int_{u_j}^{u_{j+1}} A}$, where $u_j$ and $u_{j+1}$ are all the pairs of neighboring anchoring points.  } 
\be 
\label{eq:generalized-W-loop-II}
\cU_{R, m_1}^{m_2} &\equiv  \int [dC]\, U_{R, m_1}^{m_2} \left(\cP e^{\int_\mC A }\right)  \,,\,\,\,\,\,\,\, \cU_{(\l,\, R), m_1}^{m_2} \equiv  \int [dC]\, e^{i m\int_\mC ds \sqrt{g_{\mu \nu} \dot x^\mu \dot x^\nu} } U_{R, m_1}^{m_2} \left(\cP e^{\int_\mC A }\right)\,,
\ee
where $U_{R, m_1}^{m_2}(h)$ is the a matrix element of the $R$ representation. 

The first operators \eqref{eq:generalized-W-loop} and \eqref{eq:generalized-W-loop-II} can be associated to the worldline path integrals of massless particle charged in the $R$ representation, while the second corresponds to the worldline of a massive particle.  Because of this connection, we refer to these operators as ``quark worldline operators''.
In \eqref{eq:generalized-W-loop-II}, we not only specify the representation $R$ but we also specify the states $m$ and $n$ within the representation $R$ in which the quark should be at the two end-points on the boundary;  \eqref{eq:generalized-W-loop} is insensitive to the states of the particle at the end-points as long as the two are the same. When the worldlines are boundary anchored and the end-points of the contours $C$ are both kept fix to $u_1$ and $u_2$, we denote such operators by $\cW_{\l, R}(u_1, u_2)$ or by $\cU_{R, m_1}^{m_2} (u_1, u_2)$. 

For simplicity, in this paper, we solely focus on the expectation values of the quark worldline operators when the theory is in the weak gauge coupling limit. Moreover, we take the contours associated to the worldlines to be anchored at two fixed points on the boundary and to be smoothly contractable onto the boundary segment in between the two anchoring points.

\subsection{Local operators}

To start, we consider correlation functions of local operators first on surfaces with disk topology, then in the genus expansion, and, in both cases, we determine the equivalent observables on the boundary side. 

In section \ref{sec:equiv-bdy-theory}, we have proven that $
Z_{\substack{\text{JTYM}}}^\text{disk}(\Phi_b, \b, h) = 
Z_{\substack{\text{Schw}\rtimes G}}(\b, h)
$ for both Dirichlet and mixed boundary conditions, for any choice of holonomy of the gauge field $\cA_u$. Given this equality, it is straightforward to determine how to reproduce boundary correlators of $G$-symmetry charges from the bulk perspective. By using functional derivatives with respect to the background gauge field on the boundary side and derivatives with respect to the gauge field $\cA_u$ appearing in the boundary condition for the bulk gauge field, we find the following match: 
\be
\label{eq:mathcing-currents}
\frac{\delta^k Z_{\substack{\text{JTYM}}}^\text{disk}(\Phi_b, \b, \cP e^{\int_{\partial \cM} \cA})}{\delta \cA_u^{a_1}(u_1) \dots \delta \cA_u^{a_k}(u_k)}\longleftrightarrow  \frac{\delta^k Z_{\substack{\text{Schw}\rtimes G}}(\b, \cP e^{\int_{\partial \cM} \cA})}{\delta \cA_u^{a_1}(u_1) \dots \delta \cA_u^{a_k}(u_k)} =  i^{k} \< \pmb \a_{a_1}(u_1) \dots \pmb \a_{a_k}(u_k)\>\,.
\ee
The equivalence above holds when choosing both Dirichlet or mixed boundary conditions for the bulk gauge field and, as presented in subsection \ref{sec:equiv-bdy-theory}, when choosing the appropriate boundary theory. 
Note that since $\pmb \a(u)$ is not invariant under background gauge transformations, in \eqref{eq:mathcing-currents} we should fix $\cA_u(u)$ at every point and not only its overall holonomy for any choice of gauge field boundary conditions.  

Similarly, we find a match between the conserved $G$ quadratic Casimir in Yang-Mills theory and the conserved $G$ quadratic Casimir on the boundary side:
\be
\Tr \phi^2  \longleftrightarrow \Tr \,\pmb\alpha^2  \,.
\ee
The correlators or such operators are obtained by inserting the $G$ quadratic Casimir in the path integral, to find that\footnote{For brevity, we use $\propto $ to denote the solution to correlators, un-normalized by the partition function in the associated theories. }
\be
\label{eq:exp-value-with-mixed-bc-YM}
\<\Tr \phi^2(x_1) \dots \Tr \phi^2(x_k)\>(h)   &\propto \sum_{R} \dim(R) \chi_{R}(h) (2\,C_2(R))^n \left(\frac{\tilde \Phi_b(R) }{  \beta}\right)^{3/2}\nn \\ &\times e^{\frac{\pi^2\tilde \Phi_b(R)}{\beta} + \tilde e \tilde C_2(R) - \tilde e_b \beta C_2(R)}  =\nn \\ & =  \< \Tr \,\pmb\alpha^2 (u_1) \dots \Tr \,\pmb\alpha^2 (u_n)\>\,,
\ee
where we note that the correlator is independent of the bulk insertion points $x_1$, \dots, $x_n$ and of the boundary insertion points $u_1$, \dots, $u_n$.\footnote{The factor of $2$ in front of the Casimir comes from the normalization $\cN\equiv 1/2$. } Following the same reasoning, the correlation functions of any gauge invariant operators match:
\be
\label{eq:matching-general-V}
\hat V(\phi)\longleftrightarrow \hat V(\,\pmb\alpha) \,. 
\ee
Correlation functions such as $\<\hat V_1(\phi(u_1)) \dots \hat V_n(\phi(u_n)) \>$ can be matched by replacing the factor of the Casimir $(C_2(R))^n$ in \eqref{eq:exp-value-with-mixed-bc-YM} by $V_1(R)\dots V_n(R)$. Since all diffeomorphism and gauge invariant operators are of the form \eqref{eq:matching-general-V} we conclude that the correlation functions of local operators on surfaces with disk topology match those in the boundary theory \eqref{eq:boundary-theory-for-YM}.

We now consider such correlators in the genus expansion of orientable surfaces.  
With mixed boundary conditions for the gauge field in the gravitational gauge theory, such correlators are given by
\be
\<\Tr \phi^2(x_1) &\dots  \Tr \phi^2(x_k) \>(\Phi_{b_j},\,\beta_j,\, h_j)\propto \sum_R \chi_R(h_1) \dots \chi_R(h_n)  e^{-\frac{ \tilde e_b  C_2(R) \sum_{j=1}^n \beta_j}{2}} \nn \\& \times\left[\sum_{g=0}^\infty \left(\dim(R)e^{ \tilde e \tilde C_2(R)} e^{S_0}\right)^{\chi(\cM_{g,n})}  \left(2 C_2(R)\right)^k  Z_{g,n}^{\left(\Phi_{b_j}(R)\right)}(\b_j)\right]\,,
\ee
when considering surfaces with $n$-boundaries. For simplicity we assume $\e \to 0$ such that we take $\tilde \Phi_{b_j}(R) =\Phi_{b_j}$.  This result can be reproduced from the random matrix ensemble \eqref{eq:matrix-model-YM-dual} by considering correlators of the operator
\be
\label{eq:trace-Tr-phi-sq-matrix-model}
\chi_{\Tr \phi^2\,, h}(e^{-\b_j H}) &\equiv  \int d\tilde h  \, \<\Tr \phi^2\>_{\substack{\text{BF}\\\text{mixed}}}^\text{(0, 2)}(\tilde h ^{-1}, h) \sum_{i=1}^N \left(e^{-\b_j H}\right)_{i,i}(\tilde h) \nn \\ &= \int d\tilde h \sum_R  \chi_R(\tilde h^{-1}) \chi_R(h)   (2\,C_2(R))   e^{-\frac{\tilde e_b \beta C_2(R)}{2}}\sum_{i=1}^N \left(e^{-\b_j H}\right)_{i,i}(\tilde h) \nn \\ &= \sum_R  \chi_R(h) (-C_2(R)) e^{-\frac{\tilde e_b \beta C_2(R)}{2}}  \Tr_{(\dim R)N }(e^{-\b_j H_R}) \,,
\ee
where  $\<\Tr \phi^2\>_{\substack{\text{BF}\\ \text{mixed}}}^\text{(0, 2)}$ is the expectation value of the operator $\Tr \phi^2$ on the cylinder ($\cM_{(0,2)}$) in the BF-theory with the mixed boundary condition \eqref{eq:diff-boundary-cond-A} on one of the sides of the cylinder and with Dirichlet boundary conditions on the other. Plugging the above into the ``glued'' matrix integral correlator, we indeed find that\footnote{Once again we omit to specify the holonomies associated to the traces $\chi_{\Tr \phi^2}(\dots)$. See footnote\footref{footnote:why-we-ommit-holo}.} 
\be 
\<\Tr \phi^2(x_1) \,&\dots \, \Tr \phi^2(x_k) \>(\Phi_{b_j},\,\beta_j,\, h_j) = \<\chi_{\Tr \phi^2\,}(e^{-\b_1 H})\,\dots \,\chi_{\Tr \phi^2\,}(e^{-\b_k H})\>_\text{conn.}^\text{glued}(\Phi_{b_j},\,\beta_j,\, h_j)
\ee
Similarly, by modifying the trace function in \eqref{eq:trace-Tr-phi-sq-matrix-model} by replacing $\Tr \phi^2$ by the arbitrary function $V(\phi)$, we can prove that for all gauge and diffeomorphism invariant observables on the boundary side one can construct the equivalent set of operators on the matrix integral side.  

\subsection{Quark worldline operators in the weakly coupled limit}
\label{sec:wilson-loops-in-BF-theory}

Since we have discussed the correlators of all gauge-invariant local operators, we can now move-on to computing the expectation value of the aforementioned quark worldline operators \eqref{eq:generalized-W-loop} and \eqref{eq:generalized-W-loop-II}. As previously stated, in this subsection we solely consider boundary anchored quark worldlines in the weakly coupled topological limit, with the mixed boundary conditions studied in section \ref{sec:preliminaries}. We again start by studying surfaces with disk topology and then discuss correlators of such operators in the genus expansion. For higher genus manifolds, we only consider massless quark worldline operators whose contours have both endpoints on the same boundary. Moreover,  we solely consider contours that can be smoothly contractible to a segment on the boundary when keeping these boundary endpoints fixed.  

Considering the weak gauge coupling limit offers two advantages. 

The first is that the expectation value of operators with  self-intersecting contours $C$ is the same as the  expectation value of operators with contours $\tilde C$ that have the same endpoints and are not self-intersecting; i.e., there is a smooth transformation taking $C$ and $\tilde C$ which vanishes at the endpoints.\footnote{As previously discussed, when quantizing BF-theory each patch has an associated irreducible representation $R$. As we will summarize shortly, for each Wilson line intersection, one associates a $6j$-symbol of the group $G$ which includes the four representation associated to the patches surrounding the intersection and the two representations associated to the two lines. When the line is self-intersecting, one instead uses two copies of the representation associated with that line. The fact that Wilson lines with the contour $\cC$ and $\tilde \cC$ have the same expectation value follows from orthogonality properties of the $6j$-symbol.  } Therefore, in the weak gauge coupling limit, we only have to consider the expectation value of lines that are not self-intersecting. 

The second advantage of the weak gauge coupling limit is that on surfaces with disk topology, the contribution of the gauge field in the worldline operators \eqref{eq:generalized-W-loop} and \eqref{eq:generalized-W-loop-II} can be factorized: 
\be 
\cU_{(\l,\, R), m_1}^{m_2} =  \int [dC]\, e^{i m\int_C ds \sqrt{g_{\mu \nu} \dot x^\mu \dot x^\nu} } U_{R, m_1}^{m_2} \left(\cP e^{\int_\mC A }\right) =  \left(\int [dC]\, e^{i m\int_C ds \sqrt{\dot x_\mu \dot x^\mu} }\right) U_{R, m_1}^{m_2} \left(\cP e^{\int_{\tilde C}A }\right).\ee
The above equation holds for any  contour choice $\tilde C$ which has the same end-points as the contours $C$. Correlators of $\cO_\l(C) \equiv  \left(\int [dC]\, e^{i m\int_C ds \sqrt{g_{\mu \nu} \dot x^\mu \dot x^\nu} }\right)$ have been studied in pure JT gravity on disk topologies in \cite{Blommaert:2018oro,Iliesiu:2019xuh}. Such operators were shown to be equivalent to Wilson lines in a BF theory with $\mathfrak{sl}(2, \mR)$ gauge algebra. In turn, the expectation value of such lines were shown to match correlation functions of bi-local operators in the Schwarzian theory  \cite{Mertens:2017mtv, Lam:2018pvp, Mertens:2018fds, Iliesiu:2019xuh},
\be
\label{eq:Schwarzian-bilocal-op}
\cO_\l(\cC) \equiv  \left(\int [dC]\, e^{i m\int_C ds \sqrt{g_{\mu \nu} \dot x^\mu \dot x^\nu} }\right)\,\,\,\, \leftrightarrow \,\,\,\,  \cO_\l(u_1, u_2) \equiv \left(\frac{F'(u_1) F'(u_2)}{|F(u_1)-F(u_2)|^2}\right)^\l\,,
\ee
where $F(u)$ is the Schwarzian field and $u_1$ and $u_2$ are the locations of the end-points for the countours $C$. 

Thus, by using the correlator functions of Wilson lines in  $\mathfrak{sl}(2, \mR)$ BF theory,\footnote{Or, equivalently, the expectation value of bi-local operators in the Schwarzian theory \cite{Mertens:2017mtv}. } together with the expectation value of boundary anchored non-intersecting Wilson lines in $G$-BF theory,\footnote{The expectation value of boundary anchored Wilson lines in the more general Yang-Mills theory with gauge group $G$ were studied in \cite{Blommaert:2018oue, Blommaert:2018rsf, Blommaert:2018oro}.} we determine arbitrary correlators of quark worldlines in the weak gauge coupling limit on surfaces with  disk topology. Using closely related techniques, we then move-on to the genus expansion when setting the mass of the quark to $m=0$.

\subsubsection*{A single line on the disk}

When fixing the boundary conditions for the gauge field to be given by \eqref{eq:diff-boundary-cond-A}, the expectation value of a boundary anchored quark worldline operator can be computed in two different ways. 

The first follows the reasoning presented in subsection \ref{sec:genus-zero-BF-plus-defect}: we reduce the bulk path integral in the presence of a quark worldline operator to a boundary path integral. Such a reduction was studied in the case of pure BF  theory in \cite{Blommaert:2018oue, Blommaert:2018rsf, Blommaert:2018oro}. As mentioned in subsection \ref{sec:genus-zero-BF-plus-defect} the path integral over the zero-form field $\phi$ imposes a restriction to the space of flat connections, $A = q^{-1} d q$. For such configuration the path-ordered integral that appears in the Wilson-line becomes $\cP e^{\int_C A} = q^{-1}(u_2) q(u_1)$, for any contour $C$ whose end-points are $u_1$ and $u_2$. Similarly, one can show that the the Wilson line in the $\mathfrak{sl}(2, \mR)$ BF-theory reduces to the bi-local operator \eqref{eq:Schwarzian-bilocal-op}. Thus in the boundary path-integral \eqref{eq:introduce-Lagr-multiplier-a}, we need to insert the operator $U_{R, m_1}^{m_2}(q^{-1}(u_2) q(u_1))$:\footnote{Note that because the action in the first path integral in \eqref{eq:expectation-value-U-boundary-p-integral} is invariant under 1d diffeomorphisms, one can equivalently use the $AdS_2$ coordinate  given by the Schwarzian field $F(u)$ to parametrize the boundary and the anchoring points.  }
\be 
\label{eq:expectation-value-U-boundary-p-integral}
\< \cU_{(\l, R), m_1}^{m_2}(u_1, u_2)\> &\propto \left[\int Dq D\pmb \ma\, U_{R, m_1}^{m_2}(q^{-1}(u_2) q(u_1))\,\,e^{\int du  \left(i \tr (\pmb \ma\, q^{-1}D_\cA q) + \sqrt{g_{uu}} \frac{\e\tilde e_b}2 \tr \,\pmb \ma^2 \right)}\right]\nn \\ &\qquad\qquad \times \left[\int DF \, \cO_\l(u_1, u_2) e^{\int_0^\b du \Sch(F, u)}\right]\,.
\ee
The path integral in the first parenthesis was computed in \cite{Blommaert:2018oro, Blommaert:2018oue} when the background gauge field $\cA_u = 0$. Nevertheless, we follow the same reasoning as in  \cite{Blommaert:2018oro, Blommaert:2018oue} to solve the path integral for an arbitrary background. By using the quantization procedure from subsection \ref{sec:quantization-particle-on-G} and using $U_{R, m_1}^{m_2}(q^{-1}(u_2) q(u_1)) = U_{R, p}^{m_2}(q^{-1}(u_1)) U_{R, m_1}^p(q(u_2)) $, we find that the first square parenthesis can be rewritten as \cite{ Blommaert:2018oro, Blommaert:2018oue}
\be
\label{eq:how-to-solve-first-square-paranthesis}
\< U_{R, m_1}^{m_2}\>_{G}  \equiv \tr_{\cH^G}  U_{R, p}^{m_2}(q^{-1}(u_1)) h_{12} e^{-u_{12} H} U_{R, m_1}^p(q(u_2)) h_{21} e^{-u_{21} H}\,,
\ee
where $h_{12} = \cP e^{\int_{u_1}^{u_2} \cA}$ and $h_{21}$ is given by the integral along the complementary segment. Furthermore, we have simplified notation by denoting $u_{ij}  = |u_i - u_j|$ for $i>j$ and $u_{ji} = |\beta - u_j + u_i|$.  By inserting the complete basis of eigenstates of the Hamiltonian $H$ at various locations in \eqref{eq:how-to-solve-first-square-paranthesis} one can easily compute the expression above \cite{Blommaert:2018oue, Blommaert:2018oro}.

Before, discussing the final result \eqref{eq:expectation-value-U-one-op} of the path integral in \eqref{eq:expectation-value-U-boundary-p-integral}, we briefly summarize how one can compute the expectation value of $\cU_{(\l, R), m_1}^{m_2}(u_1, u_2)$ by directly performing the bulk path integral. By using the fact that the mixed boundary conditions are equivalent to the insertion of the boundary condition changing defect \eqref{eq:defect-action-simple}, we find that the contribution of the gauge field is given by the gluing formula 
\be 
\< U_{R, m_1}^{m_2}\>_{G}   = \int dh Z_{\substack{\text{BF}\\\text{mixed}}}^{(0, 1)}(u_{12}, h_{12}h) U_{R, m_1}^{m_2}(h) Z_{\substack{\text{BF}\\\text{mixed}}}^{(0, 1)}(u_{21}, h_{21}h^{-1})\,.
\ee
This, or equivalently \eqref{eq:how-to-solve-first-square-paranthesis}, yields\footnote{Here we have normalized the $3-j$ symbol following \cite{Blommaert:2018oro}, such that 
	\be
	\int dh U_{R_1, n_1}^{m_1}(h) U_{R_2, n_2}^{m_2}(h) U_{R_3, n_3}^{m_3}(h^{-1}) =\left(\begin{matrix}R_1 &  R_2 &R_3\\m_1 & m_2 & -m_3 \end{matrix}\right) \left(\begin{matrix}R_1 &  R_2 &R_3\\n_1 & n_2 & -n_3 \end{matrix}\right)\,. 
	\ee}
\be
\label{eq:final-answer-U-BF-theory}
\< U_{R, m_1}^{m_2}\>_{\substack{\text{BF}\\\text{mixed}}} &=  \< U_{R, m_1}^{m_2} \>_{G} = \sum_{R_1, \, R_2} (\dim R_1) (\dim R_2)\, e^{-\frac{1}2 \tilde e_b\left( u_{12} \,C_2(R_1) + u_{21}\, C_2(R_2)\right)} \nn \\ &\times\sum_{\substack{p_j,\, q_j=1\\j=1, 2}}^{\dim R_j} \left(\begin{matrix}R_1 &  R &R_2\\ p_1 & m_1 & -p_2 \end{matrix}\right) \left(\begin{matrix}R_1 &  R &R_2\\ q_1 & m_2 & -q_2 \end{matrix}\right) U_{R_1, p_1}^{q_1}(h_{12}) U_{R_2, p_2}^{q_2}(h_{21})\,, 
\ee
where $\left(\begin{matrix}R_1 & \tilde R &R_2\\ p_1 & m_1 & -p_2 \end{matrix}\right)$ is the 3j-symbol for the representations $R_1$, $R$ and $R_2$ of the group $G$. 

Putting this together with the result for the expectation value of the bi-local operator in the Schwarzian theory \cite{Mertens:2017mtv} or, equivalently, for the expectation value of a Wilson line in an $\mathfrak{sl}(2, \mR)$ BF-theory \cite{Blommaert:2018oro, Iliesiu:2019xuh}, we find that 
\be 
\label{eq:expectation-value-U-one-op}
&\<\cU_{(\l, R), m}^n(u_1, u_2)\> \propto \int ds_1  \rho_0(s_1)\, ds_2\rho_0(s_2)  \tilde  N^{s_2}{\,}_{s_1, \l} \sum_{R_1, \, R_2} (\dim R_1) (\dim R_2)\, e^{- u_{12}\left( \frac{s_1^2}{2\Phi_b}+\frac{\tilde e_b C_2(R_1)}{2}\right)}\nn \\ &\times e^{- u_{21}\left(\frac{s_2^2}{2\Phi_b}+\frac{\tilde e_b C_2(R_2)}{2}\right)}  \sum_{\substack{p_j,\, q_j=1\\j=1, 2}}^{\dim R_j} \left(\begin{matrix}R_1 &  R &R_2\\ p_1 & m_1 & -p_2 \end{matrix}\right) \left(\begin{matrix}R_1 &  R &R_2\\ q_1 & m_2 & -q_2 \end{matrix}\right) U_{R_1, p_1}^{q_1}(h_{12}) U_{R_2, p_2}^{q_2}(h_{21}), \hspace{-0.2cm}
\ee
where $\tilde N^{s_2}{\,}_{s_1, \l}$ can be viewed as the fusion coefficient for the principal series repesentations $\l_1 = 1/2+i s_1$ and $\l_2= 1/2+i s_2$ and the discrete series representation $\l$ in $\SL2$, given by\footnote{For details about the computation of \eqref{eq:sl2-fusion-coeff}, see \cite{Iliesiu:2019xuh}. }
\be
\label{eq:sl2-fusion-coeff}
\tilde N^{s_2}{\,}_{s_1, \l} = \frac{|\Gamma(\l+ is_1 - is_2)\Gamma(\l+ is_1 + is_2) |^2}{\Gamma(2\l)} = \frac{\Gamma(\l\pm is_1 \pm is_2)}{\Gamma(2\l)} \,. 
\ee
A simplifying limit for \eqref{eq:expectation-value-U-one-op} appears when considering the operator $\cW_{\l, \,\tilde R}(u_1, u_2)$, with $\cA_u = 0$ all along the boundary ($h_{12} = h_{21}  = e$):
\be 
\<\,\cW_{\l, \, R}&(u_1, u_2)\,\> \propto  \sum_{R_1, \,R_2} (\dim R_1) (\dim R_2) \int ds_1  \rho_0(s_1)\, ds_2\rho_0(s_2) \nn \\ &\times   N^{R_2}\,_{R_1,  R} \,\tilde  N^{s_2}{\,}_{s_1, \l} e^{-\frac {1}{2 \Phi_b} \left[(u_2 - u_1)\left( \frac{s_1^2}{2\Phi_b}+\frac{\tilde e_b C_2(R_1)}{2}\right) + (\beta - u_2 + u_1)\left(\frac{s_2^2}{2\Phi_b}+\frac{\tilde e_b C_2(R_2)}{2}\right)\right]} \,,
\ee
where $ N^{R_2}\,_{R_1, \tilde R} $ is the fusion coefficient for the  tensor product of representations, $R_1\otimes \tilde R \to N^{R_2}\,_{R_1, \tilde R} \,R_2\,$. 

Following the same techniques presented so far we can compute correlation functions of an arbitrary number of quark worldline operators, $\cU_{(\l, R), m}^n(u_j, u_{j+1})$ or $\cW_{\l, \,\tilde R}(u_j, u_{j+1})$, by  performing the bulk path integral directly; alternatively, we can compute the expectation value of operators such as $U_{R, m}^n(q^{-1}(u_j) q(u_{j+1})) \cO_\l(u_j, u_{j+1})$ on the boundary side. To better exemplify the power of these techniques we give results for two other examples of quark worldline correlators on surfaces with disk topology.

\subsubsection*{Time-ordered correlators}

First we consider the case of multiple boundary anchored lines whose end-points are $u_{j}$ and $u_{j+1}$ (with $j= 1, 3, \dots 2n-1$) and the points are ordered as
$u_1 \leq u_2 \leq \dots \leq u_{2n}$. In such a configuration, we find that when setting $\cA_u=0$, the correlation function of $\cW_{m_{i},\tilde R_{i}}$ is given by
\be 
\label{eq:time-ordered-correlator}
&\<\,\prod_{i=1}^n \cW_{m_{i},\tilde R_{i}}(u_{2i-1}, u_{2i})  \,\> \propto \sum_{\substack{R_1, \dots\,,R_n, \\ R_0}} \int ds_0 \rho(s_0) (\dim R_0) \left(\prod_{i=1}^n ds_i \rho(s_i) (\dim R_i)\right)\nn \\ &\times \left(\prod_{i=1}^n  N^{R_0}_{R_i, \tilde R_i} \tilde N^{s_0}{}_{s_i, \l_i}\right) e^{-\left(\sum_{i=1}^{n} u_{2i,2i-1} \left(\frac{s_i^2}{2\Phi_b} + \frac{\tilde e_b C_2(R_i)}{2}\right)  \right) -  \left(\beta-\sum_{i=1}^{n} u_{2i, 2i-1}\right)\left(\frac{s_0^2}{2\Phi_b} + \frac{\tilde e_b C_2(R_0)}{2}\right) } \,.
\ee
This case corresponds to studying time-ordered correlators of the equivalent boundary operators, $\chi_R(h^{-1}(u_j) h(u_{j+1})) \cO_\l(u_j, u_{j+1})$.

\subsubsection*{Multiple intersecting lines and out of time-ordered correlators}

As our second example we consider the case of two set of boundary anchored worldlines whose end-points are $u_{1}$, $u_2$ and $u_3$, $u_4$ and the points are ordered as $u_1 \leq u_3 \leq u_2 \leq u_4$. The Wilson lines associated to the two quark worldlines operators are in a configuration that is homotopically equivalent (when fixing the endpoints) to the case in which the contours of the two lines intersect solely once. Therefore, we solely consider this latter configuration to compute the contribution of the gauge degrees of freedom to the correlator. Once again, we find that when setting $\cA_u=0$ the result simplifies. In particular, the correlation function is given by:
\be
\label{eq:OTO-four-pt-function-Wilson-loops}
\langle \cW_{m_1,\tilde R_1}&(u_1, u_2) \cW_{m_2,\tilde R_2}(u_3, u_4)  \rangle \propto \sum_{R_1,\, \dots\,, R_4}\int \left(\prod_{i=1}^4 ds_i \rho_0(s_i) \dim R_i \right) \\ &\times  \sqrt{\tilde N^{s_4}{}_{\l_1, s_1}  \tilde N^{s_3}{}_{\l_1, s_2} \tilde N^{s_3}{}_{\l_2, s_1}  \tilde N^{s_4}{}_{\l_2, s_2} }  R_{s_3\, s_4}\! \left[\; {}^{s_2}_{s_1} \;{}^{\l_2}_{\l_1}\right]  \left\{\begin{matrix}
R_3 & R_2 & \tilde R_2 \\
R_4 & R_1 & \tilde R_1
\end{matrix} \right\}^2 \nn\\ & \times\, e^{-\left[\left(\frac{s_1^2}{2\Phi_b} + \frac{\tilde e_b C_2(R_1)}{2}\right) u_{13} +\left(\frac{s_3^2}{2\Phi_b} + \frac{\tilde e_b C_2(R_3)}{2}\right)   u_{32} + \left(\frac{s_2^2}{2\Phi_b} + \frac{\tilde e_b C_2(R_2)}{2}\right) u_{24} +\left(\frac{s_4^2}{2\Phi_b} + \frac{\tilde e_b C_2(R_4)}{2}\right)  u_{41}\right] }   \,,\nn 
\ee 
where $\left\{\begin{matrix}
R_3 & R_2 & \tilde R_2 \\
R_4 & R_1 & \tilde R_1
\end{matrix} \right\}$ is the $6-j$ symbol for the representations of the group $G$ and $R_{s_3\, s_4}\! \left[\; {}^{s_2}_{s_1} \;{}^{\l_2}_{\l_1}\right] $ is the $6-j$ symbol for 4 principal and two discrete series representation of $\SL2$.\footnote{Once again, for details about the appearance of the $\SL2$ $6j$-symbol, see \cite{Iliesiu:2019xuh, groenevelt2003wilson, groenevelt2006wilson}. }

\subsubsection*{Moving to higher genus: massless quark worldlines in the genus expansion}
To conclude our discussion about non-local operators in the gravitational gauge theory, we move away from the disk topology and compute an example of a quark worldline correlator on the bulk-side. Finally, we again show how this correlator can be reproduced through a matrix integral. Specifically, we consider a boundary anchored quark  massless ($m=0$ and, consequently $\l =0$ or $1$) worldline operators with homotopically trivial contours in the weak coupling.\footnote{The reason we are solely considering correlation functions of massless field is due to the divergence observed in \cite{Saad:2019lba} when considering correlation functions of matter fields on higher genus surfaces for which the length of the closed geodesic along which the trumpet is glued has $b \to 0$. } By using the gluing procedure described above we find that the correlator for a single quark worldline on a surface with $n$-boundaries is given by,
\be
\label{eq:one-quark-worldline-genus-expansion}
\<&\cU_{(0, R), m}^n(u_1, u_2)\>(h_{12}, h_{21}, h_2, \dots, h_n)  \propto \sum_{g=0}^\infty Z_{g, n} e^{S_0 \chi(\cM_{g, n})}  \int dh Z_{\substack{\text{BF}\\ \text{mixed}}}^{(0,1)}( h_{12}h)  Z_{\substack{\text{BF}\\ \text{mixed}}}^{(g, n)}( h^{-1}h_{21})  \nn \\ &\times U_{R, m}^n(h) = \sum_{g=0}^\infty Z_{g, n} \sum_{R_1, R_2} (\dim R_1) \left(\dim R_2\, e^{S_0}\right)^{\chi(\cM_{g, n})} \chi_R(h_2) \dots \chi_R(h_n) e^{- \frac{ C_2(R) \sum_{j=2}^n e_{b_j} \beta_j }{2}} \nn 
\\ &\times e^{- \frac{\tilde e_b  u_{12} C_2(R_1)}{2}- \frac{\tilde e_b u_{21}C_2(R_2)}{2}}   \sum_{\substack{p_j,\, q_j=1\\j=1, 2}}^{\dim R_j} \left(\begin{matrix}R_1 &  R &R_2\\ p_1 & m_1 & -p_2 \end{matrix}\right) \left(\begin{matrix}R_1 &  R &R_2\\ q_1 & m_2 & -q_2 \end{matrix}\right) U_{R_1, p_1}^{q_1}(h_{12}) U_{R_2, p_2}^{q_2}(h_{21})\,.
\ee
Here, when $g>0$, the contours are contractible to the segment of the boundary whose length is $u_{12}$, with $u_{12}+u_{21} = \b$. Once again, while on the disk the the contribution of the gauge and gravitational degrees of freedom are factorized,  the two theories which are topological in the bulk are  once again coupled through the genus expansion. The gluing procedure  in \eqref{eq:one-quark-worldline-genus-expansion} is easily generalized for any number of quark worldlines whose contours are each contractible to a boundary segment. Specifically, results for time-ordered and out-of-time order correlators easily follow from \eqref{eq:time-ordered-correlator} and \eqref{eq:OTO-four-pt-function-Wilson-loops}, respectively.

It is instructive to understand how such correlators can be reproduced from matrix integrals. For simplicity, we focus on reproducing \eqref{eq:one-quark-worldline-genus-expansion} for a single boundary ($n=1$). Once again, we rely on modifying the trace of of operator  $e^{-\b H}$ that we have previously used in the correlator of matrix integrals. Therefore we define
\be 
\label{eq:trace-quark-worldline}
\chi_{U_{R,m_1}^{m_2},\, h_{12}, \, h_{21}}(e^{-\b H}) &\equiv \int d\tilde h  \, \<\cU_{R,m_1}^{m_2}\>_{\substack{\text{BF}\\\text{mixed}}}(h_{12}, h_{21}\tilde h^{-1}) \sum_{i=1}^N \left(e^{-\b H}\right)_{i,i}(\tilde h)  \nn \\ &= \sum_{R_1, R_2}  e^{-\frac{\tilde e_b}2\left(  u_{12} C_2(R_1)+u_{21}C_2(R_2)\right)} \Tr_{(\dim R_2)N}\left( e^{-\b H_{R_2}} \right) \\ &\times  \sum_{\substack{p_j,\, q_j=1\\j=1, 2}}^{\dim R_j} \left(\begin{matrix}R_1 &  R &R_2\\ p_1 & m_1 & -p_2 \end{matrix}\right) \left(\begin{matrix}R_1 &  R &R_2\\ q_1 & m_2 & -q_2 \end{matrix}\right) U_{R_1, p_1}^{q_1}(h_{12}) U_{R_2, p_2}^{q_2}(h_{21})\,,\nn
\ee
where $\<\cU_{R,m}^n\>_{\substack{\text{BF}\\\text{mixed}}}(h_{12}, h_{21}\tilde h^{-1})$ is the expectation value of the boundary anchored Wilson line $\cU_{R,m}^n(h)$ inserted in a $G$-BF theory with the mixed boundary conditions \eqref{eq:diff-boundary-cond-A}. Using the matrix integral whose action is given by \eqref{eq:matrix-model-YM-dual}, it quickly follows that
\be 
\<&\cU_{(0, R), m_1}^{m_2}(u_1, u_2)\>_{\substack{\text{JTBF}\\\text{mixed}}}^{\,\,n=1}(h_{12}, h_{21}) = \<\chi_{U_{R,m_1}^{m_2},\, h_{12}, \, h_{21}}(e^{-\b H}) \>\,.
\ee

The construction of the traces in \eqref{eq:trace-Tr-phi-sq-matrix-model}, corresponding to the insertion of the local operator $\Tr \phi^2$, and the trace \eqref{eq:trace-quark-worldline}, corresponding to the insertion of the massless quark worldline operator suggest the general prescription needed in order to reproduce any gauge theory observable in the weak gauge coupling limit. For an operator $\cO$, that can be entirely contracted to the boundary of the gauge theory, one can schematically construct the operator 
\be
 \chi_\cO(e^{-\b H}) = \int d\tilde h \<\cO\>_{\substack{\text{BF}\\ \text{mixed}}}(\tilde h^{-1}) \sum_{i=1}^N \left( e^{-\b H} \right)_{i, i} (\tilde h)\,,\qquad \<\cO\>_{\substack{\text{JTBF}\\\text{mixed}}} = \<\chi_\cO(e^{-\b H})\>\,.
 \ee 
  Of course, it would be interesting to extend this construction and the analysis performed in this subsection to worldline operators which cannot necessarily be contracted to the boundary and when the gauge theory is not necessarily weakly coupled. We hope to report  in the future on progress in this direction.

\section{Discussion}
\label{sec:discussion}

We have managed to quantize JT gravity coupled to Yang-Mills theory, both through the metric and through the dilaton field, when the theory has an arbitrary gauge group $G$ and arbitrary gauge couplings. 

When solely looking at surfaces with disk topology, we have found that the theory is equivalent to the Schwarzian coupled to a particle moving on the gauge group manifold. Explicitly, we have computed a great variety of observables in the gravitational gauge theory, ranging from the partition functions presented in section \ref{sec:genus-zero-part-function},  to correlators of quark worldline operators discussed in section \ref{sec:observables}. We matched each of them with the proper boundary observable. This boundary theory (the Schwarzian coupled to a particle moving on a group manifold) is expected to arise in the low-energy limit of several disordered theories and tensor models that have a global symmetry $G$;  the argument primarily relies on the fact that the resulting effective theory needs to have an $SL(2, \mR)\times G$ symmetry \cite{Sachdev:2015efa, Davison:2016ngz, Gross:2016kjj, Fu:2016vas, Narayan:2017qtw, Yoon:2017nig, Narayan:2017hvh, Klebanov:2018nfp, Liu:2019niv}. Nevertheless, it would be interesting to understand whether one can derive the potential and coupling to the Schwarzian theory that we have encountered for the particle moving on the group manifold $G$ directly from a specific disordered theory or a particular tensor model.\footnote{We thank G.~Tarnopolskiy for useful discussions about this direction.} 

In parallel to our analysis of surfaces with disk topology, we also computed the same correlators in the genus expansion, when considering orientable surfaces with an arbitrary number of boundaries. For all such correlators, we have found two equivalent matrix integral descriptions. In both, the starting point was to consider the matrix integral description of the $(2, p)$ minimal string, in the $p \to \infty$ limit. In the first matrix integral description, one promotes the matrix elements $H_{i, j}$ from complex numbers to complex group algebra elements in $\mathbb C[G]$. Keeping the couplings in the associated matrix model to be the same, but redefining the traces appearing in the model, after some algebraic manipulation, we obtain the second equivalent matrix integral description. 

This description is given by a collection of random matrix ensembles, where each matrix is Hermitian,  is associated to a unitary irreducible representation $R$ of $G$, and has its size is simply proportional to the dimension of the irreducible representation $R$. Using this latter matrix description, we have found that the genus expansion of correlators in the gravitational gauge theory on surfaces with $n$ boundaries matches the expectation value of $n$ operator insertions $e^{-\b H}$ in the matrix integral ensemble. Depending on which operators we include in the correlator on the gravitational side, we have shown that one can construct the appropriate trace for the operator $e^{-\b H}$ on the matrix integral side.

As discussed in \cite{Saad:2019lba}, the random matrix statistics encountered when studying pure JT gravity only qualitatively describe some aspects of the SYK model. Similarly, the random matrix ensembles that we have encountered when analyzing the gravitational gauge theory reproduce the same features of SYK models with global symmetries but do not adequately describe the disordered theory. One example in which the matrix integral provides a qualitative description is for the ramp saddle point encountered in SYK \cite{Saad:2018bqo} which was found to be analogous to the double trumpet configuration from pure JT gravity. When studying an SYK model with global symmetry, one expects similar ramp saddle points in each representation sector; as can be inferred from our results, the contribution of each representation sector to the double trumpet configuration in the gravitational gauge theory indeed reproduces the linearly growing ``ramp'' contribution to the spectral form factor. 

Besides considering correlators in the gravitational gauge theory defined on orientable surfaces, we have also briefly discussed the computation of the partition function of the theory on both orientable and unorientable surfaces. In this case, we have recovered the partition function from a GOE-like matrix integral with matrix elements in $\mathbb R[G]$. It would, of course, be interesting to analyze the same more general correlators as those studied in this paper, both in this gravitational gauge theory and in its associated random matrix ensemble. However, as mentioned in \cite{Stanford:2019vob},  when studying unorientable surfaces, all computations are limited by the logarithmic divergence encountered due to small cross-cap geometries.

One significant development in the study of 2d Yang-Mills has been its reformulation as a theory of strings \cite{Gross:1993hu, Gross:1993yt}. Furthermore, as presented in \cite{Saad:2018bqo} and \cite{Stanford:2019vob}, and as reviewed in this paper, the genus expansion of pure JT gravity is related to the matrix integral obtained from the $(2, p)$ minimal string, in the $p \to \oo$ limit. Consequently, it is natural to ask whether, when coupling 2d Yang-Mills to JT gravity, it is possible to rewrite the partition function or the diffeomorphism and gauge-invariant correlators in this theory as a sum over the branched covers considered in \cite{Gross:1993hu, Gross:1993yt}.\footnote{In fact, investigating the behavior of 2d Yang-Mills coupled to 2d quantum gravity is an open research direction suggested in the review \cite{Cordes:1994fc}.}

Regarding the classification of all diffeomorphism and gauge-invariant operators in the gravitational gauge theory and the computation of their associated correlators, we have managed to understand all local observables coming from pure Yang-Mills theory and have computed their expectation values. For non-local operators we have defined a set of quark worldline operators which generalize the Wilson lines from pure Yang-Mills theory. The purpose of this generalization was to obtain observables which are diffeomorphism invariant. We have, however, only studied these operators when considering worldlines that are boundary anchored and are smoothly contractible to a segment on the boundary. It would, of course, be interesting to understand how to perform computations for more general topological configurations. This brings up two problems. The first is to determine a way to assign weights in the path integral to the different homotopy classes in which the contours of the boundary anchored worldlines can belong. Such an assignment is well known for worldline path integrals in quantum mechanics \cite{laidlaw1971feynman}, however, considering worldlines in the genus expansion in $2d$ quantum gravity adds a layer of complexity. This is because the first homotopy group for surfaces with different genera is, of course, different. The second problem with studying worldline path integrals with topologically non-trivial contours is that for certain homotopy classes such contours are necessarily self-intersecting.\footnote{For instance, consider a closed curve on the torus $\cM_{1,0} $, for which $\pi_1(\cM_{1, 0}) = \mathbb Z\times \mathbb Z$. Consider a curve that winds $p$ times around one cycle and $q$ times around the other with $(p, q)\in \pi_1(\cM_{1, 0}) $. Then the minimum number of self-intersections for such a curve is $\gcd(p, \, q)-1$ \cite{de2009contractibility} for $p, q>0$. } Consequently, one needs to develop a bookkeeping device for tracking the $6j$-symbols associated with each intersection that would necessarily appear in the genus expansion. 

A further research direction that would lead to a better understanding of quark worldline operators would be to compute their associated correlators beyond the weak gauge coupling limit. Perhaps one can use diffeomorphism invariance to simplify this computation. For instance, by working in a diffeomorphism gauge where the metric determinant $\sqrt{g}\,$ is concentrated around the boundary and is almost vanishing in the bulk, it might be possible to reduce the computation at arbitrary gauge coupling to the computation at weak gauge coupling in the presence of a boundary condition changing defect.

Finally, going back to our introductory motivation of studying near-extremal black holes by using the gravitational gauge theory as an effective model for the the near-horizon region \cite{Almheiri:2016fws, Anninos:2017cnw, Sarosi:2017ykf, Nayak:2018qej, Moitra:2018jqs, Hadar:2018izi, Moitra:2019bub, Sachdev:2019bjn}, it would be interesting to understand  the consequences of the results in this paper for the low-energy behavior of such black holes. As mentioned in the introduction, including gauge degrees of freedom allows the analysis of this behavior beyond the S-wave sector. In that regard, a semi-classical analysis in the presence of gauge fields has already been performed in \cite{Moitra:2018jqs}, where it was found that the contribution of the massless gauge fields was on par with the contribution of the gravitational degrees of freedom. This can be seen from our disk computation when setting mixed boundary conditions for the gauge field: in that case the contribution of the gauge fields is not suppressed in the cut-off parameter, $\epsilon$, or in the value of the dilaton at extremality, $\Phi_0$, and each representation, $R$, contributes to the partition function. Beyond the partition function, our analysis of quark worldline correlators is also relevant for understanding the low-energy behavior of near-extremal black holes. For instance for Reissner-Nordstr\"{o}m  black holes in $AdS_4$, the charged lines under the gauge group $G= \text{SO}(3)$ or $\text{SU}(2)$, represent particles that have angular momentum on the internal space, $S^2$. Thus, the quark worldline correlators discussed in this paper are related to scatterings of spinning particles close to the near-horizon region. Such particles are created by the (infinite number of) Kaluza-Klein fields appearing from the dimensional reduction along the internal space. 
While the contribution of each such field was shown to be subdominant to the gravitational and gauge degrees of freedom, it is possible that when considering the cumulative effect of the massive Kaluza-Klein states, which arise from the higher partial waves on $S^2$, the effect can be significant \cite{Moitra:2018jqs}. Perhaps the analysis of quark worldline correlators in the gravitational gauge theory with arbitrary gauge coupling could shed some light in that direction. We hope to report on progress in this direction soon.

\section*{Acknowledgements}
I am especially thankful to Daniel Kapec, Raghu Mahajan, and Herman Verlinde for many useful comments throughout several stages of this project. I am also especially thankful to Jorrit Kruthoff for comments about the draft. I also thank  Petr Kravchuk, Igor Klebanov, Silviu Pufu, Juan Maldacena, Mark Mezei, Grigory Tarnopolskiy and Edward Witten for valuable discussions.  This research is supported in part by the US NSF under Grant No. PHY-1820651.

\bibliographystyle{ssg}
\bibliography{Biblio}

\end{document}